\documentclass[
	reprint,
	amsmath,
	amssymb,
	aps,
	prapplied,
]{revtex4-2}

\usepackage{amsmath}
\usepackage{amssymb}
\usepackage{amsthm}      
\usepackage{bbm}         
\usepackage{enumitem}
\usepackage{graphicx}
\usepackage{amsthm}      
\usepackage{tikz}        
\usepackage{fp}          

\usepackage{mathrsfs}    
\usepackage{mathtools}   
\usepackage{subcaption}  
\usepackage{xparse}

\usepackage[
	colorlinks,
	final,
	hyperfootnotes = false,
	linkcolor      = blue,
	citecolor      = blue]{hyperref}

\usepackage[capitalize]{cleveref}  

\def\gcnspaper{}

\usetikzlibrary{arrows}                     
\usetikzlibrary{patterns}                   
\usetikzlibrary{decorations.pathreplacing}  
\usetikzlibrary{decorations.markings}       
\usetikzlibrary{calc}                       
\usetikzlibrary{matrix}                     

\makeatletter
\usepackage{pgfkeys}
\pgfkeys{/generic-reusable/.unknown/.code={%
	\expandafter\edef\csname gr-arg-\pgfkeyscurrentkey\endcsname{\unexpanded\expandafter{\pgfkeyscurrentvalue}}%
}}

\def\gr@markincluded#1{\global\expandafter\let\csname gr-file-#1-alreadyincluded\endcsname\empty}
\def\gr@ifincluded#1{\@ifundefined{gr-file-#1-alreadyincluded}}

\def\OneTimeOnly{%
	\gr@ifincluded{\gr@filename}%
	{\@firstofone}%
	{\@gobble}%
}

\def\OnlyFor#1{%
	\gr@OnlyFor@#1,\gr@nil
}

\def\gr@OnlyFor@#1,{%
	\def\gr@temp{#1}%
	\def\next{\@ifnextchar\gr@nil{\@gobbletwo}{\gr@OnlyFor@}}%
	\ifx\gr@temp\Instance
	\let\next\gr@OnlyFor@dobody 
	\fi
	\next
}

\def\GetKeyValue#1{\csname gr-arg-/generic-reusable/#1\endcsname}

\def\gr@OnlyFor@dobody#1\gr@nil{\@firstofone}

\def\InsertModule#1#2{%
	\@ifnextchar[{\gr@InsertModule@{#1}{#2}}{\gr@InsertModule@{#1}{#2}[]}%
}
\def\gr@InsertModule@#1#2[#3]{%
	\pgfqkeys{/generic-reusable}{#3}
	\def\Instance{#2}%
\def\gr@filename{#1}%

\newcommand {\gcfrefdimincm} {5}

\begin{tikzpicture}[x=\gcfrefdimincm cm, y=\gcfrefdimincm cm]
	\newcommand {\cvswd}    {1.2}   
	\newcommand {\cvsht}    {1.0}   
	\newcommand {\codespcx} {0.28}  
	\newcommand {\codespcy} {0.72}  
	\newcommand {\nzcomx}   {0.75}  
	\newcommand {\nzcomy}   {0.25}  
	\newcommand {\neutralx} {0.6}   
	\newcommand {\neutraly} {0.4}   

	\definecolor {zeroedregion} {rgb} {0.627,0.627,0.627}  
	\definecolor {ncdividercol} {rgb} {0.38,0.38,0.38}     

	\tikzstyle {matrixborder} = [line width = 1.6pt]  

	\tikzstyle {zeroed} = [  
		line width    = 1.2pt,
		color         = zeroedregion,
		fill          = zeroedregion,
		pattern       = north east lines,
		pattern color = zeroedregion]

	\tikzstyle {ncdivider} = [  
		line width = 1.6pt,
		color      = ncdividercol]

	\tikzstyle {ncdividerfaded} = [  
		opacity    = 0.3,
		line width = 1.6pt,
		color      = ncdividercol]

	\tikzstyle {csdivider} = [  
		line width   = 0.8pt,
		dash pattern = on 2pt off 2pt]

	\tikzstyle {projbrace} = [  
		decorate,
		decoration = {brace, amplitude = 8pt, mirror},
		xshift     = 4pt]

	\tikzstyle {projnode} = [  
		black,
		midway,
		anchor = west,
		xshift = 8pt]

	\tikzstyle {nsdivider} = [  
		blue,
		line width   = 1.2pt,
		dash pattern = on 4pt off 2pt]

	\newcommand {\drawmatrixborder}  
		{\draw [matrixborder] (0,0) rectangle (1,1);}

	\newcommand {\drawncdivider} [1] {  
		\draw [##1] (\nzcomx,0) -- (\nzcomx,1);
		\draw [##1] (0,\nzcomy) -- (1,\nzcomy);
	}

	\newcommand {\drawcsdivider} {  
		\draw [csdivider] (\codespcx,0) -- (\codespcx,1);
		\draw [csdivider] (0,\codespcy) -- (1,\codespcy);
	}

	\newcommand {\drawneutral} {  
		\fill [zeroed] (\neutralx,\neutraly) rectangle (1,1);
		\fill [zeroed] (\neutralx,\neutraly) rectangle (0,0);

		\draw [nsdivider] (0,1) rectangle (\neutralx,\neutraly);
		\draw [nsdivider] (1,0) rectangle (\neutralx,\neutraly);
	}

	\newcommand {\labelprojection} [3] {  
		\draw [projbrace] (1,{##1}) -- (1,{##2}) node [projnode] {##3};
	}

	\newcommand {\labelblock} [5] {  
		\draw ({##1/2+##2/2},{##3/2+##4/2}) node[anchor = center] {##5};
	}

	\useasboundingbox (0,0) rectangle (\cvswd,\cvsht);

	\OnlyFor{CQF} {
		\drawmatrixborder
		\drawncdivider {ncdivider}

		\labelprojection {\nzcomy}{1}       {$\NCproj$}
		\labelprojection {0}      {\nzcomy} {$\NCprojperp$}

		\labelblock {0}      {\nzcomx}{\nzcomy}{1}       {$\ump\GQaa$}
		\labelblock {\nzcomx}{1}      {\nzcomy}{1}       {$\GQab$}
		\labelblock {0}      {\nzcomx}{0}      {\nzcomy} {$-\GQab^\dagger$}
		\labelblock {\nzcomx}{1}      {0}      {\nzcomy} {$\GQbb$}
	}

	\OnlyFor{C3F} {
		\drawmatrixborder
		\drawncdivider {ncdivider}
		\drawcsdivider

		\labelprojection {\codespcy}{1}         {$\CSproj$}
		\labelprojection {\nzcomy}  {\codespcy} {$\NCproj - \CSproj$}
		\labelprojection {0}        {\nzcomy}   {$\NCprojperp$}

		\labelblock {0}        {\codespcx}{\codespcy}{1}         {$\ump\Gaa$}
		\labelblock {\codespcx}{\nzcomx}  {\codespcy}{1}         {$\Gab$}
		\labelblock {\nzcomx}  {1}        {\codespcy}{1}         {$\Gac$}
		\labelblock {0}        {\codespcx}{\nzcomy}  {\codespcy} {$-\Gab^\dagger$}
		\labelblock {\codespcx}{\nzcomx}  {\nzcomy}  {\codespcy} {$\Gbb$}
		\labelblock {\nzcomx}  {1}        {\nzcomy}  {\codespcy} {$\Gbc$}
		\labelblock {0}        {\codespcx}{0}        {\nzcomy}   {$-\Gac^\dagger$}
		\labelblock {\codespcx}{\nzcomx}  {0}        {\nzcomy}   {$-\Gbc^\dagger$}
		\labelblock {\nzcomx}  {1}        {0}        {\nzcomy}   {$\Gcc$}
	}

	\OnlyFor{G-C3F} {
		\drawneutral
		\drawmatrixborder
		\drawncdivider {ncdividerfaded}
		\drawcsdivider

		\labelprojection {\codespcy}{1}         {$\CSproj$}
		\labelprojection {\neutraly}{\codespcy} {$\Pproj - \CSproj$}
		\labelprojection {0}        {\neutraly} {$P_{\perp}$}

		\labelblock {0}        {\codespcx}{\codespcy}{1}         {$\ump\Gaa$}
		\labelblock {\codespcx}{\neutralx}{\codespcy}{1}         {$\Gab$}
		\labelblock {0}        {\codespcx}{\neutraly}{\codespcy} {$-\Gab^\dagger$}
		\labelblock {\codespcx}{\neutralx}{\neutraly}{\codespcy} {$\Gbb$}
		\labelblock {\neutralx}{1}        {0}        {\neutraly} {$\Gcc$}
	}
\end{tikzpicture}

\relax
\gr@markincluded{#1}
}
\makeatother

\crefformat{equation}{(#2#1#3)}
\newcommand* {\eqn}   [1] {\cref{#1}}
\newcommand* {\eqns}  [1] {\cref{#1}}

\newcommand* {\eqnrg} [2] {\crefrange{#1}{#2}}

\newcommand {\FvM}   [2] [\Fld] {\mathcal{M}_{#2}({#1})}   

\newcommand {\Rl}               {\mathbb{R}}               
\newcommand {\Rp}               {\mathbb{R}^+}             
\newcommand {\Rnn}              {\mathbb{R}^{0,+}}         
\newcommand {\Rn}           [1] {\mathbb{R}^{#1}}          
\newcommand {\RM}           [1] {\FvM[\Rl]{#1}}            
\newcommand {\RMG}          [2] {\FvM[\Rl]{#1\times #2}}   

\NewDocumentCommand {\Rne} {om} {\vec{\mathbbm{e}}\IfNoValueF {#1} {\sp{#1}}\sb{#2}}
	\providecommand {\Rne} [2] [] {}  

\NewDocumentCommand {\RMe} {omm} {\mathbbm{e}\IfNoValueF {#1} {\sp{#1}}\sb{#2,#3}}
	\providecommand {\RMe} [3] [] {}  

\NewDocumentCommand {\Z}    {o} {\IfNoValueTF {#1} {\mathbb{Z}} {\mathbb{Z}_{\geq #1}}}
	\providecommand {\Z} [1] [] {}  
\newcommand {\Zp}               {\mathbb{N}}           
\newcommand {\Znn}              {\mathbb{Z}_{\geq 0}}  

\newcommand {\Cx}               {\mathbb{C}}              
\newcommand {\Cn}           [1] {\mathbb{C}^{#1}}         
\newcommand {\CM}           [1] {\FvM[\Cx]{#1}}           

\NewDocumentCommand {\Sn} {sm} {\IfBooleanTF{#1}{\overline{\mathbb{S}}^{#2}}{\mathbb{S}^{#2}}}
	\providecommand {\Sn} [1] {\mathbb{S}^{#1}}  

\newcommand {\eye}          [1] {{\mathbbm 1}_{#1}}  
\newcommand {\ZM}           [1] {0_{#1}}             
\newcommand {\ZMG}          [2] {\ZM{#1\times #2}}   
\newcommand {\ZV}           [1] {\vec{0}_{#1}}       

\newcommand {\OneTo}        [1] {[\![{#1}]\!]}       
\newcommand {\IntRg}        [2] {[\![{#1},{#2}]\!]}  
\newcommand {\inoneto}  [2] [1] {= #1,\dotsc,#2}     
\newcommand {\nnrg}         [1] {\in\OneTo{#1}}      

\newcommand {\expe}         [1] {\txt{e}^{#1}}       

\newcommand {\frowny}       [1]  
	{\smash{\overset{\lower0.5em\hbox{$\smash{\scriptscriptstyle\frown}$}}{#1}}}
\newcommand {\overscript}   [2]  
	{\smash{\overset{\lower0em\hbox{$\smash{\scriptstyle{#2}}$}}{#1}}}

\newcommand {\nsub}         [1] {n_\txt{#1}}                         
\newcommand {\lil}          [1] {{\scriptscriptstyle\txt{#1}}}       
\newcommand {\lilsf}        [1] {{\scriptscriptstyle\mathsf{#1}}}    

\newcommand {\isdefas}          {\triangleq}                         
\newcommand {\adj}          [1] {#1^\dagger}                         
\newcommand {\adjv}         [1] {#1^{\,\dagger}}                     
\newcommand {\tpose}        [1] {#1^{\lilsf T}}                      
\newcommand {\bigO}         [1] {\mathcal{O}\left(#1\right)}         
\newcommand {\Forall}           {\mathchoice{\;}{}{}{}\;\forall\;}   
\newcommand {\Foreach}          {\forall\;}                          
\newcommand {\Mapset}       [2] {\operatorname{Map}({#1}\xp {#2})}   
\newcommand {\st}               {\mid}                               
\newcommand {\sttext}           {\;\;\text{s.t.}\;\;}                
\newcommand {\xp}               {;\;}                                
\newcommand {\ket}          [1] {\left| #1\right\rangle}             
\newcommand {\fxr}          [2] {\left.#1\right|_{#2}}               
\newcommand {\zdots}                                                 
	{\vdots\,\raisebox{0.15em}{$\scriptstyle 0$}}

\NewDocumentCommand {\wapproxtx} {e{_^}}                             
	{\tx w\IfNoValueF {#1} {\sb{\txi{#1}}}\IfNoValueF {#2} {\sp{\txi{#2}}}}
	\providecommand {\wapproxtx} {\tx w}  

\newcommand {\stacked} [2]                
	{\begin{array}{c} {\scriptstyle #1} \\ {\scriptstyle #2}\end{array}}
\newcommand {\LGivenR} [2] {{
	\renewcommand{\arraystretch}{0.75}%
	\setlength{\arraycolsep}{1mm}%
	\begin{array}{r|l}%
		\begin{array}{r} #1 \end{array} & \begin{array}{l} #2 \end{array}%
	\end{array}%
}}

\newcommand {\ifother} [3] {             
	\begin{cases}
		#1 & \txt{if} \; #2 \\
		#3 & \txt{otherwise}
	\end{cases}
}

\newcommand {\setvar}       [1] {\mathsf{#1}}    
\newcommand {\tuplevar}     [1] {\mathsf{#1}}    
\newcommand {\setbld}       [2]                  
	{\mathchoice{\left\{\mskip-6mu\LGivenR{#1}{#2}\mskip-6mu\right\}}{\{#1\st #2\}}{}{}}
\newcommand {\setbldd}      [2]                  
	{\left\{\mskip-6mu\LGivenR{#1}{#2}\mskip-6mu\right\}}
\NewDocumentCommand {\seti} {smm}  
	{\{#3\}\IfBooleanT{#1}{\sb{#2}}}

\ifx\gcnspaper\undefined
	\let\spacevarfmt\mathscr
\else
	\let\spacevarfmt\mathcal
\fi
\NewDocumentCommand {\spacevar} {smo}  
	{\spacevarfmt{#2}%
		\IfBooleanT{#1}{\sp{\prime}}%
		\IfNoValueF{#3}{\sb{#3}}}

\newcommand {\ump} {\hphantom{-{}}}  

\NewDocumentCommand {\pvec} {sm}                      
	{\IfBooleanTF {#1} {#2} {\vec{#2}}\sp{\,\prime}}
	\providecommand {\pvec} [1] {\vec{#1}\sp{\,\prime}}  

\newcommand {\speqinset}        {\hspace{10mm}}  
\newcommand {\speqsidebyside}   {\hspace{6mm}}   
\newcommand {\sppredtostmt}     {\hspace{3mm}}   

\newcommand {\repart}           {\operatorname{Re}}          
\newcommand {\im}               {\operatorname{im}}          
\newcommand {\tr}               {\operatorname{tr}}          
\newcommand {\nullspc}          {\operatorname{null}}        
\newcommand {\rank}             {\operatorname{rank}}        
\newcommand {\spanof}       [1]                              
	{\operatorname{span}\mathchoice{\left\{{#1}\right\}}{\{#1\}}{}{}}
\newcommand {\spanover}     [2]                              
	{\mathop{\operatorname{span}}\limits_{#1}\left\{ {#2}\right\}}
\newcommand {\maxover}      [1] {\max\limits_{#1}}           
\newcommand {\inprod}       [2] {\langle{#1 | #2}\rangle}    

\newcommand {\vtor}         [1] {\operatorname{vec}\mathchoice{\left({#1}\right)}{(#1)}{(#1)}{(#1)}}
\newcommand {\vtorinv}      [1] {\operatorname{vec}^{-1}\mathchoice{\left({#1}\right)}{(#1)}{(#1)}{(#1)}}

\newcommand {\convop}  {\operatorname{conv}}  
\newcommand {\convclr}                        
	{\vphantom{\underset{\mathring{X}\shortto \hat Y}{\convop}}}

\NewDocumentCommand {\conv} {omm} {  
	\smash{\underset{#2\shortto #3}
		{\IfNoValueTF {#1} {\convop} {\overset{\raisebox{0.15em}{$\scriptstyle #1$}}{\convop}}}}
}
	\providecommand {\conv} [3] [] {\smash{\underset{#2\shortto #3}{\overset{#1}{\convop}}}}  

\ExplSyntaxOn
\NewDocumentCommand {\norm} {som} {
	\IfBooleanTF {#1} {\left\|#3\right\|} {\|#3\|}%
	\IfNoValueF {#2} {\sb{%
		\str_case:nnF {#2} {
			{F}    {\txt F}
			{inf}  {\infty}
			{L1}   {L^1}
			{L2}   {L^2}
			{Linf} {L^\infty}
		}
		{#2}%
	}}%
}
\ExplSyntaxOff
	\providecommand {\norm} [2] [] {\|#2\|_{#1}}  

\newcommand {\stdmat}    [2] [c] {\left[\begin{array}{*{20}{#1}}#2\end{array}\right]}  
\newcommand {\compactcv} [1]                                                           
	{{\tpose{\begin{bmatrix} #1 \end{bmatrix}}}}

\newcommand {\stdmatx} [3] [\x] {\stdmat{  
	#1 11    & #1 12    & \cdots & #1 1{#3} \\
	#1 21    & #1 22    & \cdots & #1 2{#3} \\
	\vdots   & \vdots   & \ddots & \vdots   \\
	#1 {#2}1 & #1 {#2}2 & \cdots & #1 {#2}{#3}}}

\newcommand {\stdvecx} [2] [\x] {\stdmat{  
	#1 1   \\
	#1 2   \\
	\vdots \\
	#1 {#2}}}

\newcommand {\compactcvx} [2] [\x]  
	{\compactcv{ #1 1 & #1 2 & \cdots & #1 {#2} }}

\newcommand {\tx}           [1] {\mathbf{\lowercase{#1}}}       
\newcommand {\txgreek}      [1] {\boldsymbol{\lowergreek{#1}}}  
\newcommand {\txi}          [1] {({#1})}                        
\newcommand {\colabel}      [1] {\langle{#1}\rangle}            
\newcommand {\coordinal}    [1] {[{#1}]}                        

\newcommand {\txcx}         [1] {\sb{#1}}                       
\newcommand {\txCx}         [1] {\sp{#1}}                       

\NewDocumentCommand {\txC} {om}  
	{\sp{\txi{#2}}\IfNoValueF {#1} {\sb{#1}}}
	\providecommand {\txC} [2] [] {\sp{\txi{#2}}}  

\NewDocumentCommand {\txc} {om}  
	{\IfNoValueTF {#1} {\sb{\txi{#2}}} {\sb{#1,\txi{#2}}}}
	\providecommand {\txc} [2] [] {\sb{\txi{#2}}}  

\NewDocumentCommand {\txCc} {omm}  
	{\sp{\txi{#2}}\IfNoValueTF {#1} {\sb{\txi{#3}}} {\sb{#1,\txi{#3}}}}
	\providecommand {\txCc} [3] [] {\sp{\txi{#2}}\sb{\txi{#3}}}  

\NewDocumentCommand {\txlbl} {me{_^}} {  
	\left[#1\right]
	\IfNoValueF {#2} {\txcx{\colabel{#2}}}
	\IfNoValueF {#3} {\txCx{\colabel{#3}}}
}
	\providecommand {\txlbl} [1] {\left[#1\right]}  

\NewDocumentCommand {\txdelta} {e{_^}} {  
	\txgreek\delta
	\IfNoValueF {#1} {\txc{#1}}
	\IfNoValueF {#2} {\txC{#2}}
}
	\providecommand {\txdelta} {}  

\NewDocumentCommand {\txpermords}{me{_^}} {
	#1
	\IfNoValueF {#2} {\txcx{\coordinal{#2}}}
	\IfNoValueF {#3} {\txCx{\coordinal{#3}}}
}
	\providecommand {\txpermords} [1] {#1}

\newcommand {\metaset}          {\setvar{S}}       
\newcommand {\metasetn}     [1] {\setvar{S}_{#1}}  
\newcommand {\metasetalt}       {\setvar{T}}       

\newcommand {\liealg}       [1] {\mathfrak{#1}}                 
\newcommand {\liegrp}       [1] {\txt{#1}}                      
\newcommand {\liecls}       [1]                                 
	{\mathchoice{\mathscr{L}\left[{#1}\right]}{\mathscr{L}[{#1}]}{}{}}
\newcommand {\liesu}        [1] {\liealg{su}({#1})}             
\newcommand {\lieu}         [1] {\liealg{u}({#1})}              
\newcommand {\liesp}        [1] {\liealg{sp}({#1})}             
\newcommand {\lieglc}       [1] {\liealg{gl}({#1},\mathbb{C})}  
\newcommand {\lieSU}        [1] {\liegrp{SU}({#1})}             
\newcommand {\lieU}         [1] {\liegrp{U}({#1})}              
\newcommand {\lieSp}        [1] {\liegrp{Sp}({#1})}             

\newcommand {\cstarcls}     [1]                                 
	{\mathchoice{\mathscr{C}^*\left[{#1}\right]}{\mathscr{C}^*[{#1}]}{}{}}

\newcommand {\ers}          [2] {\operatorname{ers}_{#1}{#2}}   
\newcommand {\ext}          [2] {\operatorname{ext}_{#1}{#2}}   

\newcommand {\txt} [1] {\text{\textnormal{#1}}}  

\newcommand {\lowergreek} [1] {\begingroup  
	\let\Gamma\gamma
	\let\Delta\delta
	\let\Lambda\lambda
	\let\Theta\theta
	\let\Pi\pi
	\let\Sigma\sigma
	\let\Upsilon\upsilon
	\let\Phi\phi
	\let\Psi\psi
	\let\Omega\omega
	#1
	\endgroup}

\makeatletter  
\ifx\setspacecompat\undefined
\setbox0\hbox{$\xdef\scriptratio{\strip@pt\dimexpr\numexpr(\sf@size*65536)/\f@size sp}$}
	\newcommand {\shortto} [1] [3pt] {{%
		\hbox{\rule[\scriptratio\dimexpr\fontdimen22\textfont2-.2pt\relax]
			{\scriptratio\dimexpr#1\relax}{\scriptratio\dimexpr.4pt\relax}}%
		\mkern-4mu\hbox{\let\f@size\sf@size\usefont{U}{lasy}{m}{n}\symbol{41}}}}
\else
	\DeclareRobustCommand {\shortto} {\mathrel{\mathpalette\short@to\relax}}
	\newcommand {\short@to} [2] {%
		\mkern2mu
		\clipbox{{.5\width} 0 0 0}{$\m@th#1\vphantom{+}{\shortrightarrow}$}%
	}
\fi
\makeatother

\newcommand {\Bord}      {\bar d}            
\ifx\gcnspaper\undefined
	\newcommand {\Bdim}  {N}                 
\else
	\newcommand {\Bdim}  {d}                 
\fi

\newcommand {\Ftx}       {\tx{f}}            
\NewDocumentCommand {\Ftxi} {smmm}           
	{\IfBooleanTF {#1} {\Ftx\txCc{#2#3}{#4}} {\Ftx\txCc{#3}{#2#4}}}
	\providecommand {\Ftxi} [3] {\Ftx\txCc{#2}{#1#3}}  

\newcommand {\Ftxp}                          
	{\prescript{\backprime}{}{\tx{f}}}
\NewDocumentCommand {\Ftxpi} {smmm}          
	{\IfBooleanTF {#1} {\Ftxp\txCc{#2#3}{#4}} {\Ftxp\txCc{#3}{#2#4}}}
	\providecommand {\Ftxpi} [3] {\Ftxp\txCc{#2}{#1#3}}  

\ifx\gcnspaper\undefined
	\newcommand {\Hilspc} {\mathcal{H}}      
\else
	\newcommand {\Hilspc} {\mathscr{H}}      
\fi

\newcommand {\LindH}      {\txt{L}_\txt{H}}      
\newcommand {\LindD}      {\txt{L}_\txt{D}}      

\newcommand {\Etx} {\tx{E}}  

\NewDocumentCommand {\Etxi} {smmm}  
	{\IfBooleanTF {#1} {\Etx\txCc{#2#3}{#4}} {\Etx\txCc{#3}{#2#4}}}
	\providecommand {\Etxi} [3] {\Etx\txCc{#2}{#1#3}}  

\newcommand {\RtoV}    {\conv{\rho}{\vec v}}  
\newcommand {\HtoG}    {\conv{\hat H}{G}}     

\newcommand {\dyson}  [2] {\mathscr{D}_{#1}\{ #2\}}   

\newcommand {\BonH}     {\spacevar{B}}  
\newcommand {\HonH}     {\spacevar{H}}  
\newcommand {\DonH}     {\spacevar{S}}  

\newcommand {\lieham} {\liealg{h}}  

\ExplSyntaxOn
\newcommand {\Bell} [1] {
	\str_case:nnF {#1} {
		{10+01} {\ensuremath{\ket{\Psi^+}}}
		{01+10} {\ensuremath{\ket{\Psi^+}}}
		{00+11} {\ensuremath{\ket{\Phi^+}}}
		{11+00} {\ensuremath{\ket{\Phi^+}}}
		{10-01} {\ensuremath{\ket{\Psi^-}}}
		{01-10} {\ensuremath{\ket{\Psi^-}}}
		{00-11} {\ensuremath{\ket{\Phi^-}}}
		{11-00} {\ensuremath{\ket{\Phi^-}}}
	}
	{\ensuremath{\ket{\blacksquare}}}
}
\ExplSyntaxOff

\NewDocumentCommand {\pauli} {mo} {  
	\hat\sigma
	\IfNoValueTF {#2} {\sb{\txt{#1}}} {\sb{\txt{#1},#2}}
}
	\providecommand {\pauli} [2] [] {\sigma_{\txt{#1},#2}}  

\newcommand {\Pauli} {\Sigma}  

\newcommand {\CS}        {\tuplevar{C}}             
\newcommand {\csp}       {{\scriptscriptstyle\CS}}  
\newcommand {\CPtuple}   {(\CS,\Pproj)}             
\newcommand {\CSHilspc}  {\Hilspc_{\csp}}           
\newcommand {\CSU}       {U_{\csp}}

\newcommand {\CSbastx}   [1][\csp] {\prescript{}{\raisebox{0.11em}{${\scriptstyle #1}$}}{\tx{e}}}

\newcommand {\kord}      [1][k] {\bar{d}_{#1}}      
\newcommand {\kordmax}          {\kord[\txt{max}]}  

\NewDocumentCommand {\BonHC} {s}  
	{\IfBooleanTF{#1}{\spacevar*{B}[\csp]}{\spacevar{B}[\csp]}}
\NewDocumentCommand {\HonHC} {s}  
	{\IfBooleanTF{#1}{\spacevar*{H}[\csp]}{\spacevar{H}[\csp]}}
\NewDocumentCommand {\DonHC} {s}  
	{\IfBooleanTF{#1}{\spacevar*{S}[\csp]}{\spacevar{S}[\csp]}}
\newcommand {\VonHC}            {\setvar{V}_{\csp}}
\newcommand {\VonPproj}         {\setvar{V}_{\Pproj}}  
\newcommand {\GonHC}            {\setvar{G}_{\csp}}
\newcommand {\GonPproj}         {\setvar{G}_{\Pproj}}

\newcommand {\CSproj} [1] [\csp] {\Pi_{#1}}

\newcommand {\CSdim}             {d_{\csp}}                     
\newcommand {\CSord}             {\bar{d}_{\csp}}               
\newcommand {\NC}                {\mathcal{A}}                  
\newcommand {\NCproj}            {\Pi_\NC}                      
\newcommand {\NCprojperp}        {\Pi_\NC^\perp}                
\newcommand {\NCdim}             {n_\NC}                        
\newcommand {\NCperpdim}         {n_\NC^\perp}                  
\newcommand {\Pproj}             {P}                            
\newcommand {\Pproja}            {\frowny{P}}
\newcommand {\Pprojperp}         {P_\perp}                      
\newcommand {\Pdim}              {\nsub{P}}
\newcommand {\Pperpdim}          {\nsub{P}^\perp}               
\newcommand {\PnotCSdim}         {\nsub{ext}}                   

\newcommand {\kproj}      [1][k] {\Pi_{#1}}                      

\newcommand {\zu}         {\vec{z}_{\txt{u}}}       
\newcommand {\Zu}         {Z_{\txt{u}}}             

\newcommand {\znu}        {\vec{n}_{\txt{u}}}                     
\newcommand {\zNu}        {N_{\txt{u}}}                           

\newcommand {\CSRtoV}            {\conv{\sigma}{\vec v}}

\NewDocumentCommand {\CSHtoG} {o} {\IfNoValueTF {#1} {\conv{\hat J}{G}} {\conv[#1]{\hat J}{G}}}
	\providecommand {\CSHtoG} [1] [] {\conv[#1]{\hat J}{G}}  
\NewDocumentCommand {\CSGtoH} {o} {\IfNoValueTF {#1} {\conv{G}{\hat J}} {\conv[#1]{G}{\hat J}}}
	\providecommand {\CSGtoH} [1] [] {\conv[#1]{G}{\hat J}}  

\newcommand {\liehamCS} {\liealg{h}_{\csp}}  

\NewDocumentCommand {\Lp} {o} {\Theta\IfNoValueF {#1} {\sb{#1}}}
	\providecommand {\Lp} [1] [] {\Theta_{#1}}  

\NewDocumentCommand {\lp} {o} {\hat\theta\IfNoValueF {#1} {\sb{#1}}}
	\providecommand {\lp} [1] [] {\theta_{#1}}  

\NewDocumentCommand {\lcp} {o} {\hat\Pi\IfNoValueF {#1} {\sb{#1}}}
	\providecommand {\lcp} [1] [] {\hat\Pi_{#1}}  

\NewDocumentCommand {\Mp} {o} {\Theta\IfNoValueTF {#1} {\sb{\txt{mp}}} {\sb{\txt{mp},#1}}}  
	\providecommand {\Mp} [1] [] {\Theta_{\txt{mp},#1}}  

\NewDocumentCommand {\lprv} {o} {\vec{p}\IfNoValueF {#1} {\sb{#1}}}
	\providecommand {\lprv} [1] [] {\vec{p}_{#1}}  

\NewDocumentCommand {\lprvphase} {o} {\varphi\IfNoValueF {#1} {(#1)}}
	\providecommand {\lprvphase} [1] [] {\varphi(#1)}  

\newcommand {\nc}            {n_{\txt{c}}}     

\newcommand {\uin}           {\vec{u}}         
\newcommand {\uink}      [1] {\vec{u}_{(#1)}}  
\newcommand {\uinknv} [2] [] {#1{u}_{#2}}      
\newcommand {\uinvec}        {\vec{\upsilon}}  
\newcommand {\uinalt}        {\tilde{u}}       

\newcommand {\uinspc}        {\mathbb{K}}             
\newcommand {\ueffspc}       {\mathbb{K}'}            
\ifx\gcnspaper\undefined
	\newcommand {\uinvecspcs}  {\mathcal{K}_{\conscheme}|_t}
	
	\newcommand {\uinspcs}     {\uinspc_{\conscheme}}        
\else
	\newcommand {\uinvecspcs}  {\mathcal{K}_t}               
	\newcommand {\uinspcs}     {\uinspc_{\Pproj}}            
\fi

\newcommand {\Hcx}           {\hat{H}_{\txt{c}}}        
\newcommand {\Hc}        [1] {\hat{H}_{\txt{c},#1}}     

\newcommand {\HZtxi} [2]  
	{\tx{H}\txCc[0]{#1}{#2}}

\newcommand {\Hctxi} [3]  
	{\tx{H}\txCc[\txt{c}]{#2}{#1#3}}

\newcommand {\Gcx}       {G_{\txt{c}}}        
\newcommand {\Gc}    [1] {G_{\txt{c},#1}}     

\newcommand {\GZtxi} [2]  
	{\tx{G}\txCc[0]{#1}{#2}}

\newcommand {\Gctxi} [3]  
	{\tx{G}\txCc[\txt{c}]{#2}{#1#3}}

\newcommand {\Jcx}       {\hat{J}_{\txt{c}}}     
\newcommand {\Jc}    [1] {\hat{J}_{\txt{c},#1}}  
\ifx\gcnspaper\undefined
	\newcommand {\CJ}    {\setvar{j}}            
\else
	\newcommand {\CJ}    {\setvar{J}}            
\fi

\newcommand {\conscheme}        {\mathcal{S}}           
\newcommand {\Crch}             {\mathcal{C}}           
\newcommand {\Crchs}            {\Crch_{\conscheme}}    

\newcommand {\Psscheme}         {\conscheme_{\Pproj}}  
\newcommand {\Pstuple}          {(\CS,\Pproj,\CH)}     

\NewDocumentCommand {\Pswgspc} {smmo} {  
	\IfBooleanTF {#1} {\overline{\mathbb{V}}} {\mathbb{V}}
	\sb{#2,#3}
	\IfNoValueF {#4} {[#4]}
}

\NewDocumentCommand {\Crchsi} {mmo} {
	\Crchs%
	(#1,#2\IfNoValueF {#3} {\xp #3})
}

\newcommand {\PswC}         [2] {C}                     
\newcommand {\PswCsub}      [1] {\PswC{}{}_{\txt{#1}}}  
\newcommand {\PswCg}        [2] {\PswCsub g}            
\newcommand {\PswCm}        [2] {\PswCsub m}            
\newcommand {\PswCms}       [2] {\PswCsub{m,s}}         
\newcommand {\PswCmb}       [2] {\PswCsub{m,b}}         
\newcommand {\PswCq}        [2] {\PswCsub q}            
\newcommand {\PswCqa}       [2] {\PswCsub{q,34}}        
\newcommand {\PswCqb}       [2] {\PswCsub{q,5}}         

\newcommand {\PswCS}        [1] {\CS_{#1}}              
\NewDocumentCommand {\DonHCn} {sm}                      
	{\IfBooleanTF{#1}{\spacevar*{S}[\PswCS{#2}]}{\spacevar{S}[\PswCS{#2}]}}
\newcommand {\CSprojn}      [1] {\CSproj[\PswCS{#1}]}

\newcommand {\ueff}             {\pvec{u}}             
\newcommand {\nceff}            {\nc'}                 
\newcommand {\ueffvectx}                               
	{\prescript{\backprime}{}{\txgreek{\upsilon}}}
\newcommand {\ueffknv}   [2] [] {#1{u}'_{#2}}          

\newcommand {\Heff}             {H'}                   
\newcommand {\Heffc}        [1] {H'_{\txt{c},#1}}      

\ifx\gcnspaper\undefined
	\newcommand {\CHeff}        {\setvar{h}'}      
	\newcommand {\HeffZtx}                         
		{\prescript{\backprime}{}{\tx H}_0}
	\newcommand {\HeffZtxi} [2]                    
		{\prescript{\backprime}{}{\tx H}\txCc[0]{#1}{#2}}
	\newcommand {\Heffctx}                         
		{\prescript{\backprime}{}{\tx H}_{\txt{c}}}
	\newcommand {\Heffctxi} [3]                    
		{\prescript{\backprime}{}{\tx H}\txCc[\txt{c}]{#2}{#1#3}}
\else
	\newcommand {\CHeff}        {\setvar{H}'}      
\fi

\ifx\gcnspaper\undefined
	\newcommand {\CH}       {\setvar{h}}        

	\newcommand {\Gs}     {\Gon{\uinvecspcs}}
\else
	\newcommand {\CH}       {\setvar{H}}
	
	\newcommand {\Gs}       {\setvar{G}[\uinvecspcs]}
\fi

\newcommand {\crpaset}  {\overline{\CH}}           

\newcommand {\crpaH}    {\setvar{h}_{\lilsf H}}     

\NewDocumentCommand {\G} {e{_^}} {
	G
	\IfNoValueF {#1} {\sb{\mathrlap {#1} \hphantom{99}}}
	\IfNoValueF {#2} {\sp{#2}}
}

\newcommand {\xkern} {\mkern-2mu{\times}}
\newcommand {\sbQaa} {11}              
\newcommand {\sbQab} {1\xkern}         
\newcommand {\sbQbb} {\xkern\!\xkern}  

\newcommand {\GQaa}              {G_{\sbQaa}}             
\newcommand {\GQab}              {G_{\sbQab}}             
\newcommand {\GQbb}              {G_{\sbQbb}}             
\newcommand {\vQa}  [1] [\Pproj] {\vec{v}_{#1}}           
\newcommand {\vQb}  [1] [\Pproj]                          
	{\vec{v}_{\vphantom{\Pi}_{#1}^\perp}}
\newcommand {\dvQa} [1] [\Pproj] {\dot{\vec{v}}_{#1}}     
\newcommand {\dvQb}                                       
	{\dot{\vec{v}}_{\vphantom{\Pi}_{\NC}^\perp}}

\newcommand {\CQaa} {C_{\sbQaa}}  
\newcommand {\CQbb} {C_{\sbQbb}}  

\newcommand {\sbaa} {\txt{cc}}        
\newcommand {\sbab} {\txt{ce}}        
\newcommand {\sbac} {\txt{c}\xkern}   
\newcommand {\sbbb} {\txt{ee}}        
\newcommand {\sbbc} {\txt{e}\xkern}   
\newcommand {\sbcc} {\xkern\!\xkern}  

\newcommand {\Gaa}     {\G_{\sbaa}}            
\newcommand {\Gab}     {\G_{\sbab}}            
\newcommand {\Gac}     {\G_{\sbac}}            
\newcommand {\Gbb}     {\G_{\sbbb}}            
\newcommand {\Gbc}     {\G_{\sbbc}}            
\newcommand {\Gcc}     {\G_{\sbcc}}            
\newcommand {\va}      {\vec{v}_{\csp}}        
\newcommand {\vb}      {\vec{v}_{\txt{ext}}}   
\newcommand {\vc} [1] [\Pproj]                 
	{\vec{v}_{\vphantom{\Pi}_{#1}^\perp}}
\newcommand {\dva}     {\dot{\vec{v}}_{\csp}}  

\newcommand {\Caa}   {C_{\sbaa}}      
\newcommand {\Cbcbc} {C_{\neg\sbaa}}  
\newcommand {\Cbb}   {C_{\sbbb}}      
\newcommand {\Ccc}   {C_{\sbcc}}      

\newcommand {\lieGx}   [1] {\liealg{g}_{#1}}  

\newcommand {\koppthesis} {\cite{cite:koppthesis}}

\newcommand {\gcnsgate} [1] {\ifmmode{\scriptstyle\txt{#1}}\else #1\fi}

\NewDocumentCommand{\gcnsnewtheorem}{ommm}{  
	\IfNoValueTF{#1}
		{\newtheorem {#2} {#3} [section]}
		{\newtheorem {#2} [#1] {#3}}

	\crefname {#2} {#4} {\MakeLowercase{#3}}
	\Crefname {#2} {#4} {#3}
	\expandafter\def\csname #2Long\endcsname{#3}
	\expandafter\def\csname #2long\endcsname{\MakeLowercase{#3}}
	\expandafter\def\csname #2short\endcsname{#4}
}

\newcounter{gcnsdef}
\NewDocumentEnvironment{gcnsdef}{ss} 
	{%
		\stepcounter{gcnsdef}%
		\vspace{8pt}
		\noindent {\bfseries\itshape Definition \arabic{gcnsdef}.}\hspace{5pt}%
	}{%
		\IfBooleanF{#2} {%
			~$\square$%
			\IfBooleanF{#1} {\vspace{8pt}}%
		}%
	}

\newcommand {\Hiii} [1] {\textit{\textbf{#1}}\hspace{5mm}}

\ExplSyntaxOn
\NewDocumentCommand{\proofref}{sm} {
	\let\nbsp\nobreakspace
	\def\proofrefint##1##2{##2}
	\def\proofrefext##1##2{Section\nbsp$##1$\space of\nbsp\extproofsappendix}
	\IfBooleanTF{#1}
		{\let\prx\proofrefint}
		{\let\prx\proofrefext}
	\str_case:nnF {#2} {
		{on-G0-structure}            {\prx {1}   {\cref{lem:on-G0-structure}}}
		{on-proj-res-identity}       {\prx {2.1} {Lemma\nbsp\labelcref{prop:on-C3x3-form}(a)}}
		{on-C3x3-form}               {\prx {2}   {\cref{prop:on-C3x3-form}}}
		{on-sum-of-skew-and-hurwitz} {\prx {3.1} {Lemma\nbsp\labelcref{prop:CQF-pci-necc}(a)}}
		{on-noise-decoupling}        {\prx {3.2} {Lemma\nbsp\labelcref{prop:CQF-pci-necc}(b)}}
		{CQF-pci-necc-full-G}        {\prx {3.3} {Prop.\nbsp\labelcref{prop:CQF-pci-necc}(c)}}
		{CQF-full-rhsz-equiv}        {\prx {3.4} {Lemma\nbsp\labelcref{prop:CQF-pci-necc}(d)}}
		{CQF-pci-necc}               {\prx {3}   {\cref{prop:CQF-pci-necc}}}
		{G_t-invariance}             {\prx {4.1} {Lemma\nbsp\labelcref{prop:on-CQF-with-arb-Q1}(a)}}
		{on-CQF-with-arb-Q1}         {\prx {4}   {\cref{prop:on-CQF-with-arb-Q1}}}
		{simple-pci-cond-impls}      {\prx {5.1} {Lemma\nbsp\labelcref{cor:compute-DIFS}(a)}}
		{simple-pci-condition}       {\prx {5.2} {Lemma\nbsp\labelcref{cor:compute-DIFS}(b)}}
		{compute-DIFS}               {\prx {5}   {\cref{cor:compute-DIFS}}}
		{on-k-sector-decoupling}     {\prx {6}   {\cref{prop:on-k-sector-decoupling}}}
		{control-algebras}           {\prx {7}   {\cref{thrm:control-algebras}}}
		{L-OC-cond-dim}              {\prx {8}   {\cref{cor:L-OC-cond-dim}}}
		{L-ESC-cond-dim}             {\prx {9}   {\cref{prop:L-ESC-cond-dim}}}
	}
	{\ensuremath{\blacksquare}}
}
\ExplSyntaxOff
	\providecommand {\proofref} {}  

\newcommand {\logg}          {\hat V}                     
\newcommand {\loggtgt}       {\logg_{\txt{tgt}}}          
\newcommand {\wkgspc}        {\mathbb{V}}                 

\newcommand {\cFit}         {\epsilon_{\lil{fit}}}        
\newcommand {\cFiti}    [1] {\epsilon_{\lil{fit},#1}}     
\newcommand {\cFitg}        {\cFiti{\lil g}}              
\newcommand {\cFitc}        {\cFiti{\lil c}}              

\newcommand* {\ionnu}      {\eta}
\newcommand* {\ionmu}      {\mu}
\newcommand* {\ionGammaZ}  {\Gamma_{\txt{Z}}}
\newcommand* {\ionD}       {\hat D_{\txt{Z}}}
\newcommand* {\ionq}       {q}

\newcommand* {\ionopx}     {\hat\sigma}
\newcommand* {\ionop}  [1] {\ionopx_{#1}}

\newcommand {\setbldst} [4] [\speqinset] {  
	\left\{\mskip-12mu\begin{gathered}
		\LGivenR{#2}{#3}\hfill \\
		#1\sttext #4\hfill
	\end{gathered}\mskip-6mu\right\}
}

\newcommand {\gcnsdeqitem} [3] [\label] {
	\item%
	\abovedisplayskip=0pt%
	\abovedisplayshortskip=0pt%
	~\vspace*{-\baselineskip}%
	\begin{equation}#1{#2}#3\end{equation}}

\newcommand {\rtwo}      {a_1}  
\newcommand {\rhalf}     {a_2}  
\newcommand {\rtwophalf} {a_3}  
\newcommand {\rtwopsq}   {a_4}  
\newcommand {\rtwomtwo}  {a_5}  

\newcommand {\toychimapsym} {\conv{\hat J}{\Gaa}}
\newcommand {\toychimatsym} {\chi_{\csp}}

\tikzset{
	toymatrix/.style = {  
		execute at begin picture     = \deftoymatrixstyles,
		every matrix/.style          = toymatrixstyle,
		every left delimiter/.style  = { xshift =  1.4ex },
		every right delimiter/.style = { xshift = -1.4ex },
		baseline                     = (toy.center)
	}}

\newcommand {\deftoymatrixstyles} {
	\def\-{\operatorname{-}\mkern-3mu}

	\newcommand {\toynodehsep} {7mm}
	\newcommand {\toynodevsep} {4mm}

	\definecolor {toybd2col} {rgb} {0.58,0.58,0.58}  

	\tikzstyle {toybd1style} = [  
		line width   = 0.8pt,
		dash pattern = on 2pt off 2pt,
		shorten >    = 1.2mm,
		shorten <    = 1.2mm]

	\tikzstyle {toybd2style} = [  
		line width = 0.8pt,
		color      = toybd2col,
		shorten >  = 1.2mm,
		shorten <  = 1.2mm]

	\tikzstyle {toymatrixstyle} = [  
		matrix of math nodes,
		nodes in empty cells,
		ampersand replacement         = \&,
		left delimiter                = [,
		right delimiter               = {]},
		every left delimiter/.style   = { xshift = 0.8ex },
		right delimiter/.append style = { xshift = -0.8ex },
		every node/.append style      = { anchor = center, font = \footnotesize },
		column sep                    = {\toynodehsep,between origins},
		row sep                       = {\toynodevsep,between origins}]
}

\NewDocumentCommand {\toymatrix} {oom} {
	\begin{tikzpicture}[toymatrix]
		\matrix (toy) { #3 };
		\IfNoValueF{#1}
			{\toybd {#1} 1}
		\IfNoValueF{#2}
			{\toybd {#2} 2}
	\end{tikzpicture}
}
	\providecommand {\toymatrix} [1] {}  

\newcommand {\toybd} [2] {  
	\coordinate (bdx) at ($(toy-1-#1) + 1/2*(\toynodehsep,0)$);
	\coordinate (bdy) at ($(toy-#1-1) - 1/2*(0,\toynodevsep)$);
	\draw [toybd#2style] (bdx |- toy.north) -- (bdx |- toy.south);
	\draw [toybd#2style] (toy.west |- bdy) -- (toy.east |- bdy);
}

\newcommand {\toyUi} {
	\toymatrix [4] [8] {
		     \& \-10 \&      \&     \&          \&          \&        \&        \&   1      \&          \\
		  10 \&      \&      \&     \&          \&          \&        \&        \& \-5      \&   2      \\
		     \&      \&      \&     \& \-10     \&          \&        \&        \&          \&   2      \\
		     \&      \&      \&     \&          \&          \&        \& 5      \&          \&          \\
		     \&      \&   10 \&     \&          \&          \& 5\rtwo \&        \&          \&          \\
		     \&      \&      \&     \&          \&          \&        \& 5\rtwo \&          \&   6      \\
		     \&      \&      \&     \& \-5\rtwo \&          \&        \&        \& \-3      \& \-2      \\
		     \&      \&      \& \-5 \&          \& \-5\rtwo \&        \&        \& \-1      \&          \\
		\-1  \&   5  \&      \&     \&          \&          \& 3      \& 1      \& \-5      \& \-5\rtwo \\
		     \& \-2  \& \-2  \&     \&          \& \-6      \& 2      \&        \&   5\rtwo \& \-10     \\
	}
}

\newcommand {\toyUii} {
	\toymatrix [4] [8] {
		     \&   20 \&    \&      \&         \&           \&           \&           \& \-2  \&      \\
		\-20 \&      \&    \&      \&         \&           \&           \&           \&   10 \& \-4  \\
		     \&      \&    \& \-10 \&         \&           \&           \&           \&      \& \-4  \\
		     \&      \& 10 \&      \&         \& \-20      \&           \& \-10      \&      \&      \\
		     \&      \&    \&      \&         \&           \& \-10\rtwo \&           \&      \&      \\
		     \&      \&    \&   20 \&         \&           \&           \& \-10\rtwo \&      \& \-12 \\
		     \&      \&    \&      \& 10\rtwo \&           \&           \&           \&   6  \&   4  \\
		     \&      \&    \&   10 \&         \&   10\rtwo \&           \&           \&   2  \&      \\
		  2  \& \-10 \&    \&      \&         \&           \& \-6       \& \-2       \&      \& \-5  \\
		     \&   4  \& 4  \&      \&         \&   12      \& \-4       \&           \&   5  \&      \\
	}
}

\newcommand {\toyUiii} {
	\toymatrix [4] [8] {
		     \&   10 \& \-10 \&      \&           \&           \&           \&           \& \-1  \&      \\
		\-10 \&      \&      \& \-10 \&           \&           \&           \&           \&   5  \& \-2  \\
		  10 \&      \&      \& \-15 \& \-20      \&           \&           \&           \&      \& \-2  \\
		     \&   10 \&   15 \&      \&           \& \-30      \&           \& \-5       \&      \&      \\
		     \&      \&   20 \&      \&           \&           \& \-10\rtwo \& \-5\rtwo  \&      \&      \\
		     \&      \&      \&   30 \&           \&           \&   5\rtwo  \& \-10\rtwo \&      \& \-6  \\
		     \&      \&      \&      \&   10\rtwo \& \-5\rtwo  \&           \&           \&   3  \&   2  \\
		     \&      \&      \&   5  \&   5\rtwo  \&   10\rtwo \&           \&           \&   1  \&      \\
		  1  \& \-5  \&      \&      \&           \&           \& \-3       \& \-1       \&      \&   10 \\
		     \&   2  \&   2  \&      \&           \&   6       \& \-2       \&           \& \-10 \&      \\
	}
}

\newcommand {\toyGCBDi} {
	5 \toymatrix [4] {
		    \&     \& 1 \&   \&          \&          \&          \&        \&            \&            \\
		    \&     \&   \& 1 \&          \&          \&          \&        \&            \&            \\
		\-1 \&     \&   \&   \&          \&          \&          \&        \&            \&            \\
		    \& \-1 \&   \&   \&          \&          \&          \&        \&            \&            \\
		    \&     \&   \&   \&          \&          \&   \rhalf \& \rhalf \&            \&            \\
		    \&     \&   \&   \&          \&          \& \-\rhalf \& \rhalf \&            \&            \\
		    \&     \&   \&   \& \-\rhalf \&   \rhalf \&          \&        \&            \&            \\
		    \&     \&   \&   \& \-\rhalf \& \-\rhalf \&          \&        \&            \&            \\
		    \&     \&   \&   \&          \&          \&          \&        \& \-1        \& \-\rtwopsq \\
		    \&     \&   \&   \&          \&          \&          \&        \&   \rtwopsq \& \-2        \\
	}
}

\newcommand {\toyGBDi} {
	\toymatrix [8] {
		 \&  \&   \&     \&     \&     \&  \&  \&              \&              \\
		 \&  \&   \&     \&     \&     \&  \&  \&              \&              \\
		 \&  \&   \& \-1 \& \-2 \&     \&  \&  \&              \&              \\
		 \&  \& 1 \&     \&     \& \-2 \&  \&  \&              \&              \\
		 \&  \& 2 \&     \&     \&     \&  \&  \&              \&              \\
		 \&  \&   \&   2 \&     \&     \&  \&  \&              \&              \\
		 \&  \&   \&     \&     \&     \&  \&  \&              \&              \\
		 \&  \&   \&     \&     \&     \&  \&  \&              \&              \\
		 \&  \&   \&     \&     \&     \&  \&  \& \-1          \& \-\rtwophalf \\
		 \&  \&   \&     \&     \&     \&  \&  \&   \rtwophalf \& \-2          \\
	}
}

\newcommand {\toyGBDii} {
	\toymatrix [8] {
		  \&   \& \-2 \&     \&         \&         \&         \&         \&             \&             \\
		  \&   \&     \& \-2 \&         \&         \&         \&         \&             \&             \\
		2 \&   \&     \& \-3 \& \-6     \&         \&         \&         \&             \&             \\
		  \& 2 \&   3 \&     \&         \& \-6     \&         \&         \&             \&             \\
		  \&   \&   6 \&     \&         \&         \& \-\rtwo \& \-\rtwo \&             \&             \\
		  \&   \&     \&   6 \&         \&         \&   \rtwo \& \-\rtwo \&             \&             \\
		  \&   \&     \&     \&   \rtwo \& \-\rtwo \&         \&         \&             \&             \\
		  \&   \&     \&     \&   \rtwo \&   \rtwo \&         \&         \&             \&             \\
		  \&   \&     \&     \&         \&         \&         \&         \& \-1         \& \-\rtwomtwo \\
		  \&   \&     \&     \&         \&         \&         \&         \&   \rtwomtwo \& \-2         \\
	}
}

\newcommand {\toyAi} {
	\toymatrix [4] {
		 \&  \&     \&     \&   \&   \&  \&  \\
		 \&  \&     \&     \&   \&   \&  \&  \\
		 \&  \&     \&   1 \& 2 \&   \&  \&  \\
		 \&  \& \-1 \&     \&   \& 2 \&  \&  \\
		 \&  \& \-2 \&     \&   \&   \&  \&  \\
		 \&  \&     \& \-2 \&   \&   \&  \&  \\
		 \&  \&     \&     \&   \&   \&  \&  \\
		 \&  \&     \&     \&   \&   \&  \&  \\
	}
}

\newcommand {\toyAii} {
	\toymatrix [4] {
		    \&     \& 1 \&   \&          \&          \&          \&        \\
		    \&     \&   \& 1 \&          \&          \&          \&        \\
		\-1 \&     \&   \&   \&          \&          \&          \&        \\
		    \& \-1 \&   \&   \&          \&          \&          \&        \\
		    \&     \&   \&   \&          \&          \&   \rhalf \& \rhalf \\
		    \&     \&   \&   \&          \&          \& \-\rhalf \& \rhalf \\
		    \&     \&   \&   \& \-\rhalf \&   \rhalf \&          \&        \\
		    \&     \&   \&   \& \-\rhalf \& \-\rhalf \&          \&        \\
	}
}

\newcommand {\toyAiii} {
	\toymatrix [4] {
		    \&     \&         \&   1     \& 2 \&   \&         \&         \\
		    \&     \& \-1     \&         \&   \& 2 \&         \&         \\
		    \&   1 \&         \&         \&   \&   \& \-\rtwo \& \-\rtwo \\
		\-1 \&     \&         \&         \&   \&   \&   \rtwo \& \-\rtwo \\
		\-2 \&     \&         \&         \&   \&   \&         \&         \\
		    \& \-2 \&         \&         \&   \&   \&         \&         \\
		    \&     \&   \rtwo \& \-\rtwo \&   \&   \&         \&         \\
		    \&     \&   \rtwo \&   \rtwo \&   \&   \&         \&         \\
	}
}

\newcommand {\toyAiv} {
	\toymatrix [4] {
		    \&     \&   1      \&          \&        \& \-4      \&          \&          \\
		    \&     \&          \&   1      \& 4      \&          \&          \&          \\
		\-1 \&     \&          \&          \&        \&          \&   2\rtwo \& \-2\rtwo \\
		    \& \-1 \&          \&          \&        \&          \&   2\rtwo \&   2\rtwo \\
		    \& \-4 \&          \&          \&        \&          \& \-\rhalf \& \-\rhalf \\
		  4 \&     \&          \&          \&        \&          \&   \rhalf \& \-\rhalf \\
		    \&     \& \-2\rtwo \& \-2\rtwo \& \rhalf \& \-\rhalf \&          \&          \\
		    \&     \&   2\rtwo \& \-2\rtwo \& \rhalf \&   \rhalf \&          \&          \\
	}
}

\newcommand {\toyAv} {
	\toymatrix [4] {
		        \&   1     \&  \&  \&  \&  \& \-\rtwo \& \-\rtwo \\
		\-1     \&         \&  \&  \&  \&  \&   \rtwo \& \-\rtwo \\
		        \&         \&  \&  \&  \&  \&         \&         \\
		        \&         \&  \&  \&  \&  \&         \&         \\
		        \&         \&  \&  \&  \&  \&         \&         \\
		        \&         \&  \&  \&  \&  \&         \&         \\
		  \rtwo \& \-\rtwo \&  \&  \&  \&  \&         \&         \\
		  \rtwo \&   \rtwo \&  \&  \&  \&  \&         \&         \\
	}
}

\newcommand {\toyAvi} {
	\toymatrix [4] {
		         \&        \&          \&   8\rtwo \& \-2\rtwo \&          \&     \&     \\
		         \&        \& \-8\rtwo \&          \&          \& \-2\rtwo \&     \&     \\
		         \& 8\rtwo \&          \&          \&          \&          \&   2 \&   2 \\
		\-8\rtwo \&        \&          \&          \&          \&          \& \-2 \&   2 \\
		  2\rtwo \&        \&          \&          \&          \&          \&   9 \& \-9 \\
		         \& 2\rtwo \&          \&          \&          \&          \&   9 \&   9 \\
		         \&        \& \-2      \&   2      \& \-9      \& \-9      \&     \&     \\
		         \&        \& \-2      \& \-2      \&   9      \& \-9      \&     \&     \\
	}
}

\newcommand {\toyAvii} {
	\toymatrix [4] {
		       \& \-8\rtwo \&          \&        \&          \&          \& \-2      \& \-2      \\
		8\rtwo \&          \&          \&        \&          \&          \&   2      \& \-2      \\
		       \&          \&          \& 8\rtwo \& \-2\rtwo \&          \&          \&          \\
		       \&          \& \-8\rtwo \&        \&          \& \-2\rtwo \&          \&          \\
		       \&          \&   2\rtwo \&        \&          \&   9\rtwo \&          \&          \\
		       \&          \&          \& 2\rtwo \& \-9\rtwo \&          \&          \&          \\
		2      \& \-2      \&          \&        \&          \&          \&          \& \-9\rtwo \\
		2      \&   2      \&          \&        \&          \&          \&   9\rtwo \&          \\
	}
}

\newcommand {\toyABDi} {
	\toymatrix [4] {
		    \&     \& 1 \&   \&          \&          \&          \&        \\
		    \&     \&   \& 1 \&          \&          \&          \&        \\
		\-1 \&     \&   \&   \&          \&          \&          \&        \\
		    \& \-1 \&   \&   \&          \&          \&          \&        \\
		    \&     \&   \&   \&          \&          \&   \rhalf \& \rhalf \\
		    \&     \&   \&   \&          \&          \& \-\rhalf \& \rhalf \\
		    \&     \&   \&   \& \-\rhalf \&   \rhalf \&          \&        \\
		    \&     \&   \&   \& \-\rhalf \& \-\rhalf \&          \&        \\
	}
}

\newcommand {\toyABDii} {
	\toymatrix [4] {
		    \&   \&     \& 1 \&          \&          \&        \&          \\
		    \&   \& \-1 \&   \&          \&          \&        \&          \\
		    \& 1 \&     \&   \&          \&          \&        \&          \\
		\-1 \&   \&     \&   \&          \&          \&        \&          \\
		    \&   \&     \&   \&          \&          \& \rhalf \& \-\rhalf \\
		    \&   \&     \&   \&          \&          \& \rhalf \&   \rhalf \\
		    \&   \&     \&   \& \-\rhalf \& \-\rhalf \&        \&          \\
		    \&   \&     \&   \&   \rhalf \& \-\rhalf \&        \&          \\
	}
}

\newcommand {\toyABDiii} {
	\toymatrix [4] {
		    \& 1 \&   \&     \&   \&     \&     \&   \\
		\-1 \&   \&   \&     \&   \&     \&     \&   \\
		    \&   \&   \& \-1 \&   \&     \&     \&   \\
		    \&   \& 1 \&     \&   \&     \&     \&   \\
		    \&   \&   \&     \&   \& \-1 \&     \&   \\
		    \&   \&   \&     \& 1 \&     \&     \&   \\
		    \&   \&   \&     \&   \&     \&     \& 1 \\
		    \&   \&   \&     \&   \&     \& \-1 \&   \\
	}
}

\newcommand {\toyGCi} {
	\toymatrix {
		  \&   \& \-1 \&     \\
		  \&   \&     \& \-1 \\
		1 \&   \&     \&     \\
		  \& 1 \&     \&     \\
	}
}

\newcommand {\toyGCii} {
	\toymatrix {
		  \&     \&   \& \-1 \\
		  \&     \& 1 \&     \\
		  \& \-1 \&   \&     \\
		1 \&     \&   \&     \\
	}
}

\newcommand {\toyGCiii} {
	\toymatrix {
		    \& 1 \&   \&     \\
		\-1 \&   \&   \&     \\
		    \&   \&   \& \-1 \\
		    \&   \& 1 \&     \\
	}
}

\newcommand {\toychimat} {
	\frac{1}{2}\toymatrix {
		    \&     \&     \&     \\
		\-1 \&     \&     \&   1 \\
		    \& \-i \&   i \&     \\
		    \& \-1 \& \-1 \&     \\
		  1 \&     \&     \& \-1 \\
		    \&     \&     \&     \\
		    \&   1 \&   1 \&     \\
		    \& \-i \&   i \&     \\
		    \&   i \& \-i \&     \\
		    \& \-1 \& \-1 \&     \\
		    \&     \&     \&     \\
		  1 \&     \&     \& \-1 \\
		    \&   1 \&   1 \&     \\
		    \&   i \& \-i \&     \\
		\-1 \&     \&     \&   1 \\
		    \&     \&     \&     \\
	}
}

\gcnsnewtheorem              {gcnstheorem} {Theorem}     {Theorem}
\gcnsnewtheorem[gcnstheorem] {gcnslemma}   {Lemma}       {Lemma}
\gcnsnewtheorem[gcnstheorem] {gcnsprop}    {Proposition} {Prop.}
\gcnsnewtheorem[gcnstheorem] {gcnscor}     {Corollary}   {Cor.}

\newcommand {\extproofsappendix} {\cite{cite:extproofs}}

\newcommand {\exampleappendix} {\cite{cite:extappendix}}

\newcommand{\effhamsproc} {Section~$10$ in~\extproofsappendix}  

\newcommand {\dysoncoeffeqno} {I.19}

\allowdisplaybreaks

\begin{document}

\title{Extended Controllability Tests for Quantum Decoherence-Free Subspaces}

\author{Eric B.\ Kopp and Raymond Kwong}
\affiliation{Department of Electrical and Computer Engineering\\
	University of Toronto}
\email{erickopp@ece.utoronto.ca}
\email{raymondkwong@ece.utoronto.ca}

\date{\today}

\begin{abstract}
	In this paper we develop two axiomatic tests for the controllability of subsystem codes embedded in decoherence-free subspaces of open quantum systems. The tests expand on existing control theory by considering quantum subsystems where a decoherence-protected quantum state is permitted to exit the set of logically encoded states in order to perform a broader range of computations. The tests target the class of all Lindbladian models and require no specific structure, regularity, or symmetry in system Hamiltonians or noise operators, making them ideal for control design in models lacking these features. The usefulness of the tests is demonstrated using a complete worked example for a trapped ion system subject to a nonstandard collective dephasing noise.
\end{abstract}

\maketitle

\section{Introduction}
\label{sec:intro}

Quantum information processing has been the subject of intensive study since the advent of several key algorithms in the 1990's and subsequent explosion of applications for quantum computers~\cite{cite:nisqera,cite:qcompsurv,cite:qcrev2015,cite:qctax} and quantum algorithms~\cite{cite:bigdata,cite:qcompsurv,cite:zoo}. Open quantum systems are notoriously fragile and sensitive to environmental noise, giving rise to the field of {\em information-preserving subsystems}~\cite{cite:ipsstruct}, which can be broadly divided into the categories of {\em i})~active quantum error correction~\cite{cite:qecbook,cite:qfc1qec,cite:qmemerr}, {\em ii})~decoherence-free subspaces (DFSs)~\cite{cite:dfssem1,cite:dfssem2,cite:dfsuqc,cite:dynamdfs,cite:nsreview,cite:qnschoi}, and {\em iii})~dynamical decoupling/dynamic error correction~\cite{cite:decgates,cite:dynamdfs,cite:dyndec2006,cite:dyndec2009,cite:dyndec2010,cite:dyndecopt,cite:qdot2016,cite:semdyndcpl,cite:tfopencon}.

Among these categories, DFSs, also called {\em passive, infinite-distance information preserving subspaces}~\cite{cite:ipsstruct}, constitute a particularly useful tool for preservation of information in open quantum systems since they require no active measurement, relying instead on symmetries in the system-environment dynamics to preserve information in logical---specially encoded---subspaces~\cite{cite:dfssem1}. Furthermore, because DFSs are a passive form of error avoidance, they can be used in conjunction with other types of error avoidance/correction as well as with control modalities such as quantum feedback control~\cite{cite:qfcsum2017}.

Despite these benefits, little research has been done to date on exploiting DFSs in a fully generalized and systematic way, or on incorporating DFSs into modern quantum control techniques such as quantum optimal control~\cite{cite:conbypar,cite:conland1,cite:gnliborzi,cite:gnliland,cite:gnlisem,cite:notrap2017,cite:optcconst,cite:optdiss,cite:optsupp,cite:qcopt2016,cite:qcopt2019,cite:qcrev2015,cite:qcsoft} and quantum robust control~\cite{cite:howrobust,cite:qclern2019,cite:qcsoft,cite:robust2019,cite:robustss,cite:stochlc}. Extant DFS literature either {\em i})~is concerned with specific control objectives such as state preparation/stabilization~\cite{cite:cz2008,cite:dfsiac,cite:dfsstable,cite:lyapcon17,cite:lyapcon19,cite:lyapdfs2017,cite:mkvstabdfs,cite:pcebyqfc,cite:robustss,cite:stblzqs}, {\em ii})~provides no explicit treatment of subsystem code identification and controllability testing~\cite{cite:pcebyqfc,cite:tarndfs,cite:tarndfs2,cite:tarndfs2013,cite:dfsuqc2,cite:dfsholo,cite:dfsuqcadi}, and/or {\em iii})~is limited to highly specific systems and types of noise for the sake of mathematical tractability~\cite{cite:cnotindfs,cite:collrot,cite:collrot2,cite:hotiondfs,cite:ivanov,cite:monzkim,cite:semdfscon}.

In \koppthesis, the author proposes a framework for identifying and controlling DFSs in an unsupervised and systematic way, i.e., generating a start-to-finish control strategy, from state encoding to control inputs, based solely on a model description. A systematic approach to control design in DFSs is of considerable importance due to the flexibility it affords to modeling, simulation, and experimentation. In particular, it allows control resources, noise models, and model features to be swapped in and out of the master equations governing system dynamics (typically accompanying changes in experimental hardware or setups) without need of exhaustive ad hoc approaches for designing control for the modified model(s). This flexibility in turn facilitates a more rapid and powerful approach to control design.

One critical issue arising in \koppthesis\ is the need to assess the controllability of subsystem codes~\cite{cite:subcode} when the physical state of the system is not required to maintain a strict correspondence to an encoded (i.e., {\em logical}) state under the influence of control except at specific times during its evolution. The additional control flexibility afforded by the relaxed requirements is particularly valuable when working in DFSs since a considerable degree of control freedom is inherently sacrificed to decouple the state from environmental noise. In particular, a system lacking sufficient control freedom for logical operator controllability~\cite{cite:con4bimlq,cite:qcdbook} (also referred to as ``encoded universality'' by Zanardi et al.\ in~\cite{cite:dfsuqc2} and elsewhere) may prove to be operator controllable when the requirement for logical-to-physical state isomorphism is relaxed.

In this paper, we develop two controllability tests (representing two well-known controllability standards) for subsystem codes embedded in DFSs, subject to the aforementioned relaxation. The tests expand on existing work in the quantum systems literature, most notably the work of Tarn, Ganeson et al.\ on operator-invariant control~\cite{cite:tarndfs,cite:tarndfs2,cite:tarndfs2013,cite:suppdecoh,cite:qfcwk2str}, Lidar et al.~\cite{cite:dfssem2,cite:dfsuqcadi,cite:dfsuqc}, Viola, Ticozzi et al.~\cite{cite:dfsiac,cite:dynamdfs,cite:semdfscon,cite:collrot2,cite:stblzqs,cite:dfsstable}, Zanardi et al.~\cite{cite:dfsholo,cite:dfsuqc2}, and others~\cite{cite:dfsiac,cite:dynamdfs,cite:semdfscon,cite:collrot2,cite:stblzqs,cite:dfsstable}. Additionally, we provide computationally efficient algorithms implementing the proposed tests.

\Cref{sec:background} provides an overview of relevant background work in modeling of open quantum systems, noiseless subspace theory, and open-loop control theory.

\Cref{sec:ipcontrol,sec:csops} develop the major concepts and propositions pertaining to noiseless control underlying the theorems developed in \cref{sec:pstatic}.

\Cref{sec:pstatic} presents the theory of ``$P$-static control'', including the aforementioned controllability tests for subsystem codes embedded in DFSs.

\Cref{sec:example} demonstrates the utility of the controllability tests developed in \cref{sec:pstatic} using a worked example of a trapped ion system subject to a nonstandard collective dephasing noise.

Future work and conclusions are presented in \cref{sec:future,sec:conclusions}.

\section{Background}
\label{sec:background}

\subsection{Notation}
\label{sec:bg-notation}

Throughout this paper, we denote the set of functions with domain $X$ and codomain $Y$ as $\Mapset{X}{Y}$, the image (range) of matrix operator $X$ as $\im X$, the closed unit $n$-ball as $\Sn{n}$, the $n\times n$ square zero matrix as $\ZM{n}$, the $n\times m$ zero matrix as $\ZMG{n}{m}$, the $n$-element zero column vector as $\ZV{n}$, the $n\times n$ identity matrix as $\eye{n}$, and the $n\times m$ identity matrix (having $1$'s on the principal diagonal and $0$'s elsewhere) as $\eye{n\times m}$. The set of integers from $1$ to $n$ (inclusive) is $\OneTo{n}$.

We denote the standard real matrix basis element having $1$ in the $i$th row, $j$th column as $\RMe{i}{j}$, and the standard $m$-dimensional real vector basis element having $1$ at index $i$ as $\Rne[m]{i}$.

We denote the $i$th element of vector $\vec{x}$ as $\vec{x}_{(i)}$, the $i$th element of vector field $\vec{x}(t)$ as $\vec{x}_{(i)}(t)$, and the $(i,j)$th element of matrix $X$ as $X_{(i,j)}$, distinguishing element indexes from indexes in a numbered series of vectors/matrices, which appear without surrounding brackets, e.g.~$X_1, X_2, \ldots$

Lie algebras are denoted using lowercase Fraktur characters, e.g., $\liesu{n}$, $\liesp{n}$, and their associated Lie groups are denoted using uppercase Roman characters, e.g., $\lieSU{n}$, $\lieSp{n}$.

\subsection{Open Quantum Systems}
\label{sec:bg-opensys}

We consider a separable Hilbert space $\Hilspc$ over the complex field $\Cx$, a set of bounded linear operators $\BonH$ on $\Hilspc$, the Hermitian subset $\HonH \subset \BonH$, and a time-varying state $\rho(t)$ residing in the set of unit-trace, positive semidefinite operators $\DonH \subset \BonH$. For purposes of computation, we find it convenient to assume $\BonH$ is finite-dimensional, representable by complex matrices $\in \CM{\Bord}$ given order $\Bord \in \Zp$ such that $\dim\BonH = \Bdim$ where $\Bdim = \Bord^2$. As is common for noiseless subspace treatments, we consider the class of models subject to the Born and Markov approximations~\cite{cite:oqstheory}, which are characterized by dynamics of the Lindbladian~\cite{cite:lindblad} form:
\begin{equation}
	\label{eq:lindbladian-dynamics}
	\dot\rho(t) = \LindH[\rho(t)\xp \{\uinknv{k}\}_k] + \LindD[\rho]
\end{equation}
which is the sum of a control-dependent Hamiltonian term, $\LindH$, of the form
\begin{equation}
	\label{eq:lindbladian-hamilton-term}
	\LindH[\rho(t)\xp \{\uinknv{k}\}_k] \isdefas
		-i\left[ \left( \hat{H}_0 + \sum\limits_k \uinknv{k}(t)\,\Hc{k} \right), \rho(t) \right]
\end{equation}
and a noise term, $\LindD$, of the form
\begin{equation}
	\label{eq:lindbladian-noise-term}
	\LindD[\rho] \isdefas \sum\limits_{j = 1}^{\nsub{d}} {\Gamma_j \mathcal{D}[\hat{D}_j\xp \rho]}
\end{equation}
where the superoperator $\mathcal{D}[\cdot]$ is given by
\begin{equation}
	\label{eq:lindblad-super-D-def}
	\mathcal{D}[\hat{D}\xp \rho] \isdefas
		\hat{D}\rho(t) \adj{\hat{D}} -
		\frac{1}{2}\left( \adj{\hat{D}}\hat{D}\rho(t) + \rho(t)\adj{\hat{D}}\hat{D} \right)
\end{equation}

We refer to $\hat{H}_0 \in \HonH$ as the {\em drift Hamiltonian}, operators $\{\Hc{k} \in \HonH\}_{k=1}^{\nc}$ as $\nc$ {\em control Hamiltonians}, and the $\nsub{d}$ tuples $\{\Gamma_j \in \Rl, \hat{D}_j \in \BonH\}_{j = 1}^{\nsub{d}}$ encapsulating all non-unitary dynamics induced by noise in the system as the {\em noise channels}. Throughout this paper we follow the usual conventions of setting $\hbar = 1$ and omitting the explicit time dependence of $\rho$.

The form of \eqn{eq:lindbladian-hamilton-term} assumes a linear contribution from each control Hamiltonian, with the $k$th term modulated by a time-varying function $\uinknv{k} \in \Mapset{\Rnn}{\Rl}$. Inputs of this form, called {\em bilinear} or {\em dipole approximations}, are common in quantum control literature~\cite{cite:bydbilin,cite:collrot2,cite:dfsstable,cite:gnliborzi,cite:qcopt2019,cite:qcscale,cite:tarndfs,cite:tarndfs2,cite:tarndfs2013,cite:timeoptgc}.

It is convenient for us to rewrite \eqn{eq:lindbladian-hamilton-term} in a vectorized form. Let
\begin{equation}
	\label{eq:def-control-vector}
	\uin(t) \isdefas \def\x#1{\uinknv{#1}(t)}\compactcvx{\nc}
\end{equation}
where $\uin \in \uinspc$ is called the {\em control input field} and $\uink{k} \equiv \uinknv{k}$ is called the $k$th {\em control input channel}. We assume all control input channels are continuous and bounded, and can be independently actuated. That is, the set of admissible input fields, $\uinspc$, is the linear subspace of bounded functions in $C^0(\Rnn\xp\Rn{\nc})$.

We also find it convenient to recast the dynamics of \eqn{eq:lindbladian-dynamics} into the equivalent linear differential equation:
\begin{align}
	\label{eq:cvs-dynamics}
	\dot{\vec{v}}(t) & = G(t)\vec{v}(t) \\
	\label{eq:covec-to-phys-state}
	\rho(t)          & = \sum\limits_{j = 1}^{\Bdim} \vec{v}_{(j)}(t) \, \hat{F}_j
\end{align}
where $\vec{v} \in \Sn{\Bdim}$ is the {\em coherence vector}~\cite{cite:alicki,cite:gpbasis,cite:odeform} (sometimes called the {\em generalized Bloch vector}~\cite{cite:suppdecoh,cite:gpbasis}), $G \in \Mapset{\Rnn}{\RM{\Bdim}}$ is a time-varying, real-valued matrix operator called the {\em G endomorphism} or {\em G matrix}, and $\{\hat F_j\}_j$ is a set of basis vectors
\begin{equation}
	\label{eq:cvsbasis-def}
	\{\hat F_j \in \CM{\Bord};\; j \inoneto{\Bdim}\} \sppredtostmt \begin{array}{rcl}
		\hat{F}_i                    & = & \adj{\hat{F}_i} \\
		\tr \adj{\hat{F}_i}\hat{F}_j & = & \delta _{ij}
	\end{array} \Forall i,j
\end{equation}

To convert density operators to coherence vector, we define a conversion map $\RtoV \in \Mapset{\DonH}{\Sn{\Bdim}}$, which is a linear injection, as~\footnote{Note here that $\hat{F}_k = \adj{\hat{F}_k}$.}:
\begin{align}
	\label{eq:def-RtoV}
	\RtoV(\hat{X}) & \isdefas \sum\limits_{j = 1}^{\Bdim} \inprod{\hat{X}}{\hat{F}_j}\Rne[\Bdim]{j} \\
	\nonumber
	{}             & = \sum\limits_{j = 1}^{\Bdim} (\tr \hat{X}\hat{F}_j)\; \Rne[\Bdim]{j}
\end{align}
such that $\vec v = \convclr\RtoV(\rho)$.

Owing to the linearity of \eqn{eq:lindbladian-dynamics} and \eqn{eq:lindbladian-hamilton-term}, $G$ decomposes as
\begin{equation}
	\label{eq:ode-matrix-by-component}
	G(t) = G[u(t)] \isdefas G_0 + \sum\limits_{k = 1}^{\nc} \uinknv{k}(t)\,\Gc{k}
\end{equation}
where $G_0 \in \RM{\Bdim}$ is called the {\em drift matrix}, and $\{\Gc{k} \in \RM{\Bdim}\}_{k = 1}^{\nc}$ are called the {\em control matrices}. Computationally efficient procedures for converting between the Lindbladian form of \eqnrg{eq:lindbladian-dynamics}{eq:lindblad-super-D-def} and the bilinear form of \eqns{eq:cvs-dynamics,eq:ode-matrix-by-component} in $\bigO{\Bdim\log\Bdim}$ time are provided in~\koppthesis.

It is useful in some contexts to include the control input field as an explicit parameter in variables $\vec{v}$ and $G$. That is, $\vec{v}(t\xp\uin)$ denotes the state $\vec{v}(t)$ at time $t$ subject to the dynamics of \eqn{eq:cvs-dynamics} under control field $\uin$. Likewise, $G(t\xp\uin)$ denotes the value $G(t)$ under control field $\uin$.

\subsection{Decoherence-Free Subspaces}
\label{sec:bg-dfs}

A decoherence-free subspace (DFS) is formally ``an entire subspace of the system's Hilbert space [...] invariant under the noise''~\cite{cite:ipsstruct}. In this paper, we consider the more general case of a subsystem code $\CS$ having the property of zero information loss, unconditionally, over an infinite time horizon. We characterize ``zero information loss'' as the time invariance of the {\em Helstrom measure}~\cite{cite:helstrom}, which is further described in~\cite{cite:ipsstruct}. Concisely stated, a quantum subspace is considered {\em noiseless} if and only if the system dynamics, subject to control action, preserve the maximum distinguishability of all state pairs in the subspace for all time per the Helstrom measure. Equivalently, a subspace is considered noiseless if the noise acts isometrically on it~\cite{cite:ipsstruct}.

This definition of a DFS is broader than is used in, e.g., \cite{cite:dfsiac,cite:dynamdfs,cite:ipsstruct}, and technically falls into the broader category of a {\em noiseless subsystem}. However, our treatment considers only the case where the syndrome co-subsystem is one-dimensional, which Ticozzi et al.\ refer to as a DFS in \cite{cite:dfsiac} (as contrasted to noiseless subsystems more generally, where the syndrome co-subsystem may be nontrivial). The case of a noise-invariant DFS, as well as other restrictions on the influence of noise (e.g., the need for ``$\gamma$-robustness''~\cite{cite:dfsiac} of subsystem invariance) are equally well handled by the theory and necessitate only minor changes to the proposed algorithms.

\subsection{Subsystem Codes}
\label{sec:bg-subcode}

Given an integer $\CSord \in \IntRg{2}{\Bord}$, let $\CSHilspc$ be a Hilbert space with the same inner product as $\Hilspc$, and let $\BonHC$ denote the set of bounded operators on $\CSHilspc$, representable by matrices $\in \CM{\CSord}$. For a logical state encoding to exist, it suffices~\cite{cite:qnschoi} that a non-trivial C*-algebra of physical observables $\HonHC* \subseteq \HonH$ and a *-isomorphism $\Phi_{\csp} \in \Mapset{\BonH}{\BonHC}$ exist such that the image of $\HonHC*$ under $\Phi_{\csp}$ is a set of logical observables $\HonHC = \setbld{\hat J \in \BonHC}{\hat J = \adj{\hat J}}$. The set of logical code states, denoted $\DonHC$, is the positive, unit-trace subset of $\HonHC$, and the equivalent set of physical code states, $\DonHC* \subset \HonHC*$, is the preimage of $\DonHC$ under $\Phi_{\csp}$.

$\CSord$ is called the order of the subsystem code. Logical states are represented by matrices of order $\CSord$, or, equivalently, as coherence vectors with dimension $\CSdim$ where $\CSdim = \CSord^2$.

A decoherence-free~\footnote{according to the definition of section~\ref{sec:bg-dfs}} subsystem code must necessarily be embedded in the noise commutant of the system~\cite{cite:ipsstruct}. Supposing that a nontrivial noise commutant $\NC$ exists, it is a C*-algebra having dimension $\NCdim \in \IntRg{4}{\Bdim}$. A key result from noiseless subspaces literature~\cite{cite:commstruct,cite:ipsstruct,cite:ipsstruct2,cite:nsreview,cite:qnschoi} is that $\NC$ admits a canonical decomposition of the form
\begin{equation}
	\label{eq:ns-canon-decomp}
	\NC \cong \bigoplus\limits_{k = 1}^{\nsub{k}} {\eye{m_k} \otimes \NC_k}
\end{equation}
where $\cong$ denotes unitary equivalence.

Each term $\NC_k = \CM{\kord},\; \kord \in \Zp$ is representable by a order-$\kord$ complex matrix algebra. The number $m_k \in \Zp$ is called the {\em ampliated multiplicity}~\cite{cite:commstruct} or simply ``multiplicity'' of $\NC_k$. Following~\cite{cite:ipsstruct}, we call the $\nsub{k}$ summands in \eqn{eq:ns-canon-decomp} {\em $k$-sectors}. The number $\kord$ may differ between $k$-sectors and is called the ``order'' of a $k$-sector. Computing the decomposition of \eqn{eq:ns-canon-decomp} is nontrivial, especially when attempting to minimize computational complexity. A computationally efficient approach to computing the algebraic structure of the noise commutant is presented in~\koppthesis, building on the procedure developed by Holbrook et al.\ in~\cite{cite:commstruct}.

In~\koppthesis, it is further shown that every decoherence-free subsystem code, $\CS$, is uniquely defined (up to unitary equivalence in $\BonHC$) by a set of four parameters
\begin{equation}
	\label{eq:CS-param-tuple-def}
	(k,\CSord,\CSU^*,m_{\csp})
\end{equation}
where $k \nnrg{\nsub{k}}$ is the index of a $k$-sector in the decomposition of \eqn{eq:ns-canon-decomp} having $\kord \geq 2$, called the {\em host $k$-sector} of $\CS$; $\CSord \in \IntRg{2}{\kord}$; $m_{\csp} \nnrg{\lfloor\tfrac{\kord}{\CSord}\rfloor}$; and $\CSU* \in \CM{\kord}$ is a unitary matrix. From these four parameters, a subsystem code tuple~\footnote{Data $\CSU$ and $\CSbastx$ are included in \eqn{eq:CS-tuple-def} for consistency with~\koppthesis, but are not referenced in this paper.}
\begin{equation}
	\label{eq:CS-tuple-def}
	\CS \isdefas (\CSdim, \CSord, \HonHC, \HonHC*, \DonHC, \DonHC*, \Phi_{\csp}, \CSU, \CSbastx, \CSproj)
\end{equation}
can be derived. In addition to the first seven elements of \eqn{eq:CS-tuple-def}, future sections of this paper make use of the datum $\CSproj \in \CM{\Bdim}$, called the {\em core projection} of $\CS$, which is the orthogonal projection of lowest rank whose range contains all coherence vectors corresponding to logical code states in $\DonHC$. Computationally efficient procedures for deriving $\Phi_{\csp}$ and $\CSproj$ from the four parameters in \eqn{eq:CS-param-tuple-def} are provided in~\koppthesis.

It is notable that each $k$-sector with $\kord \geq 2$ serves as subsystem code, which is the unique subsystem code having $\CSord = \kord$. When referring to ``$k$-sectors as subsystem codes'', we follow the same convention as in~\koppthesis\ and replace $_{\csp}$ with $_k$. For example, $\kproj$ denotes the core projection of a $k$-sector, and $d_k$ denotes the dimension of its matrix algebra.

\subsection{Open-Loop Quantum Control}
\label{sec:bg-olc}

Control in the bilinear form of \eqn{eq:lindbladian-hamilton-term} is realizable in a variety of experimental setups~\cite{cite:bydbilin,cite:collrot2,cite:dfsstable,cite:gnliborzi,cite:qcopt2019,cite:qcscale,cite:tarndfs,cite:tarndfs2,cite:tarndfs2013,cite:timeoptgc}, including the specific model presented in section~\ref{sec:example}. Assuming each element of the control field can be actuated independently, the controllability of a quantum system in \eqn{eq:lindbladian-dynamics} when $\LindD[\rho] \equiv 0$ is determined by the generating set
\begin{equation}
	\label{eq:def-Hset}
	\CH^+ \isdefas \{\hat{H}_0\} \cup \CH
\end{equation}
where $\CH$ is the ordered set of control Hamiltonians
\begin{equation}
	\label{eq:def-wcrs}
	\CH \isdefas \{\Hc{1}, \Hc{2}, \dotsc, \Hc{\nc}\}
\end{equation}

It is convenient to rewrite the Lindbladian dynamics of \eqn{eq:lindbladian-dynamics} in the Liouvillian form
\begin{align}
	\label{eq:liouville-general-dynamics}
	\dot{\hat{U}}(t\xp\uin) & = -i\left[ \hat{H}(t\xp\uin), \hat{U}(t\xp\uin) \right]
		\speqsidebyside \hat{U}(0\xp\uin) = \eye{\Bord} \\
	\label{eq:liouville-general-dynamics2}
	\rho(t)                 & = \hat{U}(t\xp\uin)\rho(0)\adj{\hat{U}}(t\xp\uin)
\end{align}
where $\hat{U} \in \Mapset{\Rp\times\uinspc}{\lieU{\Bord}}$ and $\lieU{\Bord}$ denotes the unitary group. The time-varying operator $\hat{U}(t\xp\uin)$ is called the {\em Heisenberg state} or simply ``state'' of the system.

It is well known~\cite{cite:con4bimlq,cite:qcdbook} that the set of all reachable states, $\mathcal{R} \isdefas \setbld{\hat{X}(t)}{t \geq 0}$, for \eqn{eq:liouville-general-dynamics} is given by~\cite{cite:qcdbook}
\begin{equation}
	\label{eq:reachable-states-lie-group}
	\mathcal{R} = \expe{\lieham}
\end{equation}
which is the connected Lie group associated with the Lie algebra $\lieham$ generated by the set of all realizable Hamiltonians.

More formally, letting $\liecls{\metaset}$ denote the closure of set $\metaset$ under the commutator bracket and vector space operations, then
\begin{equation}
	\label{eq:def-lieham}
	\lieham = \liecls{-i\CH^+} = \liecls{-i\spanover{\uinvec \in \Rn{\nc}}{\hat{H}[\uinvec]}}
\end{equation}
where
\begin{equation}
	\label{eq:bilinear-control-form}
	\hat{H}[\uinvec] = \hat{H}_0 + \sum\limits_{k = 1}^{\nc} \uinvec_{(k)}\Hc{k}
\end{equation}

In this paper we consider two specific types of controllability in noiseless subspaces: a baseline controllability condition called {\em equivalent state controllability} (ESC) and a stricter condition called {\em operator controllability} (OC) that implies ESC. These universal quantum controllability standards are central to gate synthesis and collectively cover a range of control objectives~\cite{cite:notrap2017}. Formal definitions of both standards are given in~\cite{cite:qcdbook}. OC is also commonly called ``universality'' in physics literature (see, e.g., \cite{cite:dfsuqc2}). ESC is sufficient for certain specialized tasks such as state preparation and distillation, and suffices to implement some quantum algorithms, motivating our treatment of both controllability criteria.

The formal tests for both ESC and OC consider the Lie algebra $\lieham$ generated by the system Hamiltonians. In particular, it has been shown~\cite{cite:qcdbook} that a quantum system is OC if and only if $\dim[\lieham] \geq \Bdim - 1$, which is the well-known {\em Lie algebra rank condition}.

A system is ESC~\cite{cite:qcdbook} iff any of the conditions holds: either {\em i})~the system is OC; or, {\em ii})
\begin{equation}
	\label{eq:ESC-algebra-1}
	\lieham \cong \liesp{\Bord}
\end{equation}
or, {\em iii})
\begin{equation}
	\label{eq:ESC-algebra-2}
	\lieham \cong \liesp{\Bord} \oplus \spanof{i\eye{\Bord}}
\end{equation}
where $\liesp{\Bord}$ denotes the symplectic Lie algebra of order $\Bord$.

Conditions (i) and (ii) can be jointly expressed as the single condition~\cite{cite:qcdbook}:
\begin{equation}
	\label{eq:ESC-rank-condition}
	\dim [i\RMe[\Bord]{1}{1}, \lieham] = 2(\Bord - 1)
\end{equation}
where $\RMe[\Bord]{1}{1}$ is the $\RMe{1}{1}$ basis element of $\RM{\Bord}$.

It is important to note that the above standards of controllability assume the absence of noise and concern the entire physical state space. However, controllability of the entire physical state space is impossible in a system subject to noise, motivating the development of related controllability tests for encoded subsystems in this paper.

\section{Information-Preserving Control}
\label{sec:ipcontrol}

Until now we have ignored the control input field $\uin$ while working with DFSs by assuming that $\uin(t) \equiv 0$. To induce meaningful changes on code states and realize useful computations, we require $\uin$ to take on some set of nonzero values over time. This problem can be fundamentally reduced to one of {\em admissible control trajectories}, or equivalently, constraints on control action that preserve the noise-invariant properties of all physical states $\DonHC*$ for a given subsystem code $\CS$.

Necessary and sufficient conditions for noise-invariant control have previously been derived by Ticozzi et al.\ in~\cite{cite:dfsiac}. For the purposes of this paper, we find it convenient to derive conditions from first principles using the $G$ endomorphisms of \eqns{eq:cvs-dynamics,eq:ode-matrix-by-component}. This is done for three reasons: {\em i})~the definition of ``information preservation'' in \cref{sec:bg-dfs} differs subtly but meaningfully from all types of ``Markovian invariance'' described in~\cite{cite:dfsiac}; {\em ii})~the theoretical results of \cref{sec:ipcontrol} give rise to canonical matrix forms that are useful for theorems of later sections; and {\em iii})~the $G$ endomorphisms are central to various algorithms in~\koppthesis.

\subsection{Notation and Terminology}
\label{sec:ipc-notation}

We first define several sets and operators used in upcoming sections. Given a set of control Hamiltonians $\CH$ and a basis $\{\hat F_j\}_j$ as in \eqn{eq:cvsbasis-def}, each control Hamiltonian $\Hcx \in \CH$ yields an equivalent skew-symmetric control matrix $\Gcx$ term in \eqn{eq:ode-matrix-by-component}. The explicit conversion is given by
\begin{equation}
	\label{eq:Gc-def}
	\Gc{k} = \HtoG(\Hc{k}) \Forall k \inoneto{\nc}
\end{equation}
where
\begin{equation}
	\label{eq:H-to-G-conversion}
	\HtoG(\hat{X}) \isdefas -i\sum\limits_{j,l = 1}^{\Bdim}
		{\tr\left( \left[ \hat{X},\hat{F}_l \right]\hat{F}_j \right)\RMe{j}{l}} \Forall \hat{X} \in \HonH
\end{equation}

As a consequence of the Jacobi identity, the map $\convclr\HtoG$ is Lie isomorphic in that it is injective and it preserves the commutation operation with a $90^\circ$ phase factor:
\begin{equation}
	\label{eq:cvs-ham-conv-lie-isomorphism}
	\HtoG\left([\hat{X}_1,\hat{X}_2]\right) =
		i\left[\HtoG(\hat{X}_1),\HtoG(\hat{X}_2)\right] \Forall \hat{X}_1, \hat{X}_2 \in \HonH
\end{equation}

Since each $\Gc{k}$ is a matrix operating on the coherence vector space (CVS), it is called the {\em CVS equivalent} of $\Hc{k}$. By linearity, we have
\begin{equation}
	\label{eq:H-to-G-linearity}
	\HtoG\left(\sum\limits_k \uink{k}(t)\,\Hc{k}\right) = \sum\limits_k \uink{k}(t)\,\Gc{k}
\end{equation}

Let $\uinvecspcs \subseteq \Rn{\nc}$ denote the set of admissible input vectors at time $t$. That is, given $t \in \Rnn$, let
\begin{equation}
	\label{eq:def-uinvecspcs}
	\uinvecspcs \isdefas \setbld{\uin(t)}{\uin \in \uinspc}
\end{equation}

Furthermore, let $\Gs$ denote the set of all admissible $G$ endomorphisms at time $t$, i.e.,
\begin{equation}
	\label{eq:def-Gs}
	\Gs \isdefas \setbld{G_0 + \sum\limits_{k = 1}^{\nc} \uinvec_{(k)}\,\Gc{k}}{\uinvec \in \uinvecspcs}
\end{equation}

Finally, we make use of the following definition.

\begin{gcnsdef}
	Let $K = J + \tilde{J}$ be the sum of a normal, Hurwitz stable matrix $J \in \CM{n}$ and a skew-Hermitian matrix $\tilde{J} \in \CM{n}$. We refer to $K$ as a {\em lossy matrix}.
\end{gcnsdef}

\subsection{Canonical Matrix Forms}
\label{sec:ipc-canonform}

The major results of \cref{sec:ipcontrol} express necessary and sufficient conditions for noiseless control in terms of $G$ endomorphisms~\footnote{Throughout this paper we refer to ``$G$ endomorphisms'', which is a slight abuse of terminology since $G$ is an operator-valued function $G \in \Mapset{\Rnn}{\RM{\Bdim}}$ whose {\em value} $G(t)$ is an endomorphism. Even so, we omit the implicit parameter $(t)$ for brevity. When referring to the elements of, e.g., $\Gs$, no parameter is required and the term ``$G$ endomorphism'' is accurate.} having certain canonical forms. We begin by defining these canonical forms and establishing their existence.

\begin{gcnslemma}\label{lem:on-G0-structure}
	Given a drift matrix $G_0$, there exists a unitary matrix $\Lambda \in \CM{\Bdim}$, a real diagonal matrix $D \in \RM{\NCdim}$, and a Hurwitz matrix $G_{\perp} \in \CM{\NCperpdim}$ such that:
	\begin{equation}
		\label{eq:G0-block-structure}
		G_0 = \adj{\Lambda}\stdmat{i D & 0 \\ 0 & G_{\perp}}\Lambda
	\end{equation}
	where $\NCperpdim \isdefas \Bdim - \NCdim$.
\end{gcnslemma}
\begin{proof}
	Proof is provided in~\proofref{on-G0-structure}.
\end{proof}

The block structure of \eqn{eq:G0-block-structure} induces two orthogonal projections, $\NCproj, \NCprojperp \in \CM{\Bdim}$, where
\begin{align}
	\label{eq:def-NCproj}
	\NCproj     & \isdefas \adj{\Lambda}\stdmat{\eye{\NCdim} & 0 \\ 0 & \ZM{\NCperpdim}}\Lambda =
		\adj{\Lambda} \left(\eye{\NCdim} \oplus \ZM{\NCperpdim}\right) \Lambda \\
	\NCprojperp & \isdefas \eye{\Bdim} - \NCproj
\end{align}

\noindent such that $\Cn{\Bdim} = \im\NCproj \oplus \im\NCprojperp$.

The use of subscript $\NC$ in these symbols is deliberate. The image of $\NCproj$ is precisely the smallest subspace of $\Cn{\Bdim}$ containing the noise commutant of the system, $\NC$.

\begin{gcnsprop}\label{prop:on-C3x3-form}
	Given a subsystem code tuple [ref.~\eqn{eq:CS-tuple-def}] $\CS$ and noise commutant algebra of dimension $\NCdim$, there exists a basis $\{\hat{F}'_j\}_j$ with the properties of \eqn{eq:cvsbasis-def} such that:
	\begin{enumerate}
		\item the $G$ endomorphism of \eqn{eq:cvs-dynamics} can be partitioned as
		\begin{equation}
			\label{eq:G_t-C3x3-form}
			G(t) = \stdmat[r]{
				 \Gaa(t)       &  \Gab(t)       & \Gac(t) \\
				-\adj{\Gab}(t) &  \Gbb(t)       & \Gbc(t) \\
				-\adj{\Gac}(t) & -\adj{\Gbc}(t) & \hspace{6pt}\Gcc(t)}
		\end{equation}
		with $\Gaa(t) \in \CM{\CSdim}, \Gbb(t) \in \CM{\NCdim-\CSdim}$ being skew-Hermitian $\Foreach t,\uin$ and $\Gcc(t) \in \CM{\NCperpdim}$ being lossy $\Foreach t,\uin$
		\gcnsdeqitem{eq:NC-proj-block-diag} {\NCproj = \eye{\NCdim} \oplus \ZM{\NCperpdim}}
		\gcnsdeqitem{eq:CS-proj-block-diag} {\CSproj = \eye{\CSdim} \oplus \ZM{\Bdim-\CSdim}}
	\end{enumerate}
\end{gcnsprop}
\begin{proof}
	Proof is provided in~\proofref{on-C3x3-form}.
\end{proof}

We call the block form of \eqn{eq:G_t-C3x3-form} the {\em Control $3\times 3$ Form} (C3F) ``with respect to'' the pair $(\CS,\NCproj)$. Intuitively, we can consider the nine constituent matrix blocks as representing:
\begin{itemize}
	\item dynamics circulating information within the code subspace ($\Gaa$)
	\item dynamics circulating information losslessly outside the code subspace ($\Gbb$)
	\item lossy dynamics ($\Gcc\,$)
	\item dynamics exchanging information between lossless subspaces ($\Gab$, $-\adj{\Gab}$)
	\item dynamics exchanging information between lossless and lossy subspaces ($\Gac$, $-\adj{\Gac}$, $\Gbc$, $-\adj{\Gbc}$)
\end{itemize}

In many contexts, it is useful to consider the block form with the first two rows and first two columns ``merged'', i.e.,
\begin{align}
	\label{eq:G_t-CQF-form}
	G(t)  & = \stdmat[l]{
		 \ump\GQaa   & \GQab \\
		-\adj{\GQab} & \GQbb} \\
	\label{eq:def-GQaa}
	\GQaa & \isdefas \stdmat[l]{
		 \ump\Gaa    & \Gab \\
		-\adj{\Gab}  & \Gbb} \\
	\label{eq:def-GQab}
	\GQab & \isdefas \stdmat[l]{\Gab & \Gac}
\end{align}

We refer to this related form as the {\em Control Quadrant Form} (CQF) ``with respect to'' the projection $\NCproj$.

The two forms---C3F and CQF---induce a natural partitioning of the coherence vector $\vec v(t)$:
\begin{equation}
	\label{eq:covec-c3f-cqf-parts}
	\vec{v}(t) \isdefas \overbrace{\stdmat{\mkern-10mu\stdmat{\va(t) \\ \vb(t)}\mkern-10mu \\ \vc[\NC](t)}}^{\txt{C3F}}
		\equiv \overbrace{\stdmat{\vphantom{\stdmat{\va(t) \\ \vb(t)}}\vQa[\NC](t) \\ \vQb[\NC](t)}}^{\txt{CQF}}
\end{equation}

\noindent where the subscripts reflect the relationship of the partitions to the projections $\CSproj$, $\NCproj$, and $\NCprojperp$.

\begin{figure}[tbh]
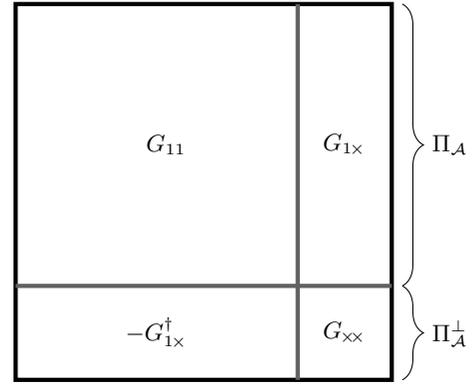
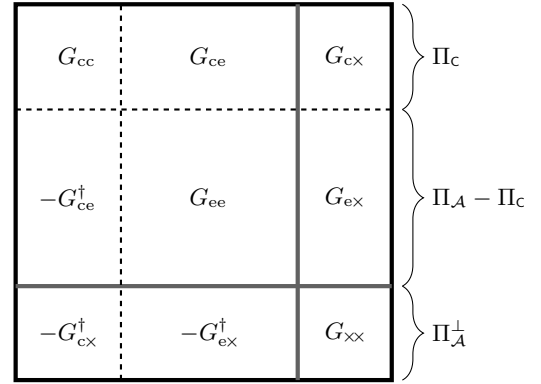
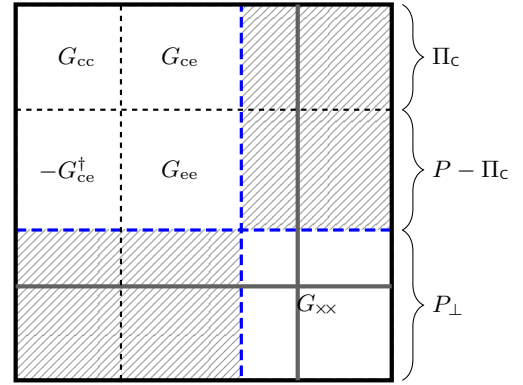

	\centering
	\begin{subfigure}[b]{\columnwidth}
		\InsertModule{figures/G-canon-forms/G-canon-forms}{CQF}
		\caption{control quadrant form (CQF)}
		\label{fig:G-canon-CQF}
	\end{subfigure}\vspace{6pt}

	\begin{subfigure}[b]{\columnwidth}
		\InsertModule{figures/G-canon-forms/G-canon-forms}{C3F}
		\caption{control $3\times3$ form (C3F)}
		\label{fig:G-canon-C3F}
	\end{subfigure}\vspace{6pt}

	\begin{subfigure}[b]{\columnwidth}
		\InsertModule{figures/G-canon-forms/G-canon-forms}{G-C3F}
		\caption{generalized control $3\times3$ form (G-C3F)}
		\label{fig:G-canon-G-C3F}
	\end{subfigure}

	\caption{Canonical block forms of $G$ matrices}
	\label{fig:G-canon-forms}
\end{figure}

The close relationship between the forms is depicted in \cref{fig:G-canon-CQF} and \cref{fig:G-canon-C3F} respectively, which also show the orthogonal projections (or equivalently, linear subspaces of $\Cn{\Bdim}$) associated with the matrix blocks.

\subsection{Propositions}
\label{sec:ipc-props}

In the propositions that follow, we assume $\uin \in \uinspc$ is a bounded, piecewise-continuous control input field, and we let $\CS$ be a decoherence-free subsystem code according to the definition of \cref{sec:bg-subcode}. The key results of \cref{sec:ipc-props} are threefold:

\begin{enumerate}
	\item \Cref{prop:CQF-pci-necc} provides a necessary and sufficient condition for information-preserving control control in the most general case, stating the condition in terms of the $G(t)$ endomorphism expressed in CQF. The condition is nonholonomic (i.e., time-varying and path-dependent), which limits its usefulness.
	\item \Cref{prop:on-CQF-with-arb-Q1} provides a time-invariant condition sufficient for information-preserving control, which is also affine, restricting $G(t)$ to a linear subspace for all time, and $\uinvecspcs$ to an affine subspace of $\Rn{\nc}$. As with \cref{prop:CQF-pci-necc}, the condition is expressed in terms of the $G(t)$ endomorphism in CQF.
	\item \Cref{cor:compute-DIFS} recasts the condition of \cref{prop:on-CQF-with-arb-Q1} into an equivalent constraint on the control input field $\uin$, introducing the notion of ``effective inputs''.
\end{enumerate}

\begin{gcnsprop}\label{prop:CQF-pci-necc}
	Let $\uin \in \uinspcs$ be a bounded, piecewise-continuous control input field, and $G(t\xp\uin)$ be a $G$ endomorphism in CQF with respect to the noise commutant projection $\NCproj$. Then a necessary and sufficient condition for state evolution with zero information loss is that
	\begin{equation}
		\label{eq:CQF-pci-necc}
		\stdmat{\ZMG{\NCperpdim}{\NCdim}\hspace{4pt} & \eye{\NCperpdim}}
			\dyson{k}{\stdmat{
				 \ump\GQaa   & \ZMG{\NCdim}{\NCperpdim}\\
				-\adj{\GQab} & \ZM{\NCperpdim}}}
			\vec{v}(t\xp\uin) = 0
	\end{equation}
	for all $t \in \Rnn, \uin \in \uinspcs, k \in \Znn$, where $\dyson{k}{\cdot},\; k \in \Znn$ are the coefficients of the Dyson series [ref.~(\dysoncoeffeqno) in~\extproofsappendix].
\end{gcnsprop}
\begin{proof}
	Proof is provided in~\proofref{CQF-pci-necc}.
\end{proof}

For brevity, we refer to any control input field satisfying the condition of \cref{prop:CQF-pci-necc} as {\em protective}.

The time-varying nature of the condition [and in particular, its dependence on $\vec{v}(t\xp\uin)$] makes it cumbersome to work with. Chap.~4 in~\koppthesis\ presents a class of protective controls, called ``P-switched control schemes'', that employ the result of \cref{prop:CQF-pci-necc} in a time-varying setting. However, a particularly desirable class of controls is one that {\em i})~can readily be shown to satisfy the full set of conditions in \eqn{eq:CQF-pci-necc} with minimal computation, and {\em ii})~imposes control input constraints that are affine and time-invariant, which greatly simplifies controllability analysis.

A straightforward approach to realizing such a class of controls is to select a single projection $\Pproj \in \CM{\Bdim}$ such that $\im\CSproj \subseteq \im\Pproj \subseteq \im\NCproj$ and contains $\vec{v}(t\xp\uin)$ for all time. That is,
\begin{equation}
	\label{eq:v-P-NC-chain}
	\vec{v}(t\xp\uin) \subseteq \im\Pproj \Forall t
\end{equation}
or equivalently,
\begin{equation}
	\label{eq:v-P-NC-chain-a}
	\vQa[\NC](t\xp\uin) \subseteq \im\Pproja \Forall t
\end{equation}
where $\Pproja \in \CM{\NCdim}$ is the upper left $\NCdim\times\NCdim$ block of $\Pproj$ in CQF such that $\Pproj = \Pproja\oplus\ZM{\NCperpdim}$.

Assuming the system state is a valid physical code state in $\CS$ at $t = 0$, then $\vec{v}(0\xp\uin) \in \im\CSproj$ and hence $\vec{v}(0\xp\uin) \in \im\Pproj$ by assumption. Having imposed the bounds on $\Pproj$, it remains to satisfy the conditions of \eqn{eq:CQF-pci-necc} for all $t$, which is addressed in the following \namecrefs{prop:on-CQF-with-arb-Q1}.

\begin{gcnsprop}\label{prop:on-CQF-with-arb-Q1}
	Let the conditions of \cref{prop:CQF-pci-necc} hold, and let $\Pproja \in \CM{\NCdim}$ and $\Pproj = \Pproja\oplus \ZM{\NCperpdim}$ be orthogonal projections such that $\im\CSproj \subseteq \im\Pproj \subseteq \im\NCproj$. Then a sufficient condition for protective control is
	\begin{align}
		\label{eq:pci-P-cond1}
		\im \GQaa\Pproja    & \subseteq \im\Pproja \Forall t \\
		\label{eq:pci-P-cond2}
		\adj{\GQab} \Pproja & = 0 \Forall t
	\end{align}
\end{gcnsprop}
\begin{proof}
	Proof is provided in~\proofref{on-CQF-with-arb-Q1}.
\end{proof}

The conditions of \cref{prop:on-CQF-with-arb-Q1} are analogous to linear subspace decoupling~\cite{cite:ccdecouple} in classical control theory---a principle seen in other noiseless quantum control proposals such as~\cite{cite:pcebyqfc,cite:tarndfs,cite:tarndfs2,cite:tarndfs2013}. In particular, we treat $\im\Pproj$---a subspace of the noise commutant---as insular and privileged, and impose a time-invariant constraint on $G(t)$ to prevent coupling between $\im\Pproj$ and its complementary space, $\im\Pprojperp$, where $\Pprojperp \isdefas \eye{\Bdim} - \Pproj$.

It follows from a proof nearly identical to \cref{prop:on-C3x3-form} that for any $\Pproj$ satisfying the conditions of \cref{prop:on-CQF-with-arb-Q1}, there exists a basis such that $G(t)$ may be written in CQF ``with respect to'' $\Pproj$ and in C3F with respect to the pair $\CPtuple$, subject to the substitutions: $\Pproj$ for $\NCproj$, $\Pprojperp$ for $\NCprojperp$, and $\Pdim \isdefas \rank\Pproj$ for $\NCdim$. When $G(t)$ is written in this form, $\Pproja = \eye{\Pdim}$, hence \eqn{eq:pci-P-cond1} is automatically satisfied, and \eqn{eq:pci-P-cond2} reduces to
\begin{equation}
	\label{eq:CQF-pci-condition}
	\adj{\GQab} = 0 \Forall t
\end{equation}
i.e., zeroing of the off-diagonal quadrants of $G(t)$ such that $G(t) = \GQaa(t) \oplus \GQbb(t) \Forall t$.

Given a tuple $\CPtuple$, we refer to the C3F block form as {\em Generalized Control $3\times 3$ Form} (G-C3F), distinguishing it from the ``standard'' form when $\Pproj = \NCproj$. \Cref{fig:G-canon-G-C3F} shows the block structure of the $G$ endomorphism in G-C3F, juxtaposed with standard C3F in \cref{fig:G-canon-C3F} to emphasize similarity. In \cref{fig:G-canon-G-C3F}, we note that the $\Gac$ and $\Gbc$ blocks (collectively, the $\GQab$ block in CQF) are grayed out to indicate nullity, and that the smaller size of the upper left $2\times 2$ block (collectively, $\GQaa$ block in CQF) relative to the standard $\GQaa$ block (shown in light gray) reflects $\Pproj$'s status as a sub-projection of $\NCproj$. The forms are otherwise identical.

{\em Generalized Control Quadrant Form} (G-CQF), which is not depicted in the figure, is identical to \cref{fig:G-canon-CQF} with the exception that $\Pproj$ is substituted for $\NCproj$, and $\Pprojperp$ for $\NCprojperp$---the same changes that distinguish G-C3F from C3F.

Having established a linear, time-invariant condition sufficient to guarantee protective control, it remains to convert this condition into an equivalent set of constraints on the control input field $\uin$. To this end, we define several terms and present a key \namecrefs{cor:compute-DIFS} to facilitate the needed conversion.

\begin{gcnsdef}*
	Any subsystem code/orthogonal projection pair $\CPtuple$ satisfying the relationship $\im\CSproj \subseteq \im\Pproj \subseteq \im\NCproj$ is called a {\em protective $\CPtuple$ pair}. The projection $\Pproj$ is called the {\em subsystem projection} of the pair, and the subspace $\im\Pproj$ is called the {\em protected subspace}.
\end{gcnsdef}

\begin{gcnsdef}
	Given a protective $\CPtuple$ pair and control resource set $\CH$, if there exists a nonempty set of control input fields such that the condition of \cref{prop:on-CQF-with-arb-Q1} is satisfied, the subsystem projection $\Pproj$ is called $\CH$-invariant. The subset of control input fields in $\uinspc$ that render $\Pproj$ $\CH$-invariant is called the {\em Decoupling Input Function Space} (DIFS) of $\Pproj$ and is denoted $\uinspcs$.

	If the zero input $\uin(t) \equiv \ZV{\nc}$ is an element of $\uinspcs$, the projection $\Pproj$ is called {\em drift-invariant}.
\end{gcnsdef}

Let $\{\vec\pi_j\}_{j = 1}^{\Pdim}$ be the eigenvectors of $\Pproj$, i.e.,
\begin{equation}
	\Pproj = \sum\limits_{j = 1}^{\Pdim} \vec\pi_{j}\adjv{\vec\pi_j}
\end{equation}
and let $\vec z(j,k)$ be a vector-valued function defined as
\begin{equation}
	\begin{gathered}
		\vec{z}(j,k) \isdefas \ifother
			{\Pprojperp G_0\vec\pi_j}   {k = 0}
			{\Pprojperp \Gc k\vec\pi_j} \\
		\speqinset \Forall k \in 0,\dotsc,\nc,\;j \inoneto{\Pdim}
	\end{gathered}
\end{equation}

Additionally, let
\begin{align}
	\label{eq:def-zu}
	\zu & \isdefas \def\x#1{\vec z(#1,0)}\stdvecx{\Pdim} \\
	\label{eq:def-Zu}
	\Zu & \isdefas \def\x#1#2{\vec z(#1,#2)}\stdmatx{\Pdim}{\nc}
\end{align}
such that $\zu \in \Cn{\Bdim\Pdim}$ and $\Zu \in \RMG{\Bdim\Pdim}{\nc}$.

We can then state the following \namecrefs{cor:compute-DIFS}.

\begin{gcnscor}\label{cor:compute-DIFS}
	The DIFS $\uinspcs$ of an $\CH$-invariant projection $\Pproj$ may be expressed as a linear-affine set of the form
	\begin{equation}
		\label{eq:DIFS-of-P}
		\uinspcs = \setbld{\uin \in \uinspc}{\uin(t) \equiv \uinalt(t) + \znu}
	\end{equation}
	where
	\begin{equation}
		\label{eq:CH-invariance-Zu-zu-cond}
		\Zu\uinalt(t) \equiv 0 \speqsidebyside \Zu\znu = -\zu
	\end{equation}

	Moreover, $\znu = 0$ is a solution if and only if $\Pproj$ is drift-invariant.
\end{gcnscor}
\begin{proof}
	Proof is provided in~\proofref{compute-DIFS}.
\end{proof}

Letting $\nceff \isdefas \nc - \rank\Zu$, it is useful to rewrite \eqn{eq:DIFS-of-P} in the form
\begin{equation}
	\label{eq:pci-input-constraints-canon}
	\uinspcs = \setbld{\zNu\ueff + \znu}{\ueff \in \ueffspc}
\end{equation}
where $\zNu \in \RMG{\nc}{\nceff}$ is a rank-$\nceff$ matrix having the property that $\im\zNu = \nullspc\Zu$, and where $\ueffspc \subset \Mapset{\Rnn}{\Rn{\nceff}}$.

We call the set $\ueffspc$ the {\em effective input function space}, and the field $\ueff$ the {\em effective input field}. Given the assumptions about the domain of $\uinspc$, it is readily seen that all fields in $\ueffspc$ are bounded and piecewise-continuous. The representation of \eqn{eq:pci-input-constraints-canon} is useful in that it permits the set $\Gs$ to be expressed in terms of a field $\ueff$ that is unconstrained within its domain.

In practice, both $\zu$ and $\Zu$ consist almost entirely of zeros, making computation of $\zNu$ and $\znu$ a modest $\bigO{\nc \Pdim n \Bdim}$ task given a sparse representation of all matrix operators. In order for a solution to exist to \eqn{eq:CH-invariance-Zu-zu-cond}, $\zu$ must fall inside the range of $\Zu$. Hence, a simple test for the $\CH$-invariance of $\Pproj$ is that
\begin{equation}
	\label{eq:CH-invariance-condition}
	\rank\stdmat{\Zu & \zu} = \rank\Zu
\end{equation}

In summary, $\uinspcs$ is determined by first constructing $\zu$ and $\Zu$, checking the condition of \eqn{eq:CH-invariance-condition}, computing $\znu$ by solving the equality of \eqn{eq:CH-invariance-Zu-zu-cond}, and finally by computing $\zNu$ (whose rows span $\nullspc\Zu$) by a standard procedure such as SVD.

\section{Operations in Subsystem Codes}
\label{sec:csops}

\subsection{Logical Operators}
\label{sec:csops-logops}

Given a subsystem code tuple $\CS$ as in \eqn{eq:CS-tuple-def}, we may consider the influence of each control Hamiltonian $\Hcx \in \CH$ on the physical states in $\DonHC*$ as having a {\em logical equivalent action} on the logical states in $\DonHC$. That is, assuming a set of admissible control inputs $\uinspcs$ can be chosen such that model dynamics under the influence of $\uinspcs$ never cause $\rho(t)$ to exit $\DonHC*$, we can exploit the *-isomorphic nature of $\Phi_{\csp}$ and map each control Hamiltonian in $\Hcx \in \CH$ to an operator $\Jcx \in \HonHC$, called a {\em logical Hamiltonian}, that operates on the logical state $\sigma(t) \in \DonHC$ in the same way as $\Hcx$ operates on a physical state.

Formally, we obtain a set of logical Hamiltonians $\{\Jc{k}\}_{k = 1}^{\nc}$ by applying $\Phi_{\csp}$ to the elements of $\CH$, i.e.,
\begin{equation}
	\label{eq:equiv-ham-on-codespace}
	\Jc{k} = \Phi_{\csp}(\Hc{k}) \Forall k \inoneto{\nc}
\end{equation}
such that the commutative relationship
\begin{gather}
	\label{eq:equiv-ham-commute-diagram}
	\begin{gathered}  
	\mbox{\begin{tikzpicture}[anchor = center, > = stealth]
		\draw (0,0)        node (rho)    {$\rho$};
		\draw (4.5cm,0)    node (drho)   {$d\rho$};
		\draw (0,-2cm)     node (sigma)  {$\sigma$};
		\draw (4.5cm,-2cm) node (dsigma) {$d\sigma$};
		\draw[->] (rho)   -- (drho)   node[pos = 0.5, above] {$-i[\Hcx,\rho]$};
		\draw[->] (rho)   -- (sigma)  node[pos = 0.5, left]  {$\Phi_{\csp}$};
		\draw[->] (sigma) -- (dsigma) node[pos = 0.5, below] {$-i[\Phi_{\csp}(\Hcx),\sigma]$};
		\draw[->] (drho)  -- (dsigma) node[pos = 0.5, right] {$\Phi_{\csp}$};
	\end{tikzpicture}}
	\end{gathered}
\end{gather}

\noindent holds subject to the condition that the physical state $\rho$ never exits $\DonHC*$.

This gives rise to a set of logical state dynamics with a logical drift Hamiltonian $\hat J_0 \isdefas \Phi_{\csp}(\hat H_0)$ and a logical Hamiltonian
\begin{align}
	\label{eq:bilinear-control-form-log}
	\hat{J}[\uinvec]  & \isdefas \hat{J}_0 + \sum\limits_{k = 1}^{\nc} \uinvec_{(k)}\Jc{k} \\
	\label{eq:hamiltonian-round-square-rel-log}
	\hat{J}(t\xp\uin) & \isdefas \fxr{\hat{J}[\uinvec]}{\uinvec = \uin(t)}
\end{align}
such that
\begin{equation}
	\label{eq:logical-state-dynamics}
	\dot \sigma(t) = -i[\hat{J}(t\xp\uin), \sigma(t)]
\end{equation}
which is a direct analog of \eqn{eq:lindbladian-hamilton-term} [or \eqn{eq:liouville-general-dynamics} in the Heisenberg picture].

Because the logical dynamics of \eqn{eq:logical-state-dynamics} are an analog of the Liouvillian dynamics of \eqn{eq:liouville-general-dynamics}, they are subject to the same controllability standards reviewed in \cref{sec:bg-olc} with only minor changes, provided $\rho(t)$ remains in $\DonHC*$ for all time.

Specifically, controllability of $\CS$ is determined by the Lie algebra $\liehamCS \subseteq \HonHC$ generated by the logical Hamiltonian set $\CJ^+$, where $\CJ^+ \isdefas \{\hat{J}_0\} \cup \{\Jc{k}\}_k$, analogously to \eqn{eq:def-Hset}. If the Lie group $\expe{\liehamCS}$ is equal to $\lieSU{\CSord}$ or $\lieU{\CSord}$, $\CS$ is designated {\em Logically Operator Controllable} (L-OC). This kind of controllability is referred to as ``encoded universality'' by Zanardi et al.\ in~\cite{cite:dfsuqc2} and elsewhere.

If the group is equal to either of the Lie groups generated by the Lie algebras of \eqns{eq:ESC-algebra-1,eq:ESC-algebra-2}, $\CS$ is designated {\em Logically Equivalent State Controllable} (L-ESC). Both notions of controllability are analogous to the strong controllability of \cref{sec:bg-olc}, differing only in that they pertain to the logical states in $\CS$ rather than to the full set of physical states for the model.

\subsection{Extended Operator Spaces}
\label{sec:csops-extss}

In the context of the $\CH$-invariant spaces introduced in \cref{sec:ipc-props}, the requirement $\rho(t) \in \DonHC* \Forall t$ is unnecessarily restrictive. Although $\rho(t)$ must fall inside $\DonHC*$ to correspond with a logical state, there is no requirement that such a correspondence be maintained at all times during the evolution of the physical state. Indeed, the only times where the correspondence is necessary are the start and end of any quantum operation, and at the time the state is measured. At all other times, $\rho(t)$ is subject only to the restriction that it reside within the noise commutant. When working in a protected subspace, this restriction is tightened to $\vec{v}(t)$ lying in the image of $\Pproj$ as a consequence of \cref{prop:on-CQF-with-arb-Q1}.

At the heart of the issue is that physical states may be subject to dynamics that preserve their noiseless character but violate the *-isomorphic relationship between physical and logical states, and can therefore be manipulated in ways that do not correspond to the action of any logical Hamiltonian. That is, control may effect reversible dynamics on the physical state for which no $\Phi_{\csp}$ map exists such that \eqn{eq:equiv-ham-commute-diagram} commutes.

Although control may be tailored to generate only physical dynamics that correspond to logical Hamiltonian dynamics, this unnecessarily limits the degree of control that can be exerted over the logical state. In particular, the set $\CJ^+$ may contain too few generators to satisfy the conditions of L-OC or L-ESC, thus rendering the logical state uncontrollable. Exploiting additional control freedom is therefore critical for ensuring systems remain controllable with fewer control resources.

To accommodate this greater set of potentially non-Hamiltonian control options, we must consider an extended state space of greater dimension than $\DonHC$, and an extended set of endomorphisms on this space. We devote the remainder of \cref{sec:csops} to rigorously defining this space, its properties, and its associated set of endomorphisms, leaving a treatment of controllability in the space to \cref{sec:pstatic}.

\Hiii{Extended State Space}%
The two requirements for this extended space are that {\em i})~there exists an injective *-homomorphic map from $\DonHC$ to states in the extended space so that reversible operations on the extended states imply reversible operations on logical states and vice versa, and {\em ii})~the extended state space resides entirely within $\im\Pproj$ in order to satisfy the conditions of \cref{prop:on-CQF-with-arb-Q1}.

\begin{figure}[htb]
	\centering

%
%

\newcommand {\xlsrefdimincm} {3.2}  

\begin{tikzpicture}[> = stealth, x = \xlsrefdimincm cm, y = -\xlsrefdimincm cm, every node/.style = {anchor = center}]
	\newcommand{\pt}{*0.03528/\xlsrefdimincm}  

	\newcommand{\Psz}     {0.6}     
	\newcommand{\CSsz}    {0.3}     
	\newcommand{\Gsep}    {0.25}    
	\newcommand{\vwid}    {24\pt}   
	\newcommand{\opspc}   {15\pt}   
	\newcommand{\juxspc}  {0.07}    
	\newcommand{\tlmarg}  {0.2}     
	\newcommand{\brpad}   {0.015}   
	\newcommand{\brsz}    {0.03}    
	\newcommand{\pbgmarg} {0.5\pt}  
	\newcommand{\eqmarg}  {2\pt}    

	\newcommand{\xvll}  {0}                    
	\newcommand{\xvlc}  {\xvll+\vwid/2}        
	\newcommand{\xvlr}  {\xvll+\vwid}          
	\newcommand{\xeqop} {\xvlr+\opspc/2}       
	\newcommand{\xGl}   {\xvlr+\opspc}         
	\newcommand{\xCSc}  {\xGl+\CSsz/2}         
	\newcommand{\xCSr}  {\xGl+\CSsz}           
	\newcommand{\xCSPc} {\xGl+\CSsz/2+\Psz/2}  
	\newcommand{\xPc}   {\xGl+\Psz/2}          
	\newcommand{\xPPc}  {(\xPr)/2+(\xGr)/2}    
	\newcommand{\xPr}   {\xGl+\Psz}            
	\newcommand{\xGr}   {\xGl+1}               
	\newcommand{\xvrl}  {\xGr+\juxspc}         
	\newcommand{\xvrc}  {\xvrl+\vwid/2}        
	\newcommand{\xvrr}  {\xvrl+\vwid}          

	\newcommand{\ytlc}  {0}                    
	\newcommand{\yGt}   {\ytlc+\tlmarg}        
	\newcommand{\yGc}   {\yGt+1/2}             
	\newcommand{\yCSc}  {\yGt+\CSsz/2}         
	\newcommand{\yCSb}  {\yGt+\CSsz}           
	\newcommand{\yCSPc} {\yGt+\CSsz/2+\Psz/2}  
	\newcommand{\yPb}   {\yGt+\Psz}            
	\newcommand{\yPPc}  {\yGt+\Psz/2+1/2}      
	\newcommand{\yGb}   {\yGt+1}               
	\newcommand{\yyGt}  {\yGt+1+\Gsep}         
	\newcommand{\yyGc}  {\yyGt+1/2}            
	\newcommand{\yyPc}  {\yyGt+\Psz/2}         
	\newcommand{\yyPb}  {\yyGt+\Psz}           
	\newcommand{\yyPPc} {\yyGt+\Psz/2+1/2}     
	\newcommand{\yyGb}  {\yyGt+1}              

	\definecolor{lightgray} {rgb} {0.75,0.75,0.75}
	\definecolor{midgray}   {rgb} {0.627,0.627,0.627}
	\definecolor{darkgray}  {rgb} {0.38,0.38,0.38}

	\tikzstyle{matvecbrace}    = [black, line width = 0.9pt]
	\tikzstyle{Gblockborder}   = [black, line width = 0.5pt]
	\tikzstyle{Pblockghosting} = [line width = 0.8pt, color = lightgray, dash pattern = on 2pt off 2pt]

	\tikzstyle{labelarrow}     = [
		->,
		line width = 0.5pt,
		color      = midgray]

	\tikzstyle{ssbrace}        = [decorate, decoration = {brace, amplitude = 6pt}, xshift = 5pt]
	\tikzstyle{ssbracemath}    = [black, midway, anchor = west, xshift = 4pt]

	\tikzstyle{injectionarrow} = [
		line cap   = butt,
		shorten >  = 3pt,
		line width = 3pt,
		color      = lightgray,
		decoration = {markings, mark = at position 1 with {\arrow[scale = 0.9, > = stealth]{>}}},
		postaction = {decorate}]

	\tikzstyle{injectionlabel} = [color = midgray, xshift = -2pt]
	\tikzstyle{extlogsunitbox} = [line width = 0.5pt, dash pattern = on 2pt off 2pt, color = lightgray]

	\newcommand {\mvbrack} [4] {
		\draw [matvecbrace] (#1-\brpad+\brsz,#3-\brpad) --
			(#1-\brpad,#3-\brpad) --
			(#1-\brpad,#4+\brpad) --
			(#1-\brpad+\brsz,#4+\brpad);

		\draw [matvecbrace] (#2+\brpad-\brsz,#3-\brpad) --
			(#2+\brpad,#3-\brpad) --
			(#2+\brpad,#4+\brpad) --
			(#2+\brpad-\brsz,#4+\brpad);
	}

	\newcommand {\nummxblock} [5] {
		\draw [Gblockborder] (#1,#3) rectangle (#2,#4);
	}

	\newcommand{\drawinjectionarrow}[2]{
		\draw [injectionarrow] (#1) -- node [injectionlabel,sloped,above] {${}_\mathsf{INJECTION}$} (#2);
	}

	\mvbrack {\xvll}{\xvlr}{\yGt}{\yGb}
	\draw (\xvlc,\yCSc)  node {$\dva$};
	\draw (\xvlc,\yCSPc) node {$\ZV{\PnotCSdim}$};
	\draw (\xvlc,\yPPc)  node {$\ZV{\Pperpdim}$};

	\mvbrack {\xvll}{\xvlr}{\yyGt}{\yyGb}
	\draw (\xvlc,\yyPc)  node {$\dvQa$};
	\draw (\xvlc,\yyPPc) node {$\ZV{\Pperpdim}$};

	\draw (\xeqop,\yGc)  node {$=$};
	\draw (\xeqop,\yyGc) node {$=$};

	\mvbrack    {\xGl} {\xGr} {\yGt} {\yGb}
	\nummxblock {\xGl} {\xCSr}{\yGt} {\yCSb}{2}
	\nummxblock {\xCSr}{\xPr} {\yCSb}{\yPb}{3}
	\nummxblock {\xPr} {\xGr} {\yPb} {\yGb}{5}
	\draw (\xCSc,\yCSc)   node {$\Gaa$};
	\draw (\xCSPc,\yCSPc) node {$\Gbb$};
	\draw ({\xPPc},\yPPc) node {$\Gcc$};

	\mvbrack    {\xGl}{\xGr}{\yyGt}{\yyGb}
	\nummxblock {\xGl}{\xPr}{\yyGt}{\yyPb}{4}
	\nummxblock {\xPr}{\xGr}{\yyPb}{\yyGb}{5}
	\draw (\xPc,\yyPc)     node {$\GQaa$};
	\draw ({\xPPc},\yyPPc) node {$\GQbb$};

	\mvbrack {\xvrl}{\xvrr}{\yGt}{\yGb}
	\draw (\xvrc,\yCSc)  node (logstatevec) {$\va$};
	\draw (\xvrc,\yCSPc) node               {$\ZV{\PnotCSdim}$};
	\draw (\xvrc,\yPPc)  node               {$\ZV{\Pperpdim}$};

	\mvbrack {\xvrl}{\xvrr}{\yyGt}{\yyGb}
	\draw (\xvrc,\yyPc)  node {$\vQa$};
	\draw (\xvrc,\yyPPc) node {$\ZV{\Pperpdim}$};

	\draw [Pblockghosting]
		(\xvrr,\yPb+\pbgmarg) --
		(\xGl-\pbgmarg,\yPb+\pbgmarg) --
		(\xGl-\pbgmarg,\yGt-\pbgmarg) --
		(\xvrr,\yGt-\pbgmarg);

	\draw [Pblockghosting] (\xPr+\pbgmarg,\yPb+\pbgmarg) -- (\xPr+\pbgmarg,\yGt-\pbgmarg);
	\draw [Pblockghosting]
		(\xvrr,\yyPb+\pbgmarg) --
		(\xGl-\pbgmarg,\yyPb+\pbgmarg) --
		(\xGl-\pbgmarg,\yyGt-\pbgmarg) --
		(\xvrr,\yyGt-\pbgmarg);

	\draw [Pblockghosting] (\xPr+\pbgmarg,\yyPb+\pbgmarg) -- (\xPr+\pbgmarg,\yyGt-\pbgmarg);

	\draw [ssbrace] (\xvrr,\yGt) -- (\xvrr,\yPb)
		node (uppersslabel) [ssbracemath] {$\in \Sn{\CSdim} \oplus \ZV{\PnotCSdim}$};

	\draw [ssbrace] (\xvrr,\yyGt) -- (\xvrr,\yyPb)
		node (lowerssbrace) [midway, anchor = west, xshift = 3pt] {};

	\draw (uppersslabel.center |- 0,\yyPc)
		node [anchor = center] (lowersslabel) {$\Sn{\Pdim}$};

	\path (lowerssbrace.east) -- node {$\in$} (lowersslabel.west);

	\drawinjectionarrow{\xPc,\yPb}{\xPc,\yyGt}
	\drawinjectionarrow{uppersslabel.south}{lowersslabel.north}

	\draw (\xGr,\ytlc) node[anchor = east] (logham) {$[\hat J,\,\cdot\,]$};
	\draw [labelarrow] (logham.west |- 0,\ytlc) --
		node [rectangle, black, fill = white, midway] {$\CSHtoG$} (\xCSc,\ytlc) ->
		(\xCSc,\yGt);

	\draw (uppersslabel.center |- 0,\ytlc) node (logstate) {$\sigma$};
	\draw [labelarrow] (logstate.west |- 0,\ytlc) --
		node [rectangle, black, fill = white, midway] {$\CSRtoV$} (\xvrc,\ytlc) ->
		(\xvrc,0 |- logstatevec.north);

	\path (\xvll,\yGb+\eqmarg)  -- node[anchor = north] {(a)} (uppersslabel.east |- 0,\yGb+\eqmarg);
	\path (\xvll,\yyGb+\eqmarg) -- node[anchor = north] {(b)} (uppersslabel.east |- 0,\yyGb+\eqmarg);
\end{tikzpicture}

	\caption{%
		Relationship between the code state space, the working state space, and endomorphisms operating on these spaces}
	\label{fig:ext-log-state}
\end{figure}
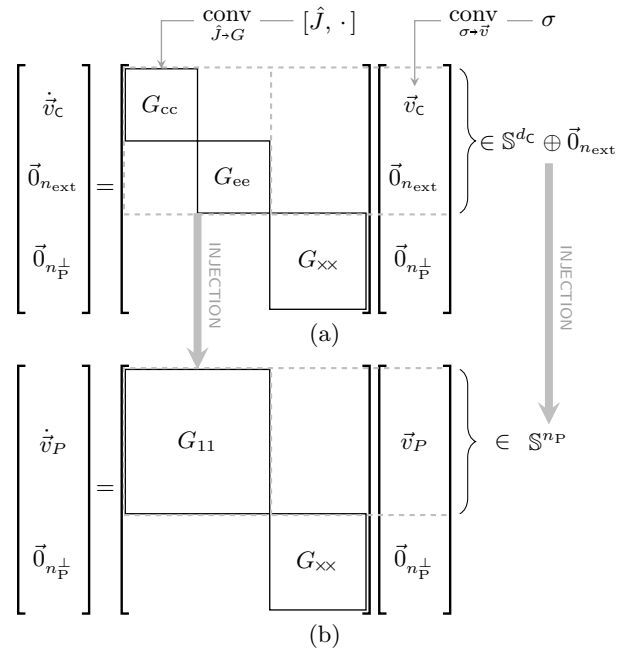

Assume that a tuple $\Pstuple$ specifying a subsystem code, subsystem projection, and control resource set has been selected via a suitable algorithm (see, e.g., Chap.~5 of~\koppthesis) such that $\im\CSproj \subseteq \im\Pproj$. Development of the extended state space is most easily handled using the coherence vectors and $G$ endomorphisms of the linear ODE model of \eqns{eq:cvs-dynamics,eq:covec-to-phys-state} with a basis selected such that $G$ endomorphisms are expressed in G-C3F with respect to $\CPtuple$, which reduces most intermediate steps to trivial operations on matrix blocks. The development follows \cref{fig:ext-log-state}, which depicts the relationships between pertinent states, maps, and endomorphisms.

Consulting the G-C3F diagram of \cref{fig:G-canon-G-C3F}, we note that the matrix is symmetrically partitioned with three row partitions and three column partitions, making nine blocks total. These are the same nine blocks as in \cref{fig:G-canon-C3F}, with four blocks explicitly grayed out to indicate nullity (which is the implication of \cref{prop:on-CQF-with-arb-Q1} for $G$ matrices in G-C3F), and with the border between the second and third partitions moved leftward/upward to signify that the protected subspace, $\im\Pproj$, is generally of lesser dimension than the noise commutant algebra and is embedded within it.

Ordered from left to right (or equivalently, top to bottom), we refer to the three G-C3F partitions as the {\em code partition}, the {\em extended partition}, and the {\em unprotected partition} respectively, with descriptions as follows:

\begin{itemize}
	\item The {\em code partition} is the top/left partition corresponding to elements of $\vec v(t)$ supported on $\CSproj$. In \cref{fig:ext-log-state}a, a coherence vector is juxtaposed with the $G$ matrix, having the form of \eqn{eq:covec-c3f-cqf-parts} with $\Pproj$ substituted for $\NC$. The juxtaposition highlights the correspondence between the G-C3F matrix partitions and the elements of \eqn{eq:covec-c3f-cqf-parts}. In particular, only the topmost $\CSdim$ elements of the vector are supported on $\CSproj$, implying that for every physical code state $\rho(t) \in \DonHC*$, the corresponding coherence vector $\vec v(t)$ lies in the domain
	\begin{equation}
		\label{eq:covec-G-C3F-parts}
		(\vec v(t) = \va(t) \oplus \ZV{\PnotCSdim} \oplus \ZV{\Pperpdim}) \in \VonHC
	\end{equation}
	where $\PnotCSdim \isdefas \Pdim - \CSdim$, where
	\begin{equation}
		\label{eq:def-VonHC}
		\begin{split}
			\VonHC & \isdefas \left(\Rn{\CSdim} \oplus \ZV{\PnotCSdim} \oplus \ZV{\Pperpdim}\right) \cap \Sn{\Bdim} \\
				{} & = \Sn{\CSdim} \oplus \ZV{\PnotCSdim} \oplus \ZV{\Pperpdim}
		\end{split}
	\end{equation}
	and where $\va(t) \in \Sn{\CSdim}$ is called the ``code part'' of $\vec v(t)$.
	\item The {\em extended partition} is the middle partition corresponding to elements of $\vec v(t)$ supported on $\Pproj-\CSproj$. The dimension of $\im\Pproj \ominus \im\CSproj$ is $\PnotCSdim$, and therefore the extended partition blocks occupy the $\PnotCSdim$ rows below the code partition (likewise, the $\PnotCSdim$ columns right of the code partition).

	In \eqn{eq:covec-c3f-cqf-parts}, the extended partition corresponds to a vector $\vb(t) \in \Sn{\PnotCSdim}$ called the ``extended part'' of $\vec v(t)$. In \cref{fig:ext-log-state}a, the extended part of $\vec v(t)$ is identically zero, which is the case for all states $\in \VonHC$.
	\item The {\em unprotected partition} is the bottom/right partition corresponding to elements of $\vec v(t)$ supported on $\Pprojperp$. In \eqn{eq:covec-c3f-cqf-parts}, the unprotected partition corresponds to a vector $\vc(t) \in \Sn{\PnotCSdim}$ called the ``unprotected part'' of $\vec v(t)$. The unprotected partition is shared by matrices in G-CQF. In both \cref{fig:ext-log-state}a (G-C3F) and \cref{fig:ext-log-state}b (G-CQF), the unprotected part of $\vec v(t)$ is identically zero.
\end{itemize}

Contrasting \cref{fig:ext-log-state}a with \cref{fig:ext-log-state}b, the code partition and extended partition are merged into a single partition in the latter. This merging of partitions, which distinguishes G-CQF from G-C3F, creates a partition having $\Pdim$ rows/columns called the {\em working partition}. The working partition corresponds to the elements of $\vec v(t)$ supported on
\begin{equation*}
	\CSproj + (\Pproj - \CSproj) = \Pproj
\end{equation*}

In \eqn{eq:covec-c3f-cqf-parts}, the working partition is host to a vector $\vQa[\NC](t) \in \Sn{\NCdim}$ called the ``working part'' of $\vec v(t)$. When considering G-CQF (rather than standard CQF), the working part of $\vec v(t)$ is denoted $\vQa(t)$, as in \cref{fig:ext-log-state}b. This partitioning of the coherence vector into working and unprotected parts yields vectors of the form
\begin{equation}
	\vec v(t) = \vQa(t) \oplus \ZV{\Pperpdim}
\end{equation}
residing in the domain
\begin{equation}
	\label{eq:def-VonPproj}
	\begin{split}
		\VonPproj & \isdefas \Sn{\Bdim} \cap (\Cn{\Pdim} \oplus \ZV{\Pperpdim}) \\
		       {} & = \Sn{\Pdim} \oplus \ZV{\Pperpdim}
	\end{split}
\end{equation}

Clearly $\VonHC \subset \VonPproj$, with the mapping between the code part $\va(t)$ and the working part $\vQa(t)$ being the trivial injection: $\vQa(t) = \va(t) \oplus \ZV{\PnotCSdim}$. The set $\VonPproj$ therefore satisfies condition (i) of the extended state space. Furthermore, since $\im\Pproj = \Cn{\Pdim} \oplus \ZV{\Pperpdim}$ in G-CQF, it follows that $\VonPproj \subset \im\Pproj$ (and that $\VonPproj$ is in fact the largest possible subset of $\Sn{\Bdim}$ with this property). Hence, $\VonPproj$ also satisfies condition (ii) of the extended state space. From these observations, we conclude that $\VonPproj$ is the largest possible set capable of serving as an extended state space with the required properties.

\Hiii{Extended Hamiltonian Space}%
The partitioned structure of G-C3F and G-CQF is likewise useful for defining the sets of endomorphisms ($G$ matrices) operating on $\VonHC$ and $\VonPproj$.

Because a logical Hamiltonian $\hat J$ acts as an endomorphism on the logical state space, its equivalent $G$ endomorphism, given by
\begin{equation}
	\label{eq:def-CSHtoG}
	G = \CSHtoG(\hat J) \isdefas \HtoG\cdot\Phi_{\csp}^{-1}(\hat J) \Forall \hat{J} \in \HonHC
\end{equation}
likewise leaves $\VonHC$ invariant, implying the $\Gab$ and $\Gac$ blocks of \cref{fig:G-canon-C3F} are null, along with counterparts $-\adj{\Gab}$ and $-\adj{\Gac}$. Since blocks $\Gbc$ and $-\adj{\Gbc}$ are also null by \cref{prop:on-CQF-with-arb-Q1}, it follows that $G$ has the block diagonal structure depicted in \cref{fig:ext-log-state}a, with support on blocks $\Gaa$, $\Gbb$ and $\Gcc$, and with block $\Gaa$ alone affecting logical states.

For convenience, we denote the image of $\HonHC$ under $\CSHtoG\convclr$, which has this block structure, as $\GonHC \subset \CM{\Bdim}$, noting that it is a linear Lie algebra. For a given element of $\GonHC$, we refer to the $\Gaa$ matrix block as the ``code block'', the $\Gbb$ matrix block as the ``extended block'', and the $\Gcc$ block as the ``unprotected block''. \Cref{fig:ext-log-state}a demonstrates how the mapping of a logical Hamiltonian, $\hat{J}$, to the code block, and a logical state, $\sigma$, to the code partition, in the case where $\vec v(t) \in \VonHC$, trivially reduces to dynamics of the form
\begin{equation}
	\label{eq:logical-cvs-dynamics}
	\dva(t) = \Gaa(t)\, \va(t)
\end{equation}
which is the ODE analog of \eqn{eq:logical-state-dynamics}.

We also note for future reference that because $\Phi_{\csp}^{-1}$ maps $\eye{\CSord}$ to a scalar multiple of the identity, i.e., $a\eye{\Bord},\; a \in \Rl$, and because $\eye{\Bord}$ lies in the kernel of $\HtoG\convclr$ [the identity operator trivially commutes with all $\hat{F}_l$'s in \eqn{eq:H-to-G-conversion}], it follows that
\begin{equation}
	\label{eq:CSHtoG-kernel}
	\CSHtoG(c\eye{\CSord}) = \ZM{\Bdim} \Forall c \in \Cx
\end{equation}

The set of endomorphisms acting on $\VonPproj$ follows a similar derivation, with the exception that this latter set of endomorphisms is not required to leave $\VonHC$ invariant. All four matrix blocks $\{\Gaa,\Gab,-\adj{\Gab},\Gbb\}$ covering the code partition and extended partition in \cref{fig:G-canon-C3F} may be nonzero, and are therefore merged into a single matrix block, $\GQaa$, called the ``working block'', in \cref{fig:G-canon-CQF}, which acts on the working partition of $\vec v(t)$.

The two requirements for this extended set of endomorphisms are that: {\em i})~it resides in the image of $\convclr\HtoG$, and {\em ii})~it conforms to the block structure of G-CQF. Because the $\GQaa$ block acts on $\VonPproj$, we denote the set of all such blocks $\GonPproj \subset \im\convclr\HtoG$. It is a linear Lie algebra and a superset of the set of all $\CH$-invariant $G$ endomorphisms (with respect to $\Pproj$). As depicted in \cref{fig:ext-log-state}, it is also a superset of $\GonHC$, with (code block, extended block) pairs mapped to working blocks via the trivial injection: $\GQaa(t) = \Gaa(t) \oplus \Gbb(t)$.

\subsection{Subspace Restrictions}
\label{sec:csops-ssrest}

As a precursor to controllability assessment, it is pertinent to ask: Given a subsystem projection $\Pproj$, assuming total control can be effected over the system Hamiltonian $\hat H$, and assuming every endomorphism $\in \GonPproj$ can be generated, are there any universal restrictions on how the code partition and extended partition can interact and exchange information while system dynamics remain $\CH$-invariant? The answer to this question is `yes', with a particularly severe restriction relating to the $k$-sectors of \cref{sec:bg-subcode}.

\begin{gcnslemma}\label{lem:c-star-matrix-blocks}
	Let $\mathcal{A} \cong \lieglc{L}$ and $\mathcal{B} \subseteq \lieglc{L_1} \oplus \lieglc{L_2} \oplus \dotsb \oplus \lieglc{L_p},\;p \in \Zp$ be C* algebras. If $L_j < L \Forall j$ then $\mathcal{A} \not\cong \mathcal{B}$.
\end{gcnslemma}
\begin{proof}
	A well-established result is that the ring of matrices $\lieglc{n}$ for $n \geq 1$ is simple.

	We proceed by induction on $p$. If $p = 1$, the proof follows by simply counting dimensions. Suppose the result holds for some $p$, and let
	\begin{equation*}
		\mathcal{B} = \lieglc{L_1} \oplus \dotsb \oplus \lieglc{L_{p+1}}
	\end{equation*}
	with $L_j < L$ for $1 \leq j \leq p+1$. To prove by contradiction, suppose $\lieglc{L} \subseteq \mathcal{B}$. Then $\lieglc{L} \cap (0 \oplus \lieglc{L_{p+1}})$ is an ideal in $\lieglc{L}$, and since $L_{p+1} < L$, by simplicity we must have $\lieglc{L} \cap (0 \oplus \lieglc{L_{p+1}}) = 0$. This results in
	\begin{equation*}
		\lieglc{L} \subset \lieglc{L_1} \oplus \dotsb \oplus \lieglc{L_{p}} \oplus 0
	\end{equation*}
	which is impossible. Since $\mathcal{A} \cong \lieglc{L}$, it follows that $\mathcal{A} \not\cong \mathcal{B}$.
\end{proof}

\begin{gcnslemma}\label{lem:subsystem-code-upper-bound}
	The $k$-sectors with order $\kord \geq 1$ in the decomposition of \eqn{eq:ns-canon-decomp} constitute {\em maximal} subsystem codes in that the order of the largest possible noiseless subsystem code, $\kordmax$, is bounded by $\kordmax \leq \max\left( \kord[1], \kord[2], \dotsc, \kord[p]\right)$.
\end{gcnslemma}
\begin{proof}
	That every $k$-sector is a subsystem code follows trivially from the observation that the $k$-sector algebra $\mathcal{A}_k$ resides in a matrix block $\cong \lieglc{\kord}$ according to the decomposition of \eqn{eq:ns-canon-decomp}. The upper bound on $\CSord$ is proved by contradiction. Assume $\CSord > \kordmax$. We require that $\BonHC \cong \lieglc{\CSord}$ and that $\BonHC*$ resides in the noise commutant, i.e., $\BonHC* \subseteq \NC$. We have that $\BonHC* \cong \BonHC$ and thus $\BonHC* \cong \lieglc{\CSord}$. However, $\NC \cong \lieglc{\kord[1]} \oplus \lieglc{\kord[2]} \oplus \dotsb \oplus \lieglc{\kord[\nsub{k}]}$ where $\kord \leq \kordmax < \CSord \Forall k$. Thus, by \cref{lem:c-star-matrix-blocks}, $\BonHC* \not\cong \lieglc{\CSord}$, a contradiction. It follows that $\CSord$ must be $\leq \kordmax$.
\end{proof}

As a consequence of \cref{lem:subsystem-code-upper-bound}, the core projection of a valid subsystem code must be a sub-projection of any one $k$-sector core projection. This same restriction need not apply to the subsystem projection $\Pproj$. For example, is it easily shown that $\Pproj = \kproj[k_1] + \kproj[k_2]$ for any two $k$-sectors $k_1$ and $k_2$ is a valid subsystem projection. Even so, regarding where an extended state can circulate, $k$-sectors prove to be partition boundaries for systems subject to open loop control, as shown in the following \namecrefs{prop:on-k-sector-decoupling}.

\begin{gcnsprop}\label{prop:on-k-sector-decoupling}
	Given the assumptions of \cref{prop:CQF-pci-necc}, let $\kproj \in \CM{\Bdim}$ be the core projection of a $k$-sector. Then
	\begin{equation}
		\label{eq:k-sector-decoup-cond-gen}
		(\NCproj - \kproj)\, G(t)\, \kproj = 0 \Forall t
	\end{equation}
\end{gcnsprop}
\begin{proof}
	Proof is provided in~\proofref{on-k-sector-decoupling}.
\end{proof}

\Cref{prop:on-k-sector-decoupling} implies that even with total control over the system Hamiltonian, in the absence of coupling between the protected and unprotected subspaces of the system, no control exists that is capable of coupling the elements of $\vQa(t)$ supported on the core projection $\kproj$ of the host $k$-sector to elements of $\vQa(t)$ not supported on $\kproj$ (i.e., those elements supported on $\NCproj - \kproj$). Put differently, a working state space $\VonPproj$ spanning more than one $k$-sector (i.e., having the property that $\kproj \vQa \neq 0$ for more than one $k$) cannot exchange information between the partitions of the state residing in different $k$-sectors.

Hence, while a subsystem projection $\Pproj$ {\em may possibly not be} a sub-projection of $\kproj$, any extension subspace $\im\Pproj \ominus \im\kproj$ is irrelevant to the controllability of subsystem codes hosted by this $k$-sector since information can never be exchanged between the subspaces inside and outside the $k$-sector.

It is important to note that this inability to exchange information between $k$-sectors is a consequence of the algebraic structure of $\NC$, and can be lifted by relaxing the restriction that $\im\Pproj \subseteq \NC$. In the event $\im\Pproj \not\subseteq \NC$, states under the influence of control will irreversibly bleed information into the bath, but if the effects of these irreversible dynamics are sufficiently small, they may be ignored or corrected via QEC.

Specifically, the eigenspace of $G(t)$ corresponding to an eigenvalue $\lambda$ with $\repart\{\lambda\} \approx 0$ can be treated as though $\repart\{\lambda\} = 0$ for the purposes of selecting $\Pproj$, provided the control dynamics are sufficiently fast. This process limits the effects of noise to a quantity dominated by $\tau$, the time spent outside the noise commutant, which can (in theory) be made arbitrarily small by applying sufficiently strong control fields---an effect is also known as {\em noise softening}~\cite{cite:qfcbymeas} (as contrasted with noise-{\em free} subspaces). We do not provide a treatment of noise softening in this paper.

\section{\texorpdfstring{$\Pproj$}{P}-Static Controllability}
\label{sec:pstatic}

\subsection{Preliminaries}
\label{sec:pstatic-prelim}

\noindent As a brief review:

\begin{itemize}
	\item \Cref{sec:ipc-canonform} defined the canonical matrix forms necessary for analysis.
	\item \Cref{sec:ipc-props} provided the most general condition for information-preserving control in \cref{prop:CQF-pci-necc}. This was simplified to a sufficient decoupling condition in \cref{prop:on-CQF-with-arb-Q1}, and recast as a set of input constraints in \cref{cor:compute-DIFS}.
	\item \Cref{sec:csops-logops} defined the L-OC and L-ESC controllability standards for a subsystem code, analogous to the OC and ESC standards of \cref{sec:bg-olc}.
	\item Finally, \cref{sec:csops-extss} defined extended state and Hamiltonian ($G$ endomorphism) spaces necessary for developing the theory of logical controllability outside the constraints of strict physical-to-logical state correspondence at all times.
\end{itemize}

All four treatments rely on an implicit subsystem code $\CS$ and an implicit control resource set $\CH$. The latter three treatments also rely on an implicit subsystem projection $\Pproj$. These three data---$\CS$, $\Pproj$, and $\CH$---collectively parameterize one of the many possible ways to represent/control noiseless information in an open quantum system. The combination of data, i.e., $\Pstuple$, is called a ``$P$-static control scheme'' in~\koppthesis. The term ``$P$-static'' derives from the use of a single, static subsystem projection $\Pproj$, as contrasted with $P$-switched schemes, which switch between multiple subsystem projections, and $P$-free schemes, which have no $\Pproj$ datum. We treat only $P$-static schemes in this paper.

Assume the $\Psscheme$ triple has been chosen. Our objective is to assess whether the subsystem code $\CS$ meets the CS-OC and/or CS-ESC controllability standard(s), given the control resources in $\CH$, subject to the linear-affine, time-invariant constraints on the control input field imposed by $\Pproj$ (via \cref{cor:compute-DIFS}). We refer to the assessment as a test for ``$\Pproj$-static controllability'', consistent with the terminology in~\koppthesis.

$P$-Static controllability analysis is carried out in the extended, partitioned operator spaces of G-C3F. When working with $G$ endomorphisms in G-C3F algebraically, we find it convenient to define several transformational operators, which are maps between sets of matrices. In each definition, assume $p, m \in \Zp$ and $m > p$.

\begin{gcnsdef}*
	Given a nonempty matrix set $\metaset \subseteq \CM{m}$, the {\em $p$-diagonal subset of $\metaset$} is the (possibly empty) subset of matrices in $\metaset$ expressible as the direct sum of a $p\times p$ matrix and an $(m-p)\times (m-p)$ matrix. A nonempty matrix set that is equal to its own $p$-diagonal subset (i.e., containing only block-diagonal elements having the specified dimensions) is called a $p$-diagonal set.

	\noindent For each matrix
	\begin{equation*}
		(A = A_1 \oplus A_2) \in \metaset
	\end{equation*}

	\noindent of a $p$-diagonal set $\metaset$, the {\em $p$-block} of $A$ is the matrix operator $A_1 \in \CM{p}$.
\end{gcnsdef}

\begin{gcnsdef}*
	Given a nonempty matrix set $\metaset \subseteq \CM{m}$, {\em $p$-erasure of $\metaset$}, denoted $\ers{p}{\metaset}$\label{defsym:perasure}, is the (possibly empty) set $\metasetalt \subseteq \CM{p}$ of all $p$-blocks of the $p$-diagonal subset of $\metaset$. That is,
	\begin{equation}
		\label{eq:def-p-erasure}
		\ers{p}{\metaset} \isdefas \setbldst{A_1}{\exists\; A_2 \in \CM{m-p}}{A_1 \oplus A_2 \in \metaset}
	\end{equation}
\end{gcnsdef}

\begin{gcnsdef}
	Given a nonempty matrix set $\metasetalt \subseteq \CM{p}$, the {\em $m$-extension of $\metasetalt$}, denoted $\ext{m}{\metasetalt}$, is the set $\metaset \subseteq \CM{m}$ of $p$-diagonal matrices whose $p$-erasure is $\metasetalt$. That is,
	\begin{equation}
		\label{eq:def-m-extension}
		\ext{m}{\metasetalt} \isdefas \setbld{A_1 \oplus A_2}{A_1 \in \metasetalt,\;A_2 \in \CM{m-p}}
	\end{equation}
\end{gcnsdef}

In literal terms, $\ers{p}{\metaset}$ filters $\metaset$ down to its block diagonal elements $\in \CM{p} \oplus \CM{m-p}$ and then projects each element onto its $p\times p$ upper left block; $\ext{m}{\metasetalt}$ acts as the inverse projection, yielding the subset of $\CM{p} \oplus \CM{m-p}$ whose $p$-erasure is $\metasetalt$.

Finally, upcoming analysis refers to a set $\wkgspc \subset \CM{\Bdim}$:
\begin{equation}
	\label{eq:def-wkgspc}
	\wkgspc \isdefas \setbldst[\speqsidebyside]{C \in \CM{\Bdim}}{\exists\;U \in \lieSU{\CSord}}
		{(\Foreach \sigma \in \DonHC)\sppredtostmt \CSRtoV(U\sigma\adj U) = C\,\CSRtoV(\sigma)\,}
\end{equation}
where $\CSRtoV\convclr \isdefas \RtoV\convclr \cdot \Phi_{\csp}^{-1}$, which is the set of all endomorphisms whose action on a coherence vector $\vec v \in \VonHC$ is equivalent to a logical gating operation. When $G$ endomorphisms are expressed in G-C3F, $\wkgspc$ is a set (in fact, a semigroup) of $\CSdim$-diagonal matrices.

\subsection{Lie Algebraic Criteria}
\label{sec:pstatic-liealg}

As established in \cref{sec:ipc-props}, the $\CH$-invariance of $\Pproj$ reduces to the condition that $\GQab(t) \equiv 0$ when $G(t)$ is in G-CQF with respect to $\Pproj$, giving rise to the set of endomorphisms $\GonPproj$. Any channel matrix $C$ generated by the dynamics
\begin{align}
	\label{eq:G_t-block-form-dynamics}
	\dot{C}(t) & = G(t)\,C(t) \speqsidebyside G(t) \in \GonPproj \Forall t \in \Rnn \\
	C(0)       & = \eye{\Bdim} \\
	\label{eq:channel-matrix-action}
	\vec v(t)  & = C(t)\vec v(0)
\end{align}
will, at time $T$, have the structure
\begin{equation}
	\label{eq:channel-matrix-P-invar}
	C(T) = \CQaa(T) \oplus \CQbb(T)
\end{equation}
with a unitary matrix $\CQaa(T) \in \CM{\Pdim}$ in the working block and a lossy matrix $\CQbb(T) \in \CM{\Pperpdim}$ in the unprotected block. The unprotected block is disposable from the standpoint of analysis since $\vec v(0) \in \VonPproj$ by assumption, and by \eqn{eq:channel-matrix-action} we have:
\begin{equation}
	\begin{split}
		\vec{v}(T) & = C(T)\,\vec{v}(0) \\
		        {} & = \left(\CQaa(T) \oplus \CQbb(T)\right)\left(\vQa(0) \oplus \ZV{\Pperpdim}\right) \\
		        {} & = \CQaa(T)\,\vQa(0) \oplus \ZV{\Pperpdim}
	\end{split}
\end{equation}
and hence the effect of $\CQbb(T)$ on the output state vanishes.

Any channel matrix corresponding to a logical gate must necessarily leave $\VonHC$ invariant and therefore has the form
\begin{equation}
	\label{eq:channel-matrix-CS-invar}
	C(T) = \Caa(T) \oplus \Cbcbc(T)
\end{equation}
with unitary matrix $\Caa(T) \in \CM{\CSdim}$ and a normal matrix $\Cbcbc(T) \in \CM{\Bdim-\CSdim}$. (The unitarity of $\Caa(T)$ follows from its generation by lossless dynamics.)

It follows that any channel matrix having the structure of both \eqn{eq:channel-matrix-P-invar} and \eqn{eq:channel-matrix-CS-invar} will be of the form
\begin{equation}
	\label{eq:channel-matrix-3x3}
	C(T) = \Caa(T) \oplus \Cbb(T) \oplus \Ccc(T)
\end{equation}
with $\Caa(T)$, $\Cbb(T)$, and $\Ccc(T)$ occupying the code block, extended block, and unprotected block of G-C3F respectively, and with $\Caa(T)$ and $\Cbb(T)$ being unitary.

The set of all matrices with the given block structure constitutes an upper bound on the set of channel matrices that leave both $\VonHC$ and $\VonPproj$ invariant, which is equivalently set of channels that both {\em i})~correspond to unitary operations on $\va(t) \in \VonHC$, via $\CSdim$-erasure, i.e.,
\begin{equation}
	\vec{v}(T) = C(T)\,\vec{v}(0) \implies \va(T) = \Caa\va(0)
\end{equation}
and {\em ii})~can be generated while keeping $\im\Pproj$ invariant $\Foreach t$, thereby satisfying \cref{prop:on-CQF-with-arb-Q1}.

The problem of testing for $P$-static controllability can therefore be solved using a four-step procedure involving G-C3F matrix mechanics:
\label{page:pstatic-test-outline}
\begin{enumerate}[label=\Roman*., ref=\Roman*]
	\item\label{item:P-stat-chan1} For a given resource set $\CH$, determine the set $\metasetn{N}$ of all generatable channel matrices of the form in \eqn{eq:channel-matrix-P-invar}. If the set is empty, conclude that $\Pproj$ is not $\CH$-invariant. Otherwise, compute the set of equivalent unitary endomorphisms $\metasetn{P}$ operating on $\VonPproj$ via $\Pdim$-erasure; that is, $\metasetn{P} = \ers{\Pdim}{\metasetn{N}}$.
	\item\label{item:P-stat-chan2} Determine the set of all channel matrix code blocks corresponding to logical gates, i.e.,
	\begin{align}
		\label{eq:P-stat-chan-all-gates}
		\metasetn{\csp} & \isdefas \ers{\CSdim}{\wkgspc} \\
		             {} & = \setbldst{\Caa \in \CM{\CSdim}}{\exists\; \Cbcbc \in \CM{\Bdim-\CSdim}}
		                    	{\Caa \oplus \Cbcbc \in \wkgspc}
	\end{align}
	where $\wkgspc$ is as defined in \eqn{eq:def-wkgspc}, and then compute the set of equivalent unitary endomorphisms $\metasetn{P}'$ operating on $\VonPproj$ via $\Pdim$-extension; that is, $\metasetn{P}' = \ext{\Pdim}{\metasetn{\csp}}$.
	\item\label{item:P-stat-chan3} Compute the set intersection $\metasetn{P}^* = \metasetn{P} \cap \metasetn{P}'$, which is the set of all generatable endomorphisms acting on $\VonPproj$ that {\em i})~leave $\VonHC$ invariant, and {\em ii})~correspond to logical gates. If the intersection is empty, conclude that the subsystem code is uncontrollable. Otherwise, compute the set of equivalent unitary endomorphisms $\metasetn{\csp}^*$ acting on $\VonHC$ via $\CSdim$-erasure; that is, $\metasetn{\csp}^* = \ers{\CSdim}{\metasetn{P}^*}$, which will be a Lie group.
	\item\label{item:P-stat-chan4} Determine whether $\metasetn{\csp}^*$ is *-isomorphic to any of the groups required to satisfy the given controllability standard (L-OC or L-ESC).
\end{enumerate}

Although this procedure is straightforward, determining channel matrix set representations such that the extension and erasure operations, as well as the intersection operation of step~\ref{item:P-stat-chan3}, are computationally tractable is problematic. Because each set is a Lie group, however, with an associated Lie algebra, and since subgroup relationships between Lie groups are preserved as subalgebra relationships between their respective Lie algebras, we may represent each group by its algebra. This reduces the analysis procedure to matrix multiplications and familiar operations on vector spaces, providing the needed computational tractability. Hence, rather than working with groups of channel matrices, we derive controllability conditions based on the $G$ endomorphisms constituting their Lie algebras.

\begin{gcnstheorem}\label{thrm:control-algebras}
	Given a set of admissible $G$ endomorphisms $\Gs$ and a protective $\CH$-invariant pair $\CPtuple$, the subsystem code $\CS$ is L-OC if and only if
	\begin{equation}
		\label{eq:L-OC-cond}
		\ers{\CSdim}{\CSHtoG(\liesu{\CSord})} \subseteq \ers{\CSdim}{\lieGx{P}}
	\end{equation}

	\noindent where
	\begin{align}
		\label{eq:control-algebras-lieGxP}
		\lieGx{P}               & \isdefas \liecls{\ers{\Pdim}{\Gs}} \\
		\CSHtoG(\liesu{\CSord}) & \isdefas \setbld{\CSHtoG(\hat F_{\csp})}{\hat F_{\csp} \in \liesu{\CSord}}
	\end{align}
\end{gcnstheorem}
\begin{proof}
	Two useful identities involving the erasure and extension operations are as follows:
	\begin{itemize}
		\item Given a nonempty set $\metaset \subseteq \CM{m}$ of $p$-diagonal matrices generating a Lie algebra, the operations of Lie closure and $p$-erasure commute. That is,
		\begin{equation}
			\label{eq:lie-cls-p-ers-commute}
			\ers{p}{\liecls{\metaset}} = \liecls{\ers{p}{\metaset}}
		\end{equation}
		\item Given matrix sets $\metaset \subseteq \CM{p}$ and $\metasetalt \subseteq \CM{m}$ with $m > p$:
		\begin{equation}
			\label{eq:p-ers-m-ext-ident2}
			\ers{p}{\left((\ext{m}{\metaset}) \cap \metasetalt\right)} = \metaset \cap \ers{p}{\metasetalt}
		\end{equation}
	\end{itemize}

	The proof follows the four-step procedure outlined above. In step~\labelcref{item:P-stat-chan1}, we consider that $\metasetn{\Bdim} = \exp{\lieGx{\Bdim}}$ where $\lieGx{\Bdim} = \liecls{\Gs}$. With $G$ endomorphisms expressed in G-C3F with respect to $\CPtuple$, every endomorphism $G(t) \in \Gs$ has the block diagonal form of \cref{fig:ext-log-state}b on account of $\CH$-invariance. Furthermore, by the definition of $\CH$-invariance, $\Gs$ is known to be nonempty. The Lie algebraic closure of block diagonal matrices is also block diagonal, hence $\lieGx{\Bdim}$ is a linear subspace $\subseteq \lieglc{\Pdim} \oplus \lieglc{\Pperpdim}$. The equivalent actions on $\VonPproj$, which we denote by the set $\lieGx{P}$, are given by $\Pdim$-erasure; that is
	\begin{equation}
		\label{eq:lie-P-def}
		\lieGx{P} = \ers{\Pdim}{\lieGx{\Bdim}}
	\end{equation}
	where it can be shown that $\lieGx{P}$ is a Lie algebra due to the block diagonal nature of $\lieGx{\Bdim}$.

	Because $\lieGx{\Bdim}$ is $\Pdim$-diagonal, the commutation identity of \eqref{eq:lie-cls-p-ers-commute} may be applied to \eqref{eq:lie-P-def}, yielding
	\begin{equation}
		\lieGx{P} = \ers{\Pdim}{\liecls{\Gs}} = \liecls{\ers{\Pdim}{\Gs}}
	\end{equation}

	In step~\labelcref{item:P-stat-chan2}, we consider that $\metasetn{\csp} = \exp{\lieGx{\csp}'}$ where $\lieGx{\csp}'$ resides in the code block of $\im\CSHtoG\convclr$, a set which comprises all $G$ endomorphisms corresponding to Hamiltonian-like evolution of $\sigma(t) \in \DonHC$. The set of all logical Hamiltonian operators acting on $\sigma(t)$ is spanned by $\lieu{\CSord}$. However, since $\lieu{\CSord}\backslash\liesu{\CSord} \cong \Rl\cdot i\eye{\CSord}$ and $i\eye{\CSord}$ lies in the kernel of $\CSHtoG\convclr$ [ref.~\eqn{eq:CSHtoG-kernel}], the $i\eye{\CSord}$ basis component of any logical Hamiltonian vanishes in $\Gaa$ and $\CSHtoG(\lieu{\CSord}) = \CSHtoG(\liesu{\CSord})$. Therefore,
	\begin{equation}
		\label{eq:CSHtoG-u-su-equiv}
		\ers{\CSdim}{\im\CSHtoG} = \ers{\CSdim}{\CSHtoG(\lieu{\CSord})} = \ers{\CSdim}{\CSHtoG(\liesu{\CSord})}
	\end{equation}
	without loss of generality.

	We thus have
	\begin{equation}
		\label{eq:def-lie-CSp}
		\lieGx{\csp}' \isdefas \ers{\CSdim}{\im\CSHtoG} = \ers{\CSdim}{\CSHtoG(\liesu{\CSord})}
	\end{equation}
	where $\ers{\CSdim}{\CSHtoG\convclr(\liesu{\CSord})}$ is the LHS of \eqn{eq:L-OC-cond}. We note that $\lieGx{\csp}'$ is guaranteed to be nonempty as a consequence of $\CS$ being a subsystem code. That is, the existence of the $\Phi_{\csp}$ *-isomorphism suffices to guarantee that $\dim \lieGx{\csp}' = \CSdim$ and therefore that $\lieGx{\csp}'$ is nonempty and nontrivial. Hence, for the condition of \eqn{eq:L-OC-cond} to hold, both $\ers{\CSdim}{\lieGx{P}}$ and $\lieGx{P}$ must be nonempty and nontrivial. If either $\lieGx{P}$ or $\ers{\CSdim}{\lieGx{P}}$ is found to be empty or trivial, the controllability test must return a ``not controllable'' outcome.

	The set of equivalent unitary actions acting on coherence vectors, called ``working CVS gates'' in~\koppthesis, is given by
	\begin{equation}
		\begin{split}
			\metasetn{P}' & = \metasetn{\csp} \oplus \lieU{\PnotCSdim} \\
			           {} & = \exp\{\lieGx{\csp}'\} \oplus \exp\{\lieu{\PnotCSdim}\} \\
			           {} & = \exp\left\{\lieGx{\csp}' \oplus \lieu{\PnotCSdim}\right\}
		\end{split}
	\end{equation}
	implying the Lie algebraic analog of $\metasetn{P}'$ should be $\lieGx{P}' = \lieGx{\csp}' \oplus \lieu{\PnotCSdim}$.

	The $\Pdim$-extension operation of step~\labelcref{item:P-stat-chan2} maps the set $\lieGx{\csp}'$ to
	\begin{equation}
		\label{eq:lieGxP-from-Pdim-ext}
		\lieGx{P}' = \ext{\Pdim}{\lieGx{\csp}'} = \lieGx{\csp}' \oplus \lieglc{\PnotCSdim}
	\end{equation}
	rather than to $\lieGx{\csp}' \oplus \lieu{\PnotCSdim}$, however.

	Consequently, the $\lieGx{P}'$ of \eqn{eq:lieGxP-from-Pdim-ext} contains elements that are not valid endomorphisms on $\VonPproj$. Not all elements of $\lieGx{P}'$ are antisymmetric or even normal, for instance, based on this construction. Even so, since $\lieGx{P}$ {\em does} consist entirely of valid $G$ endomorphisms, the intersection $\lieGx{P} \cap \lieGx{P}'$ also necessarily has this property. In practice, no basis for $\lieGx{P}'$ is ever computed (see below); it is merely a useful mathematical abstraction.

	The Lie algebraic analog to step~\labelcref{item:P-stat-chan3} is straightforward: Letting
	\begin{equation}
		\label{eq:def-lie-P-star}
		\lieGx{P}^* \isdefas \lieGx{P} \cap \lieGx{P}'
	\end{equation}
	then $\lieGx{P}^*$ is a linear Lie algebra and the Lie group $\metasetn{P}^*$ is given by $\metasetn{P}^* = \exp{\lieGx{P}^*}$. The Lie algebra $\lieGx{\csp}^*$ is subsequently computed by $\CSdim$-erasure, i.e., $\lieGx{\csp}^* \isdefas \ers{\CSdim}{\lieGx{P}^*}$, and it follows that group $\metasetn{\csp}^* = \exp{\lieGx{\csp}^*}$ comprises all $\CH$-realizable unitary channel matrices $\Caa$ operating on $\VonHC$ corresponding to valid logical manipulations of $\sigma(t)$.

	For the L-OC test, step~\labelcref{item:P-stat-chan4} reduces to the determination: Is $\lieGx{\csp}^* = \lieGx{\csp}'$? That is, given $\lieGx{\csp}^*$ is the set of all $\CH$-realizable logical Hamiltonian-like actions, and $\lieGx{\csp}'$ is the set of Hamiltonian-like actions required for encoded universality, are the two sets equal? If so, then the subsystem code $\CS$ is L-OC by the principles discussed in \cref{sec:csops-logops}.

	Combining steps~\labelcref{item:P-stat-chan2}--\labelcref{item:P-stat-chan4} therefore yields the necessary and sufficient condition for L-OC:
	\begin{equation}
		\label{eq:L-OC-cond-proto1}
		\ers{\CSdim}{\left((\ext{\Pdim}{\lieGx{\csp}'}) \cap \lieGx{P}\right)} = \lieGx{\csp}'
	\end{equation}

	By the identity of \eqref{eq:p-ers-m-ext-ident2}, equation \eqref{eq:L-OC-cond-proto1} can be rewritten
	\begin{equation}
		\label{eq:L-OC-cond-proto2}
		\lieGx{\csp}' \cap \ers{\CSdim}{\lieGx{P}} = \lieGx{\csp}'
	\end{equation}

	Finally, for any two sets $\metaset$, $\metasetalt$, the equivalence holds that
	\begin{equation}
		\label{eq:S-subseteq-T-ident}
		\metaset \subseteq \metasetalt \iff \metaset = \metaset \cap \metasetalt
	\end{equation}

	Letting $\metaset = \lieGx{\csp}'$ and $\metasetalt = \ers{\CSdim}{\lieGx{P}}$, then \eqref{eq:L-OC-cond-proto2} is equivalent to
	\begin{equation}
		\label{eq:L-OC-cond-proto3}
		\lieGx{\csp}' \subseteq \ers{\CSdim}{\lieGx{P}}
	\end{equation}
	which becomes \eqn{eq:L-OC-cond} after substitution of \eqref{eq:def-lie-CSp}, thus completing the proof.
\end{proof}

To better illustrate the roles of the many algebras, matrix block mechanics, etc.\ used in the proof of \cref{thrm:control-algebras}, \cref{sec:toyex} provides a concrete, low-dimensional `toy' example with explicit values for all stages.

The condition of \cref{thrm:control-algebras} is more conveniently expressed as a condition on dimensions, analogous to the Lie algebra rank condition of \cref{sec:bg-olc}.

\begin{gcnscor}\label{cor:L-OC-cond-dim}
	Let the assumptions of \cref{thrm:control-algebras} hold. Then $\CS$ is L-OC if and only if
	\begin{equation}
		\label{eq:L-OC-cond-dim}
		\dim\left(\ers{\CSdim}{\CSHtoG(\liesu{\CSord})} \cap \ers{\CSdim}{\lieGx{P}}\right) = \CSdim - 1
	\end{equation}
	where $\lieGx{P} \isdefas \liecls{\ers{\Pdim}{\Gs}}$.
\end{gcnscor}
\begin{proof}
	Proof is provided in~\proofref{L-OC-cond-dim}.
\end{proof}

Necessary and sufficient conditions for L-ESC of $\CS$ follow a similar derivation.

\begin{gcnsprop}\label{prop:L-ESC-cond-dim}
	Let the assumptions of \cref{thrm:control-algebras} hold. Then $\CS$ is L-ESC if either i)~$\CS$ is L-OC, or ii)
	\begin{equation}
		\label{eq:L-ESC-cond-dim}
		\dim \left[i\ers{\CSdim}{\CSHtoG(\RMe[\CSord]{1}{1})},\;\lieGx{\csp}^*\right] = 2(\CSord - 1)
	\end{equation}
	where
	\begin{align}
		\label{eq:L-ESC-cond-dim-liecsps}
		\lieGx{\csp}^* & \isdefas \ers{\CSdim}{\CSHtoG(\liesu{\CSord})} \cap \ers{\CSdim}{\lieGx{P}} \\
		\label{eq:L-ESC-cond-dim-lieGxP}
		\lieGx{P}      & \isdefas \liecls{\ers{\Pdim}{\Gs}}
	\end{align}
	and where $\RMe[\CSord]{1}{1} \in \DonHC$ is the operator $\operatorname{diag}(1,0,\dotsc,0)$.

	Furthermore, if neither (i) or (ii) holds, $\CS$ is not L-ESC.
\end{gcnsprop}
\begin{proof}
	Proof is provided in~\proofref{L-ESC-cond-dim}.
\end{proof}

\subsection{Effective Hamiltonians}
\label{sec:pstatic-effham}

The affine nature of the DIFS $\uinspcs$ in $P$-static control allows for a simplifying abstraction of $\Gs$ that builds the constraints of \cref{cor:compute-DIFS} into the L-OC and L-ESC conditions of \cref{sec:pstatic-liealg}. Specifically, the Lie algebra $\lieGx{P}$ appearing in \cref{cor:L-OC-cond-dim,prop:L-ESC-cond-dim} may be generated in a way directly analogous to $\lieham$ in \eqn{eq:def-lieham}. That is, we wish to compute a set of operators
\begin{equation}
	\label{eq:def-Heffset}
	\CHeff^+ \isdefas \{\Heff_0\} \cup \{\Heffc{k}\}_{k = 1}^{\nceff},\;\;\Heff_0, \Heff_k \in \CM{\Pdim}
\end{equation}
analogous to the set of Hamiltonian generators of \eqn{eq:def-Hset} such that we can write
\begin{equation}
	\label{eq:def-lieeffham}
	\lieGx{P} = \liecls{-i\CHeff^+}
\end{equation}
rather than relying on the more obscure calculation for $\lieGx{P}$ in \eqns{eq:control-algebras-lieGxP,eq:L-ESC-cond-dim-lieGxP}.

The ability to characterize $\lieGx{P}$ in terms of the generators in $\CHeff^+$ both simplifies the computation of a basis for $\lieGx{P}$ and renders the algebra amenable to various reductions and simplifications developed for unconstrained, noiseless open loop control (e.g., various necessary-but-not-sufficient conditions for controllability described in~\cite{cite:qcdbook}), which can also greatly improve computational performance of the tests.

Because of the form of \eqn{eq:def-lieeffham}, $\CHeff^+$ is called the set of {\em effective Hamiltonians}. The endomorphism $\Heff_0$ is called the {\em effective drift Hamiltonian}, and the indexed matrix set $\CHeff \isdefas \{\Heffc{k}\}_{k = 1}^{\nceff}$ is called the set of {\em effective control Hamiltonians}. If a model is noiseless, then $\CHeff^+ = \CH^+$ trivially.

A procedure for computing $\CHeff^+$ from the $\Zu$ and $\znu$ data of \cref{sec:ipc-props} is provided in~\effhamsproc.

\section{Example}
\label{sec:example}

\subsection{Preface}
\label{sec:ex-preface}

The utility of the controllability tests developed in \cref{sec:pstatic} is best demonstrated with a worked example of an open quantum system. However, the outcomes of the tests are of little value unless they can be supported by results showing that logical operations can be performed noiselessly on encoded states. Procuring such results requires, at the very least:

\begin{itemize}
	\item finding one or more $\Pstuple$ triples representing a candidate information encoding, set of control input constraints, and set of control resources for controllability testing
	\item efficiently computing all data needed to perform the controllability tests of \cref{sec:pstatic}, and efficiently checking the conditions of \cref{cor:L-OC-cond-dim} and/or \cref{prop:L-ESC-cond-dim}
	\item synthesizing a control input field $\uin$ able to effect a prescribed logical gating operation on an encoded state, and simulating the model dynamics under the influence of this input
	\item quantifying the ``correctness'' (i.e., closeness between simulated and desired results) in a meaningful way
\end{itemize}

To perform these tasks, we rely on the framework developed in~\koppthesis, called the ``Generalized Control of Noiseless Subspaces (GCNS) Framework''. In this paper, we refer only to the work products of the framework relating to the controllability tests of \cref{sec:pstatic}. A more extensive set of experimental results is viewable in~\exampleappendix.

\subsection{Model}
\label{sec:ex-model}

We consider the class of cold trapped ion systems first proposed by Cirac and Zoller~\cite{cite:cz1995,cite:cz2005,cite:cz2008,cite:james}, which has been employed by dozens of papers and validated in numerous experiments for qubit registers ranging from one to hundreds of qubits~\cite{cite:100sofions,cite:bench11qb,cite:consrv2016,cite:hotiondfs,cite:ion2006,cite:ion2008,cite:iongt2014,cite:iongt2016,cite:ionsim53qb,cite:iontrapdsn,cite:ivanov,cite:monzkim,cite:qfc4cool,cite:robustion,cite:scrambling}. The primary source of decoherence is a known form of collective dephasing~\cite{cite:100sofions,cite:cz2005} which admits a non-trivial noiseless subspace due to the symmetric nature of the dephasing process.

A typical $n$-qubit implementation consists of a string of $n - 1$ ions stored in a linear radiofrequency trap and cooled sufficiently so that the Coulomb forces between them are quantum mechanical rather than classical in nature. The quanta serving as qubits are then realized using internal energy levels of the ions. Although there is some choice as to which states are used, the details ultimately only affect numerical parameters in the system operators and are of little consequence to the control theory. Here we consider the ``single photon'' scheme proposed in~\cite{cite:james}.

Subject to appropriate tuning, the drift Hamiltonian of the system is expressible as
\begin{equation}
	\label{eq:ion-drift-hamiltonian}
	\hat H_{0} = \pi\ionnu\sum\limits_{j = 2}^n \pauli z[j] +
		\frac{\pi\ionmu}{2}\sum\limits_{j = 2}^n \pauli z[1]\pauli z[j]
\end{equation}
where $\ionnu,\ionmu \in \Rl$ and where
\begin{equation}
	\label{eq:def-comp-pauli-operators}
	\pauli z[j] \isdefas
		\underbrace{\pauli i}_1
		\otimes \dotsb
		\otimes \underbrace{\pauli z}_j
		\otimes \dotsb
		\otimes \underbrace{\pauli i}_n
\end{equation}

Here and elsewhere, $\pauli i, \pauli x, \pauli y, \pauli z$ denote $\eye{2}$ and the Pauli X, Y, and Z matrices scaled to unit Frobenius norm. For convenience, we define a set $\Pauli \subset \CM{2}$:
\begin{equation}
	\label{eq:def-Pauli}
	\Pauli \isdefas \{\pauli i, \pauli x, \pauli y, \pauli z\}
\end{equation}

The dominant noise in the system is characterized by a single collective dephasing channel:
\begin{equation}
	\label{eq:ion-noise-channel}
	\LindD[\rho] = \ionGammaZ\mathcal{D}[\ionD\xp\rho]
\end{equation}
where $\ionGammaZ \in \Rp$ and $\ionD \isdefas \sum_{j = 2}^n \pauli z[j]$.

To procure a reasonably rich example, we consider an $n = 5$ -qubit system, and set $\ionnu$ and $\ionmu$ to unequal nonzero values, chosen to be $\ionnu = \frac{19}{3}$ and $\ionmu = \frac{8}{5}$ so that the dynamics have no periodicity to exploit.

\Hiii{Control Resources}%
Control is exerted over the system by manipulating the intensity of lasers targeted at the component ions. Provided all intensities can be independently actuated, admissible controls are drawn from a pool of Heisenberg-type Hamiltonian terms~\cite{cite:heisham} characterized by
\begin{equation}
	\label{eq:def-crpaH}
	\crpaH[n] \isdefas \setbld{\langle\ionop 1\ionop 2\dotsm\ionop n\rangle}{\ionop k \in \Pauli \Forall k \inoneto n}
\end{equation}
where the notation $\langle\ionop 1\ionop 2\dotsm\ionop n\rangle$ is shorthand for the Kronecker product $\ionop 1 \otimes \ionop 2 \otimes \dotsb \otimes \ionop n$.

However, not all terms in $\crpaH[n]$ can be effected. Specifically, the action on the bus qubit is fully determined by the action on the remaining $n - 1$ qubits. To model field constraints such as those in~\cite{cite:100sofions,cite:cz2005,cite:dfsuqcadi,cite:hotiondfs,cite:ionsim53qb,cite:ivanov,cite:james,cite:qcscale,cite:robustion}, it is sufficient to define a map $\ionq \in \Mapset{\Pauli}{\Z}$:
\begin{equation}
	\label{eq:def-ionq}
	\ionq(\ionopx) \isdefas \left\{ \begin{array}{lll}
		\ump 1 & \text{if} & \ionopx \in \{ \pauli i, \pauli x \} \\
		\ump 0 & \text{if} & \ionopx = \pauli y \\
		    -1 & \text{if} & \ionopx = \pauli z
	\end{array} \right. \Forall \ionopx \in \Pauli
\end{equation}
and impose the condition on the bus qubit (qubit $1$):
\begin{equation}
	\label{eq:ion-bus-q-cond}
	\ionq(\ionop 1) = \sum\limits_{j = 2}^n \ionq(\ionop j)\kern16pt (\operatorname{mod} 4)
\end{equation}

Furthermore, we suppose that unwanted parasitic coupling to higher-order bus modes occurs when either $\ionop 1 = \pauli i$ or $\ionop 1 = \pauli x$, hence we impose the additional condition that $\ionq(\ionop 1) \not\equiv 1\;(\operatorname{mod} 4)$. In conjunction with \eqns{eq:def-crpaH,eq:ion-bus-q-cond}, this results in a set of potential Hamiltonian terms:
\begin{equation}
	\label{eq:ion-all-set}
	\def\LGivenR#1#2{\;\,#1}
	\crpaset = \setbldst{\langle\ionop 1\ionop 2\dotsm\ionop n\rangle \in \crpaH[n]}{}
		{\ionq(\ionop 1) \equiv \sum\limits_{j = 2}^n \ionq(\ionop j) \equiv \{-1,0\}\;(\operatorname{mod} 4)\;}
\end{equation}
having order
\begin{equation}
	\label{eq:ion-all-set-order}
	|\crpaset| = \tfrac{1}{8} 4^n + 2^{n/2 - 1}\cos(\tfrac{1}{4}\pi n)
\end{equation}
which in the case of $n = 5$ yields $|\crpaset| = 126$ potential control resources.

The control resource sets considered for testing (i.e., potential values of $\CH$) are the nonempty subsets of $\crpaset$ having the properties:

\begin{enumerate}
	\item $\CH$ may not contain more than $10$ resources or give rise to more than $4$ effective input channels; that is, $\nc \leq 10$ and $\nceff \leq 4$.
	\item Let the ``$\ionq$-signature'' $\ionq(\Hcx)$ of a control resource $\Hcx \in \crpaset$ be the list of $\ionq$ values of $\Hcx$'s $\ionopx$ operands in positions $2$ to $n$, sorted into ascending order. For example,
	\begin{align*}
		\ionq(\pauli z\pauli x\pauli y\pauli x\pauli x) & = \{0,1,1,1\} \\
		\ionq(\pauli y\pauli i\pauli z\pauli z\pauli x) & = \{-1,-1,1,1\}
	\end{align*}

	Then $\CH$ must have the property that any resource acting as $\pauli y$ on the bus qubit must have a $\ionq$-signature distinct from all other resources acting as $\pauli y$ on the bus qubit. Symbolically,
	\begin{equation}
		\label{eq:ion-bus-action-parity-rule}
		\begin{gathered}
			(\Foreach \Hcx \in \CH,\;\ionop 1(\Hcx) = \pauli y) \sppredtostmt
				\nexists\;\Hcx' \in \CH\setminus\{\Hcx\}\hfill \\
			\speqinset\sttext (\ionop 1(\Hcx') = \pauli y) \;\land\; (\ionq(\Hcx) = \ionq(\Hcx'))
		\end{gathered}
	\end{equation}
	where $\ionop 1(\Hcx)$ denotes the $\ionop 1$ operator of $\Hcx$.
\end{enumerate}

The reasoning behind the inclusion of these rules is provided in~\exampleappendix. Stated generally, the rules are included to improve the complexity and richness of the example. The number of control resource sets meeting all criteria is approximately $4.9\times 10^{22}$.

\subsection{Controllability Results}
\label{sec:ex-conres}

A $5$-qubit Bloch ball basis~\cite{cite:bbbasis} is used for the linear ODE model of \eqns{eq:cvs-dynamics,eq:covec-to-phys-state}. The order-$1{,}023$ drift matrix is computed for an arbitrary dephasing strength of $\Gamma_Z = \frac{10}{3}\pi$. The channel time is chosen to be $T = 3\Gamma_Z^{-1}$. This choice of parameters is performance-related; details are provided in~\exampleappendix.

The algebraic structure of the noise commutant is given as:
\begin{align}
	\label{eq:ion-nc-dim}
	\NCdim & = 280 \\
	\label{eq:ion-nc}
	\NC    & \sim \CM{2} \oplus \CM{2} \oplus \CM{8} \oplus \CM{8} \oplus \CM{12}
\end{align}

These $k$-sectors are enumerated $1$ to $5$ from left to right. By \cref{lem:subsystem-code-upper-bound}, the largest noiseless subsystem code has order $\CSord = 12$, holding equivalent information to two qubits and a qutrit.

The search for candidate subsystem codes is restricted to $k$-sector ID $k = 3$, which has order $\kord[3] = 8$. Additionally, we restrict the set of admissible subsystem codes to those where $4 \leq \CSord \leq 7$. We use the L-OC controllability standard when assessing controllability.

Under its standard search parameters, the GCNS framework yields $34$ controllable $\Pstuple$ triples. Among these, we select a triple where
\begingroup
\newcommand {\pppppauli} [5] {\pauli{#1}\pauli{#2}\pauli{#3}\pauli{#4}\pauli{#5}}
\begin{equation}
	\begin{gathered}
		\CH = \{\Hc{k}\}_{k = 1}^7 \\
		\Hc{1} = \pppppauli yxixi \speqsidebyside \Hc{2} = \pppppauli yxzxz \\
		\Hc{3} = \pppppauli yzzzz \speqsidebyside \Hc{4} = \pppppauli ziiyx \\
		\Hc{5} = \pppppauli ziyix \speqsidebyside \Hc{6} = \pppppauli zzyzx \\
		\Hc{7} = \pppppauli zzzyx
	\end{gathered}
\end{equation}
\endgroup
and where $\Pproj = \kproj[3]$, which is a drift-invariant subsystem projection with a DIFS
\begin{equation}
	\begin{gathered}
		\uinspcs = \setbld{\uin = \zNu\ueff}{\ueff \in \Mapset{\Rp}{\Rn{\nceff}}} \\
		\zNu = -\frac{1}{4}\left[ \begin{smallmatrix}
			\sqrt{2} & 0 &        0 &        0 \\
			\sqrt{2} & 0 &        0 &        0 \\
			       0 & 4 &        0 &        0 \\
			       0 & 0 & \sqrt{2} &        0 \\
			       0 & 0 & \sqrt{2} &        0 \\
			       0 & 0 &        0 & \sqrt{2} \\
			       0 & 0 &        0 & \sqrt{2}
			\end{smallmatrix} \right]
	\end{gathered}
\end{equation}
having $\nc = 7$ control inputs, $\nceff = 4$ effective degrees of control freedom, and with $\Pdim = 63$.

We choose this triple specifically because it is the first returned candidate where the most obvious choice of subsystem code---letting $\CS$ occupy the entire $k$-sector---proves to be neither L-OC nor L-ESC. It can be shown that the Lie algebra $\lieGx{P}$, which is equivalent to $\lieGx{\csp}$ in such a case, has dimension $36$, hence automatically fails the L-OC test of \cref{cor:L-OC-cond-dim}, and furthermore fails the L-ESC test of \cref{prop:L-ESC-cond-dim}.

Even so, several order-$4$ (two-qubit) subsystem codes residing in the $k$-sector prove to be L-OC. We designate one such subsystem code $\CS_{1368}$ and present its full tuple in~\exampleappendix. Among the pertinent parameters, the subsystem code has order $\CSord = 4$, dimension $\CSdim = 16$, and multiplicity of $1$.

As part of controllability analysis, a basis is computed for the Lie algebra $\lieGx{\csp}^*$, comprising $15$ order-$16$ matrices. From this we conclude $\dim \lieGx{\csp}^* = 15 = \CSdim - 1$, and by \cref{cor:L-OC-cond-dim}, $\CS$ is L-OC. It follows that we can perform any two-qubit operation on the logical state $\sigma \in \DonHC$ noiselessly via control through the specified input channels.

It is also worth noting that the pair $(\CS_{1368},\CSprojn{1368})$ yields a $\lieGx{\csp}$ Lie algebra of dimension $2$ and therefore is {\em not} L-OC, implying the system state must circulate in a domain strictly larger than $\VonHC$ in order for the system to be controllable. Stated differently, the physical state of the system must circulate beyond the domain where it has a valid correspondence to a logical state in order for the encoded subsystem to be controllable. To the best of our knowledge, this is the first time this kind of strict ``subcontrollability'' of a noiseless subsystem has been demonstrated in the literature.

\subsection{Simulation Results}
\label{sec:ex-simres}

As a final validation of the controllability assessment, we synthesize a control input field $\uin$ to generate an arbitrary logical gate $\loggtgt$ in $\CS_{1368}$ and quantify the difference between $\loggtgt$ and the logical gate generated by system dynamics subject to $\uin$, which is ideally $\approx 0$. The procedure for synthesizing control input fields for $P$-static control schemes is presented in Chap.~6 of~\koppthesis. In this paper, we present only an analysis of the error in the generated output. A full and detailed record of the outputs is included in~\exampleappendix.

The target gate $U_\text{tgt}$ can be any two-qubit unitary gate, and we select the well-known \gcnsgate{CNOT} (a.k.a.~controlled-NOT, \gcnsgate{CX}, controlled-X) gate, i.e.,
\begin{equation}
	\label{eq:def-CNOT-gate}
	\gcnsgate{CNOT} \isdefas \stdmat{1 & 0 & 0 & 0 \\ 0 & 1 & 0 & 0 \\ 0 & 0 & 0 & 1 \\ 0 & 0 & 1 & 0}
\end{equation}
for its familiarity.

The objective of control synthesis is to effect any transformation of the physical state such that for any initial logical state $\sigma_0 \in \DonHCn{1368}$, after evolving under the influence of control for $T$ normalized time units, the physical state is equivalent to the logical state $\sigma_T = \loggtgt\sigma_0\adj\loggtgt$, analogous to the physical transformation of \eqn{eq:liouville-general-dynamics2}. We furthermore restrict all input channels to continuous curves bounded by
\begin{equation}
	\label{eq:ion-uin-hard-limit}
	\norm[inf]{\uink k} \leq 250 \Forall k \inoneto 7
\end{equation}
to simulate hard limits on actuator magnitude.

In practice, due to limited solver precision and roundoff error, the software returns a set of positive coefficients $\{\epsilon_k\}_k$ and $4\times 4$ matrices $\{\hat{E}_k\}_k$ (called ``Kraus form'' operators in~\cite{cite:krausops}) having unit operator norm such that
\begin{equation}
	\label{eq:ion-gen-chan-expr}
	\sigma_T = \sum\limits_k \epsilon_k\adj{\hat{E}}\sigma_0 \hat{E}_k
\end{equation}
where $\adj{\hat{E}_1} \approxeq \loggtgt$, $\epsilon_1 \approxeq 1$ and $\epsilon_k \approxeq 0 \Forall k \geq 2$.

The inherent imprecision can be quantified via the operator norm error:
\begin{equation}
	\label{eq:def-cFit}
	\cFit \isdefas \tfrac{1}{2}\maxover{\sigma_0 \in \DonHC}
		\norm[2]{\loggtgt\sigma_0\adj\loggtgt\adj\sigma_T - \eye{4}}
\end{equation}
representing the maximum possible discrepancy between $\sigma_T$ and its proper (ideal) value, which we find to be $\cFit = 2.14\times 10^{-5}$ in this case~\footnote{Note that error is quantified in a more sophisticated way in~\exampleappendix\ and the figure $\cFit = 2.14\times 10^{-5}$ does not appear {\em per se}. However, for comparison purposes, $\min\{\cFitg,\cFitc\} \leq \cFit \leq \cFitg + \cFitc$.}. 

\Cref{tab:fitting-error} provides a more common and visually descriptive representation: a $4\times 4$ table displaying the target value ($\loggtgt rc$) and error ($e_{(r,c)}$) for the elements of the target operator matrix $\loggtgt$, where
\begin{equation}
	e_{(r,c)} = |\loggtgt rc - (\adj{\hat E_1})_{(r,c)}| \Forall r,c \inoneto 4
\end{equation}

The table clearly shows the structure of the \gcnsgate{CNOT} gate as well as the small magnitude of the error. This, along with explicit values for $\uin$, $T$, and the $\hat{E}_k$ operators (also provided in~\exampleappendix), confirms the controllability analysis even to the extent of synthesizing a complete solution for a specific computational objective.

\begin{figure}[htb]
	\centering
	\def\vece#1#2{$\FPeval\x{round(#1:2)}\x\textsc{e}{#2}$}  
	\begin{tabular}{
		c@{\hspace{2.5mm}}
		rl@{\hspace{3mm}}
		rl@{\hspace{3mm}}
		rl@{\hspace{3mm}}
		rl
	}
		\toprule
		{} & \multicolumn{8}{c}{{\bf Col}} \\
		{\bf Row} &
			\multicolumn{2}{c}{1} &
			\multicolumn{2}{c}{2} &
			\multicolumn{2}{c}{3} &
			\multicolumn{2}{c}{4} \\
		1  &  {\bf 1} & \vece {6.4208} {-6}  &  {\bf 0} & \vece {5.0886} {-6} &
		      {\bf 0} & \vece {8.7239} {-7}  &  {\bf 0} & \vece {1.9135} {-6} \\
		2  &  {\bf 0} & \vece {5.0886} {-6}  &  {\bf 1} & \vece {3.8966} {-6} &
		      {\bf 0} & \vece {3.8247} {-6}  &  {\bf 0} & \vece {2.1331} {-6} \\
		3  &  {\bf 0} & \vece {1.9135} {-6}  &  {\bf 0} & \vece {2.1330} {-6} &
		      {\bf 0} & \vece {6.3788} {-6}  &  {\bf 1} & \vece {7.1408} {-6} \\
		4  &  {\bf 0} & \vece {8.7242} {-7}  &  {\bf 0} & \vece {3.8247} {-6} &
		      {\bf 1} & \vece {1.9417} {-6}  &  {\bf 0} & \vece {6.3788} {-6} \\
		\botrule
	\end{tabular}
	\caption{%
		Values (bold) for all elements of the target $\loggtgt$ operator matrix, and errors ($e_{(r,c)}$) for all elements in the generated $\adj{\hat E_1}$ operator matrix%
	}
	\label{tab:fitting-error}
\end{figure}

\section{Future Work}
\label{sec:future}

As noted in \cref{sec:example}, the controllability tests developed in this paper are primarily useful as components of a larger framework for the automated design of noiseless information encoding and control in open quantum systems. In such a framework, a designer provides a model description, e.g., in the form of the Lindbladian operators of \eqn{eq:lindbladian-dynamics}, as well as operating specifications (e.g., minimum dimension of subsystem code, classes of input signal, etc.). From these data, the framework returns a control solution for the given model and specifications, without need of user supervision or model-specific subroutines.

The design of such a framework is beyond the scope of this paper. We note, however, that in addition to the controllability tests developed in this paper, three additional components are required to complete the framework:
\begin{enumerate}
	\item automated, computationally efficient procedures for procuring candidate protective $\Pstuple$ triples for $P$-static controllability tests, as well as candidate parameterizations for other classes of noiseless control modalities
	\item automated procedures for synthesizing the control input fields (i.e., $\uin$) needed to effect all required gating operations on the logical state; this is also called ``gate synthesis'', ``quantum optimal control'', and the ``global nonlinear inverse problem''~\cite{cite:gnlisem} and is the subject of extensive research~\cite{cite:conland1,cite:consrv2016,cite:gnliland,cite:gnlisem,cite:optcconst,cite:qcopt2016}
	\item procedures for evaluating accessibility~\cite{cite:lwcreply,cite:notrap2017} (also called ``global surjectivity''~[ibid.], ``reachability''~\cite[chap.~3]{cite:geoconbook}, and ``attainable sets''~\cite[\S1.6.2]{cite:bilinbook}), which is the problem of characterizing the reachable subset of gating operations under time and amplitude constraints; as well as procedures for automated calibration of input field magnitude and duration
\end{enumerate}

Future work will focus on each of these components, as well as the framework {\em in toto}. Work may also be devoted to improving control performance in the presence of model inaccuracy via quantum feedback control~\cite{cite:qfcsum2017}, which acts as an additional noise operator in the noise channel $\LindD$ of \eqn{eq:lindbladian-dynamics}~\cite{cite:1qdynfeed,cite:pcebyqfc,cite:suppdecoh}. Although quantum continuous measurement cannot be used for the purpose of manipulating the state within the noiseless subspace, it can be used to detect and counteract state drift outside the subspace~\cite{cite:mkvstabdfs,cite:dfsiac} due to model and control error.

Finally, computationally efficient algorithms for evaluating the test conditions of \cref{cor:L-OC-cond-dim,prop:L-ESC-cond-dim} as well as an analysis of the complexity of these algorithms (which is of particular importance to issues of speed and scalability) will be included in future work.

\section{Conclusions}
\label{sec:conclusions}

In this paper, we have presented two controllability tests for subsystem codes embedded in DFSs, which address the issue of controllability in the case where the physical state of the system is permitted to circulate outside of the set of logically-encoded states. This greater design freedom allows a controller to effect a greater range of noiseless computations using a given set of control resources. The tests are applicable to the set of all open quantum systems subject to the Born-Markov approximation. They are evaluated using straightforward computations on matrix algebras derived from model operators.

The tests are useful as standalone analysis tools when working with control in DFSs or as part of broader automated workflows for generating complete control solutions for any model in the supported class of quantum models. The workflows in turn facilitate greater flexibility and faster iteration in control design.

In developing the tests, we have presented a set of canonical matrix forms, developed a theory of information-preserving control from first principles using these matrix forms, and provided all mathematical proofs and major theorems underlying a class of control scheme called ``$P$-static control''.

A proof-of-concept for the proposed tests was provided based on a well-studied trapped ion model subject to nonstandard collective dephasing, which admits multiple operator controllable two-qubit noiseless subsystem codes. Data from all steps in a controllability assessment of the model using the new tests were included as part of the example.

\section{Acknowledgment}
\label{sec:ack}

The authors acknowledge and appreciate support by the National Sciences and Engineering Research Council of Canada (NSERC), which contributed to the development of this work.

\appendix
\section{\texorpdfstring{Example for \cref*{thrm:control-algebras}}{Illustrative Example}}
\label{sec:toyex}

As a supplement to the proof of \cref{thrm:control-algebras}, we provide explicit values for all stages of the proof for a illustrative `toy' system with $\Bdim = \Bord^2 = 16$. The example demonstrates all salient aspects of the theorem but is contrived to use operator algebras of dimension $\leq 7$ representable by matrix operators of order $\leq 16$.

In the case of order-$16$ matrices, only the top left $10\times 10$ submatrices are shown for the sake of compactness. This truncation does not affect the example, excepting that the rightmost and bottommost blocks would normally extend an additional six columns/rows rightward and downward. Additionally, the example begins with the $G$ endomorphisms $\{G_0,\Gc 1,\Gc 2\}$ rather than the operators of the Lindbladian representation, since the former have been contrived to have particularly compact and simple matrix representations.

For all operator definitions, let
\begin{align*}
	     \rtwo & \isdefas \sqrt{2}  &     \rhalf & \isdefas \sqrt{\tfrac{1}{2}}  &
	\rtwophalf & \isdefas \sqrt{\tfrac{1}{2}} + \tfrac{1}{2} \\
	  \rtwopsq & \isdefas \sqrt{\tfrac{1}{2}} + \tfrac{7}{4}
	                                &  \rtwomtwo & \isdefas \sqrt{2} - 2
\end{align*}

As a starting point, we consider endomorphisms $G_0$~\eqn{eq:def-toyUi}, $\Gc 1$~\eqn{eq:def-toyUii}, and $\Gc 2$~\eqn{eq:def-toyUiii}. We assert that the $\hat F_j$ basis vectors are selected such that the endomorphisms are in G-C3F with respect to a protective $\CPtuple$ pair having $\CSdim = 4$, $\Pdim = 8$.
\begin{gather}
	\label{eq:def-toyUi}   \toyUi   \\
	\label{eq:def-toyUii}  \toyUii  \\
	\label{eq:def-toyUiii} \toyUiii
\end{gather}

The equations display the $9$ matrix blocks of the G-C3F partitioning. The $\Gac$ and $-\adj{\Gac}$ blocks of all three matrices are nonzero, hence control action is needed to prevent coupling between $\VonHC$ and the unprotected subspace of the system. The most direct way to accomplish this is to select a DIFS $\uinspcs$ such that $\VonHC$ is invariant [or, equivalently, by setting $\Gac \equiv 0$ and $\Gab \equiv 0$ per \eqn{eq:CQF-pci-condition}]. However, although such a $\uinspcs$ exists, all control freedom is exhausted by the decoupling. The DIFS is the singleton set
\begin{equation}
	\uinspcs = \left\{\uin(t) \equiv \stdmat[r]{\tfrac{3}{4} \\[5pt] -\tfrac{1}{2}}\right\}
\end{equation}
and the $G(t)$ matrix is fixed at the value of \eqn{eq:def-toyGCBDi}. Hence, control is able to isolate the encoded state from noise but is not able to effect any useful transformation.

\begin{equation}
	\label{eq:def-toyGCBDi}
	\toyGCBDi
\end{equation}

A remedy is to allow the state to circulate outside of $\VonHC$ inside the broader protective subspace $\VonPproj$ where $\Pdim = \rank P = 8$ for the toy system. In this case, the DIFS decouples $\VonPproj$ from the unprotected subspace of the system [or, equivalently, by setting $\Gac \equiv 0$ and $\Gbc \equiv 0$, since all generators are in G-C3F wrt $\CPtuple$]. The required DIFS is
\begin{equation}
	\begin{gathered}
		\uinspcs = \left\{\uin(t) = \stdmat[r]{0 \\[5pt] 1} + \stdmat[r]{-1 \\[5pt] 2}\cdot\ueffknv 1(t)\;\right|
			\speqinset \hfill \\
		\hfill \left.\vphantom{\stdmat[r]{0 \\[5pt] 1}}\ueffknv 1 \in \Mapset{\Rnn}{\Rl}\right\}
	\end{gathered}
\end{equation}
resulting in a linear (and $\Pdim$-diagonal) subspace $\Gs$ spanned by the two basis vectors:
\begin{gather}
	\label{eq:def-toyGBDi}  \toyGBDi  \\
	\label{eq:def-toyGBDii} \toyGBDii
\end{gather}

We note that the $\GQaa$ (upper left) blocks of \eqns{eq:def-toyGBDi,eq:def-toyGBDii} are equal to $i\Heff_0$ and $i\Heffc 1$, i.e., the effective Hamiltonians of the system. These effective Hamiltonians generate a $7$-dimensional Lie algebra $\lieGx{P}\;(= \liecls{\ers{\Pdim}{\Gs}})$ spanned by the basis vectors:
\begin{gather}
	\label{eq:def-toyAi}   \toyAi   \\
	\label{eq:def-toyAii}  \toyAii  \\
	\label{eq:def-toyAiii} \toyAiii \\
	\label{eq:def-toyAiv}  \toyAiv  \\
	\label{eq:def-toyAv}   \toyAv   \\
	\label{eq:def-toyAvi}  \toyAvi  \\
	\label{eq:def-toyAvii} \toyAvii
\end{gather}

The Lie sub-algebra of $\lieGx{P}$ that leaves $\VonHC$ invariant is given by the $\CSdim$-diagonal subspace of $\lieGx{P}$, which is a $3$-dimensional subspace spanned by basis vectors:
\begin{gather}
	\label{eq:def-toyABDi}   \toyABDi   \\
	\label{eq:def-toyABDii}  \toyABDii  \\
	\label{eq:def-toyABDiii} \toyABDiii
\end{gather}

Since the code ($\Gaa$) blocks of \eqnrg{eq:def-toyABDi}{eq:def-toyABDiii} are clearly linearly independent, the Lie algebra $\lieGx{\csp}$, which is the $\CSdim$-erasure of $\lieGx{P}$, is clearly also $3$-dimensional, and spanned by the $\Gaa$ blocks, i.e.,
\begin{align}
	\label{eq:def-toyGCi}   & \toyGCi   \speqsidebyside
	                          \toyGCii  \\
	\label{eq:def-toyGCiii} & \toyGCiii
\end{align}
such that $\dim\lieGx{\csp} = 3$.

As a final stage, $\lieGx{\csp}$ is intersected with the Lie algebra $\lieGx{\csp}'$ [as defined in \eqn{eq:def-lie-CSp}] of logical Hamiltonian-like dynamics to obtain $\lieGx{\csp}^*$, for use in \cref{cor:L-OC-cond-dim}. This ordinarily requires the $\CSHtoG\convclr$ map, which is representable as a $\CSord\times\CSord \times \Bdim\times\Bdim$ ($= 1{,}024$-element) $4$-tensor---too large to include in this paper. Instead, we provide the map
\begin{equation}
	\label{eq:def-toychimapsym}
	\toychimapsym \hat X \isdefas \ers{\CSdim}{\CSHtoG \hat X}
\end{equation}
which can efficiently be obtained by, e.g., the procure described in~\koppthesis, such that $\lieGx{\csp}' = \toychimapsym\left(\liesu{\CSord}\right)$.

In the case of the toy system, the required map is expressible as
\begin{align}
	\label{eq:def-toychimap}
	\toychimapsym \hat X & = \vtorinv{\toychimatsym\vtor{\hat X}} \Forall \hat X \in \HonHC \\
	\label{eq:def-toychimat}
	\toychimatsym        & = \toychimat
\end{align}
where $\vtor{\cdot}$ denotes the standard vectorization of a square, order-$n$ matrix into an $n^2$-element vector, and $\vtorinv{\cdot}$ is the inverse operation.

By passing a basis for $\liesu{\CSord}$ into~\eqn{eq:def-toychimap}, it can readily be determined that $\lieGx{\csp}^* = \lieGx{\csp}' = \lieGx{\csp}'$. It follows that $\dim \lieGx{\csp}^* = 3 = \CSdim - 1$, and therefore, by \cref{cor:L-OC-cond-dim}, that the toy system meets the L-OC controllability standard for the proposed $P$-static control scheme.

\vspace{6mm}  
\section*{References}
\label{sec:refs}
%


\end{document}